\documentclass[12pt]{article}
\pdfoutput=1
\usepackage{amssymb}
\usepackage{amsmath}
\usepackage{latexsym}
\usepackage[usenames]{color}
\usepackage{cite}
\usepackage{setspace}
\usepackage{mathtools}
\usepackage{simplewick}
\numberwithin{equation}{section}
\definecolor{darkblue}{cmyk}{0.9,0.9,0,0}
\definecolor{darkred}{rgb}{0.6,0,0.3}
\usepackage{graphicx} 
\usepackage{epsfig}
\usepackage[setpagesize=false,pagebackref=false, linktocpage, bookmarksopen=true, colorlinks=true, linkcolor=darkblue,citecolor=darkblue,urlcolor=darkblue]{hyperref}
\usepackage{hyperref}

\newcommand{\tr}{{\rm tr}}

\def\del{\partial}

\def\fn#1{\footnote{#1}}
\def\nn{\nonumber}
\def\eqref#1{(\ref{#1})}
\def\comma{\,,}
\def\period{\,.}
\def\bb{\hat{O}}
\def\dd{\mathcal{O}}
\def\llangle{\langle\!\langle}
\def\rrangle{\rangle\!\rangle}
\def\dmatrix#1#2{\left| 
\begin{array}{#1}
#2\end{array} 
\right|}
\def\NO#1{:\!#1\!:\,}
\def\NOB#1{\typecolon #1\typecolon}
\newcommand{\beq}{\begin{equation}}
\newcommand{\eeq}{\end{equation}}
\DeclareFontEncoding{LS1}{}{}
\DeclareFontSubstitution{LS1}{stix}{m}{n}
\DeclareSymbolFont{symbols2}{LS1}{stixfrak}{m}{n}
\DeclareMathSymbol{\typecolon}{\mathbin}{symbols2}{"25}
\begin{document}
\thispagestyle{empty}

\renewcommand{\thefootnote}{\fnsymbol{footnote}}
\setcounter{page}{1}
\setcounter{footnote}{0}
\setcounter{figure}{0}
\begin{flushright}
PUTP-2572 \\
\end{flushright}
\begin{center}
$$$$
{\large
{More Exact Results in the Wilson Loop Defect CFT: Bulk-Defect OPE, \\Nonplanar Corrections and Quantum Spectral Curve}\par}

\vspace{1.6cm}

\textrm{Simone Giombi\fn{\tt sgiombi AT princeton.edu}, Shota Komatsu\fn{\tt shota.komadze AT gmail.com}}
\\ \vspace{2cm}
\footnotesize{\textit{
$^{\ast}$ Department of Physics, Princeton University, Princeton, NJ 08544, USA
\vspace{1mm} \\
$^{\dagger}$ School of Natural Sciences, Institute for
Advanced Study, Princeton, NJ 08540, USA
}  
\vspace{4mm}
}

\par\vspace{1.5cm}

\textbf{Abstract}\vspace{2mm}
\end{center}
We perform exact computations of correlation functions of $1/2$-BPS local operators and protected operator insertions on the 1/8-BPS Wilson loop in $\mathcal{N}=4$ SYM. This generalizes the results of our previous paper \href{http://arxiv.org/abs/1802.05201}{{\tt arXiv:1802.05201}}, which employs supersymmetric localization, OPE and the Gram-Schmidt process. In particular, we conduct a detailed analysis for the $1/2$-BPS circular (or straight) Wilson loop in the planar limit, which defines an interesting nontrivial defect CFT. We compute its bulk-defect structure constants at finite 't Hooft coupling, and present simple integral expressions in terms of the $Q$-functions that appear in the Quantum Spectral Curve---a formalism originally introduced  for the computation of the operator spectrum. The results at strong coupling are found to be in precise agreement with the holographic calculation based on perturbation theory around the AdS$_2$ string worldsheet, where they correspond to correlation functions of open string fluctuations and closed string vertex operators inserted on the worldsheet. Along the way, we clarify several aspects of the Gram-Schmidt analysis which were not addressed in the previous paper. In particular, we clarify the role played by the multi-trace operators at the non-planar level, and confirm its importance by computing the non-planar correction to the defect two-point function. We also provide a formula for the first non-planar correction to the defect correlators in terms of the Quantum Spectral Curve, which suggests the potential applicability of the formalism to the non-planar correlation functions.
\noindent

\setcounter{page}{1}
\renewcommand{\thefootnote}{\arabic{footnote}}
\setcounter{footnote}{0}
\setcounter{tocdepth}{2}
\newpage
\tableofcontents

\parskip 5pt plus 1pt   \jot = 1.5ex

\newpage
\section{Introduction\label{sec:intro}}
Wilson loops are important observables in gauge theories: They describe the coupling between a heavy probe particle and gauge fields, and are an efficient tool for distinguishing different phases of the theory. In $\mathcal{N}=4$ supersymmetric Yang-Mills theory, one can also consider supersymmetric generalizations of the Wilson loop which couple to the scalar field as well as to the gauge field. Such Wilson loops---in particular the 1/2-BPS Wilson loop, which preserves a maximal amount of supersymmetries---played a significant role since the early days of the AdS/CFT correspondence \cite{Maldacena:1998im,Rey:1998ik}. On the gauge theory side, the exact expectation value of the $1/2$-BPS Wilson loop was computed first by resumming a class of ladder diagrams in \cite{Erickson:2000af,Drukker:2000rr}. This computation was later justified by a rigorous argument based on supersymmetric localization in \cite{Pestun:2009nn}. On the string theory side, the leading strong coupling behavior can be computed by evaluating the regularized area of the minimal worldsheet surface anchored on the Wilson loop at the boundary \cite{Berenstein:1998ij,Drukker:1999zq}. The perfect matching with the strong coupling limit of the exact result \cite{Erickson:2000af,Drukker:2000rr}  provided one of the first important evidences for the existence of the holographic gauge/string duality. Subleading corrections at strong coupling may also be computed by evaluating the partition function for the quantum fluctuations around the classical string solution \cite{Drukker:2000ep}. Very recently, a precise match between the localization prediction and the one-loop term on the string theory side was obtained \cite{Medina-Rincon:2018wjs}.  

The study of the $1/2$-BPS Wilson loop recently gained new interest also from the point of view of conformal defects. Indeed the $1/2$-BPS Wilson loop, being defined on a straight line or circular contour, preserves a SL$(2,R)$ subgroup of the full conformal group \cite{Drukker:2006xg}. Owing to this fact, it can be viewed as an example of defect conformal field theory and has been studied from various perspectives. At weak coupling, the correlation functions of insertions on the Wilson loop were computed in \cite{Cooke:2017qgm,KiryuToAppear}, while the correlators of ``defect-changing operators'' which change the scalar coupling of the $1/2$-BPS Wilson loop were analyzed in \cite{Kim:2017sju}. At strong coupling, an extensive study of the correlators of the insertions was performed in \cite{Giombi:2017cqn} by using perturbation theory around the AdS$_2$ worldsheet. Furthermore the generalization to the non-supersymmetric loop which interpolates between the $1/2$-BPS Wilson loop and the standard Wilson loop was explored in \cite{Beccaria:2017rbe,Beccaria:2018ocq,Correa:2018fgz} both at weak and strong coupling.

In the previous paper \cite{Giombi:2018qox}, we showed that a certain class of correlators on this defect CFT\fn{The computation in \cite{Giombi:2018qox} applies to more general $1/8$-BPS Wilson loops. However, the relation to the defect CFT exists only for the $1/2$-BPS Wilson loop.} can be computed exactly by using the combination of supersymmetric localization, OPE and the Gram-Schmidt process. The results of the computation depend nontrivially on the coupling constants and give an infinite family of defect CFT data including the structure constants of the defect BPS primaries of arbitrary lengths. Such data would provide crucial inputs for performing further analysis, for instance in the context of the conformal bootstrap \cite{Liendo:2016ymz,Liendo:2018ukf}. 

The results described above mostly concern the correlators of insertions inside the Wilson loop. However, from the point of view of the defect CFT, there is yet another important class of observables: the correlation functions between the ``bulk'' local operators defined outside the Wilson loop and the defect operators inserted on the Wilson loop. Such correlators play a central role in formulating the defect crossing equation \cite{Billo:2016cpy,Gadde:2016fbj}, and allow us to connect the defect CFT data and the CFT data in the bulk. A special example of this, which has already been studied before, is the case of correlation functions of local operators and supersymmetric Wilson loop with no insertions \cite{Berenstein:1998ij, Semenoff:2001xp, Zarembo:2002ph, Bassetto:2009rt, Giombi:2009ds, Giombi:2012ep}: in the defect CFT language, this gives the bulk-defect OPE coefficient of the bulk operator and the identity insertion on the defect.  

The main goal of this paper is to generalize the analysis in \cite{Giombi:2018qox} to such bulk-defect correlators: More precisely we consider the correlators of the scalar insertions $\tilde{\Phi}^{L}$ on the Wilson loop and a single-trace operator ${\rm tr}[\tilde{\Phi}^{J}]$ defined outside the Wilson loop (the case of several bulk insertions can also be obtained from our methods). The scalar $\tilde{\Phi}$ is a position-dependent linear combination of the scalar fields which is chosen so that the correlator becomes independent of the positions. It has another important property that the insertion of single $\tilde{\Phi}$ is related via localization to an infinitesimal deformation of the Wilson loop. This property allows us to compute the correlators involving $\tilde{\Phi}$'s by the area derivatives of the expectation values of the Wilson loop, or of the correlator of the Wilson loop and the local operators, both of which are computable from localization \cite{Pestun:2009nn,Giombi:2009ds,Giombi:2012ep}. As we explain in this paper, the computation can then be generalized to the correlators involving insertions of higher charges $\tilde{\Phi}^{L}$'s ($L\geq 2$) with the help of OPE and the Gram-Schmidt analysis.

Although the general framwork we present can be applied to the correlators at finite $N_c$, in this paper we mostly focus on the leading large $N_c$ limit of the bulk-defect correlators. In the planar limit, we find that the results for the bulk-defect correlators can be expressed simply in terms of integrals,
\beq
\begin{aligned}
\langle \mathcal{W}[\prod_{k=1}^{n}\tilde{\Phi}^{L_k}]{\rm tr}[\tilde{\Phi}^{J}]\rangle\sim \oint d\mu \,\,B_J(x)\prod_{k=1}^{n}Q_{L_k}(x)\comma
\end{aligned}
\eeq
where the definitions of the quantities in the formula are given in section \ref{subsec:integral}. This formula is the generalization of the one found in our previous paper \cite{Giombi:2018qox}. As was pointed out there, the function $Q_L(x)$ that appears in the formula coincides with the so-called Q-function in the Quantum Spectral Curve formalism \cite{Gromov:2013pga}. This appearance of the Q-function is unexpected and strongly suggests that the Quantum Spectral Curve (QSC), which was originally invented for computing the spectrum of the operators, can be useful also for analyzing the correlation functions. 
Using this integral representation, we expand the results at weak and strong coupling. At weak coupling, the results match the perturbative answers $\mathcal{N}=4$ SYM, while at strong coupling they reproduce the correlation functions of fluctuations of the string coordinates and the vertex operator on the AdS$_2$ worldsheet, which we explicitly compute to leading order in the $\alpha' \sim 1/\sqrt{\lambda}$ expansion.

In the course of the computation, we also clarify several aspects of the Gram-Schmidt analysis which were not discussed in our previous paper. Most importantly we point out the necessity of including the multi-trace-like operators $\mathcal{W}[\tilde{\Phi}^{L}]\prod_{k=1}^{m}{\rm tr}[\tilde{\Phi}^{J_k}]$, which may be viewed from the dual perspective as bound states of open strings and closed strings, which have to be included in the defect CFT spectrum. Such operators are negligible in the planar limit but can affect the computation at the non-planar level. To confirm this effect and check the validity of our formalism, we compute the non-planar correction to the defect two-point function explicitly and check the results against the direct perturbative computations. We also provide an integral representation for the first non-planar correction in terms of the Quantum Spectral Curve (see \eqref{eq:integralnonplanarQSC}), which suggests the potential applicability of the QSC formalism to the nonplanar corrections.

The rest of the paper is organized as follows: In section \ref{sec:setup}, we review the definitions of the BPS Wilson loops and the results of the supersymmetric localization. After doing so, we introduce the correlators that we analyze and discuss their relations to the defect CFT data. We then explain in section \ref{sec:construction} how to compute such correlators using OPE and the Gram-Schmidt analysis. We first present a general formalism for constructing higher-charge operators applicable to finite $N_c$, and then discuss the simplification at large $N_c$. Using the results in section \ref{sec:construction}, we then evaluate the bulk-defect correlators in the planar limit in section \ref{sec:largeN}, deriving the integral expression and obtaining the weak- and strong-coupling expansions. These results are in perfect agreement with the direct perturbative computations at weak and strong coupling in section \ref{sec:comparison}. We then discuss the non-planar corrections to the defect two-point functions in section \ref{sec:nonplanar}. Finally in section \ref{sec:conclusion} we conclude and discuss future directions. Several appendices are included to collect some explicit results of the computation.
\section{Set up\label{sec:setup}}
Before delving into the computation, let us first explain the set up by reviewing the supersymmetric subsector of $\mathcal{N}=4$ SYM and showing its relation to the defect OPE data.
\subsection{Supersymmetric subsector of $\mathcal{N}=4$ SYM\label{subsec:subsector}}
\begin{figure}
\centering
\includegraphics[clip,height=5.5cm]{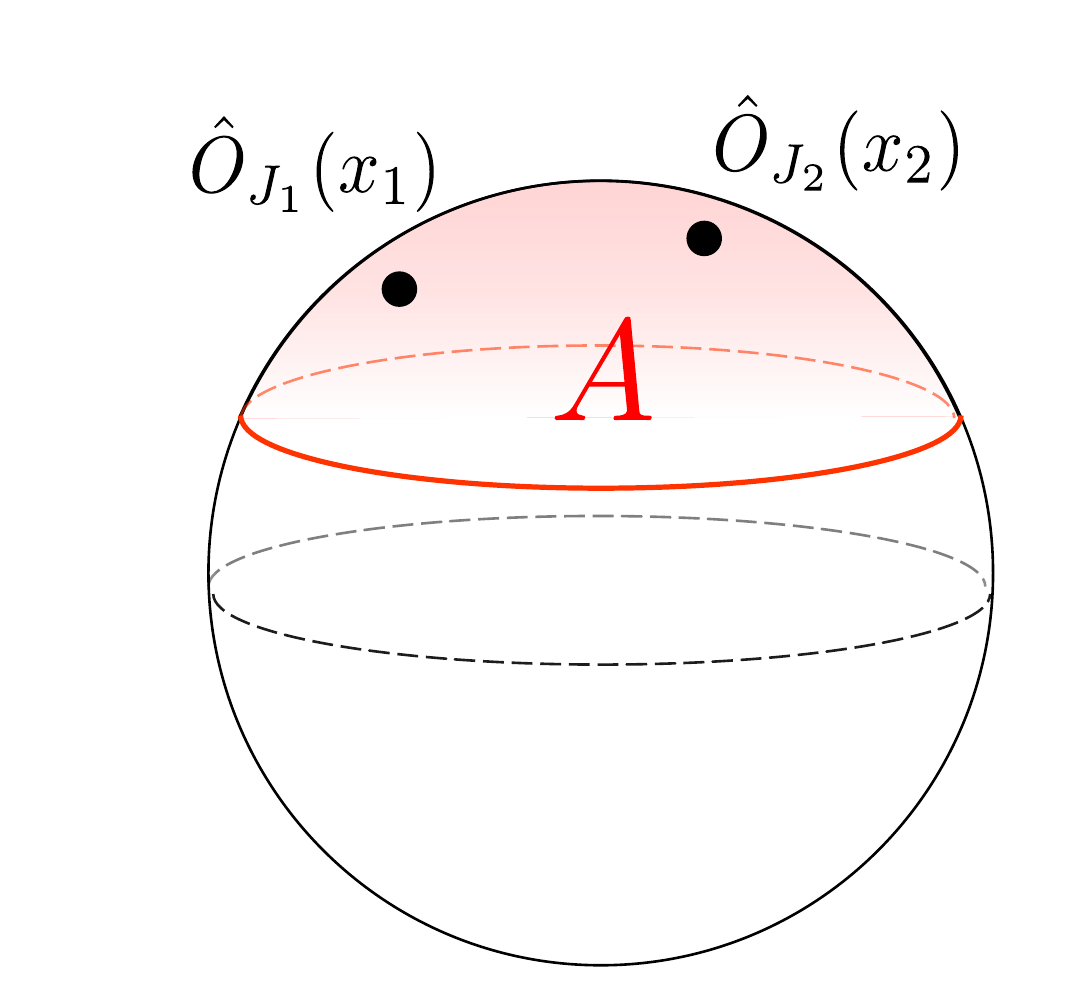}
\caption{Correlators of the $1/8$-BPS Wilson loop and local operators. The $1/8$-BPS Wilson loop, denoted by a red curve, can be defined on a arbitrary contour on $S^2$ and its expectation value depends only on the area $A$ of the subregion inside the loop (shown as the red-shaded region). In addition to the $1/8$-BPS Wilson loop, one can consider the bulk single-trace operators $\bb_{J_i}$. In this paper, we always assume that the bulk single-trace operators are on the same side of the Wilson loop, namely in the red-shaded region in the figure.}
\label{fig:fig0}
\end{figure}

The central object in this paper is the $1/8$-BPS Wilson loop defined by
\beq
\mathcal{W}=\frac{1}{N_c}{\rm tr}\,{\rm P} \exp \left[\oint_{C}\left(iA_j+\epsilon_{kjl}x^{k}\Phi^{l}\right)dx^{j}\right]\qquad (i,j,k=1,2,3)\,.
\eeq
Here the Wilson loop couples to three out of  the six real scalars $\Phi^{l}$ and the contour $C$ is placed on a $S^2$ subspace of $R^4$ (see figure \ref{fig:fig0}). In the rest of this paper, we choose $S^2$ to be of unit radius (namely $x_1^{2}+x_2^{2}+x_3^{2}=1$) and consider only the Wilson loop in the fundamental representation. Owing to the specific choices that we made for the contour and the scalar couplings, this Wilson loop preserves four supercharges and therefore is $1/8$-BPS.

One can also add local operators without breaking two of the four supercharges. A prototypical example is the single-trace operator which is given by
\beq
\bb_J (x)\equiv \mathcal{N}_J\times {\rm tr} [\tilde{\Phi}^{J}(x)] \qquad x\in S^2\comma
\eeq
where $\tilde{\Phi}$ is a position-dependent linear combination of the scalar fields,
\beq
\tilde{\Phi}(x)=x_1 \Phi^{1}+x_2 \Phi^{2}+x_3 \Phi^{3}+i\Phi^4\comma
\eeq
and we chose the standard normalization for the chiral primary operators\fn{Note that the convention here is slightly different from the one in \cite{Giombi:2009ds,Giombi:2012ep}: In that paper, the gauge group generators are {\it anti-}Hermitian while in this paper we stick to a more standard convention in physics in which the generators are Hermitian. This explains the extra $(-i)^{J}$ factor in \cite{Giombi:2009ds,Giombi:2012ep}.} (see \cite{Giombi:2009ds,Giombi:2012ep}),
\beq\label{eq:Nj}
\mathcal{N}_J=2^{J/2}\frac{(2\pi)^J}{\lambda^{J/2}\sqrt{J}}\period
\eeq
In addition to single-trace operators, there are also multi-trace operators defined by
\beq
\bb_{J_1,\ldots, J_n}(x)=\prod_{k=1}^{n}\bb_{J_k}(x)\qquad x\in S^2\period
\eeq

These operators, together with the $1/8$-BPS Wilson loops, form a supersymmetric subsector of $\mathcal{N}=4$ SYM.
Based on perturbation theory and AdS/CFT, it was conjectured in \cite{Drukker:2007yx,Drukker:2007qr} that the correlators in this subsector can be computed by the bosonic two-dimensional Yang-Mills theory on $S^2$ (in the zero-instanton sector) with the coupling constant
\beq
g_{\rm 2d}^{2} =-\frac{g_{\rm 4d}^2}{2\pi}\period
\eeq 
The relation between the observables in $2d$ and $4d$ is given by
\beq
\mathcal{W} \quad \leftrightarrow \quad {\tr} {\rm P}e^{\oint A_{\rm 2d}}\comma\qquad \bb_J (x)\quad \leftrightarrow\quad {\rm tr}\left(i\ast_{\rm 2d}F_{\rm 2d}\right)^{J}\period
\eeq
The conjecture was supported later by supersymmetric localization \cite{Pestun:2009nn} and tested against a number of nontrivial checks \cite{Giombi:2009ms,Giombi:2009ek,Giombi:2009ds,Giombi:2012ep,Bassetto:2009rt,Bassetto:2009ms,Bonini:2014vta,Bonini:2015fng}.

Thanks to its topological property, the computation in the two-dimensional Yang-Mills theory can be further reduced to simple Gaussian (multi-)matrix models \cite{Giombi:2009ds,Giombi:2012ep}, and the results depend only on the area of the region inside the Wilson loop. For instance, the expectation value of the Wilson loop is given by
\beq
\langle \mathcal{W}\rangle= \frac{1}{Z} \int [dX]\frac{1}{N_c} {\rm tr}(e^{X})e^{-\frac{(4\pi)^2}{2A(4\pi -A)g_{\rm YM}^2}{\rm tr}\left(X^2\right)}\comma
\eeq
where $A$ is the area of the subregion on $S^2$ surrounded by the loop $C$  (see figure \ref{fig:fig0}). Similarly the correlators of the Wilson loop and single-trace local operators are given by
\beq
\begin{aligned}
\langle \mathcal{W} \bb_J\rangle=& \frac{\mathcal{N}_J}{Z}\int [dX][dY]\frac{1}{N_c}{\rm tr}(e^{X}){\rm tr}(Y^{J})e^{-\frac{(4\pi)^2}{2g_{\rm YM}^2(4\pi-A)}{\rm tr}\left(AY^2+2X Y\right)}\comma\\
\langle \mathcal{W} \bb_{J_1} \bb_{J_2}\rangle=&\frac{\mathcal{N}_{J_1}\mathcal{N}_{J_2}}{Z}\int [dX][dY_1][dY_2]\frac{1}{N_c}{\rm tr}(e^{X}){\rm tr}(Y_1^{J_1}){\rm tr}(Y_2^{J_2})e^{-S_{W\bb\bb}}\comma
\end{aligned}
\eeq
with
\beq
S_{W\bb\bb}=\frac{(4\pi)^2(4\pi -A)}{2g_{\rm YM}^2 (8\pi-A)}{\rm tr}\left[\frac{X^2}{(4\pi-A)^2} +(Y_1^2+Y_2^2)-\frac{8\pi Y_1 Y_2}{4\pi-A}+\frac{2 (Y_1+Y_2)X}{4\pi -A}\right]\period
\eeq
Here we assumed that both operators are on the same side of the Wilson loop (see figure \ref{fig:fig0}). These results can be straightforwardly generalized to the multi-trace operators. For instance the correlator of the Wilson loop and a double-trace operator is given by\fn{Note that the correlator $\langle \mathcal{W} \bb_{J_1} \bb_{J_2}\rangle$ is different from the correlator $\langle \mathcal{W} \bb_{J_1,J_2}\rangle$. The former is the correlator of two single-trace operators inserted at two separate points in the bulk while the latter is the correlator of a double-trace operator inserted at a single point in the bulk. Diagrammatically, the former includes the contractions between two single-traces while the latter does not.}
\beq
\langle \mathcal{W} \bb_{J_1,J_2}\rangle= \frac{\mathcal{N}_{J_1}\mathcal{N}_{J_2}}{Z}\int [dX][dY]\frac{1}{N_c}{\rm tr}(e^{X}){\rm tr}(Y^{J_1}){\rm tr}(Y^{J_2})e^{-\frac{(4\pi)^2}{2g_{\rm YM}^2(4\pi-A)}{\rm tr}\left(AY^2+2X Y\right)}\period
\eeq

In the large $N_c$ limit, these correlators can be expanded in terms of the modified Bessel functions. For instance we have
\beq\label{eq:WOWW}
\begin{aligned}
\langle \mathcal{W} \rangle=&\frac{2}{\sqrt{\lambda^{\prime}}}I_1 (\sqrt{\lambda^{\prime}})+\frac{\lambda^{\prime}}{48 N_c^2}I_2 (\sqrt{\lambda^{\prime}})+O(1/N_c^3)\comma\\
\langle \mathcal{W}\bb_J\rangle =&\frac{2^{-J/2}\sqrt{J}}{N_c}\left(\frac{2\pi-a}{2\pi+a}\right)^{J/2}I_J(\sqrt{\lambda^{\prime}})+O(1/N_c^2)\comma
\end{aligned}
\eeq
where $\lambda^{\prime}$ and $a$ are given by
\beq
a\equiv A-2\pi\comma\qquad \lambda^{\prime}\equiv \lambda \left(1-\frac{a^{2}}{4\pi^2}\right)\comma\qquad \lambda\equiv g_{\rm YM}^{2}N_c\period
\eeq

On the other hand, the large $N_c$ expansion of the correlator $\langle \mathcal{W} \bb_{J_1}\bb_{J_2}\rangle$ consist of two terms; the disconnected term and the connected term $\langle \mathcal{W} \bb_{J_1} \bb_{J_2}\rangle=\langle \mathcal{W} \bb_{J_1} \bb_{J_2}\rangle_{\rm disc}+\langle \mathcal{W} \bb_{J_1} \bb_{J_2}\rangle_{\rm conn}$. For the purpose of this paper, we only need the disconnected term since the connected term is subleading. Written explicitly, the disconnected term is given by
\beq\label{eq:bulktwoptdisconnected}
\begin{aligned}
\langle \mathcal{W} \bb_{J_1} \bb_{J_2}\rangle_{\rm disc}=&\left(-\frac{1}{2}\right)^{J_1}\delta_{J_1,J_2}\langle \mathcal{W}\rangle\\
&=\left(-\frac{1}{2}\right)^{J_1}\delta_{J_1,J_2}\left[\frac{2}{\sqrt{\lambda^{\prime}}}I_1 (\sqrt{\lambda^{\prime}})+\frac{\lambda^{\prime}}{48 N_c^2}I_2 (\sqrt{\lambda^{\prime}})+O(1/N_c^3)\right]\period
\end{aligned}
\eeq
The explicit expression for the connected term can be found in Appendix \ref{ap:bulk2pt}.
\subsection{Operators on the Wilson loop and the area derivatives\label{subsec:area}}
In addition to the operators discussed in the previous subsection whose correlators can be computed by localization rather directly, there is another interesting class of operators which are obtained by inserting scalars inside a Wilson loop trace:
\beq
\mathcal{W}[\NO{\tilde{\Phi}^{L_1}}\NO{\tilde{\Phi}^{L_2}}\cdots \NO{\tilde{\Phi}^{L_n}}]\equiv \frac{1}{N_c}{\rm tr}{\rm P}\left[\NO{\tilde{\Phi}^{L_1}(\tau_1)} \cdots \NO{\tilde{\Phi}^{L_n} (\tau_n)}e^{\oint_{C}\left(iA_j+\epsilon_{kjl}x^{k}\Phi^{l}\right)dx^{j}}\right]\period
\eeq
Here we parametrize the loop by $\tau \in [0,2\pi]$. Note also that we put a normal-ordering symbol to $\tilde{\Phi}^{L}$ in order to emphasize the absence of self-contractions inside each operator. At this point this may seem unnecessary complication, but the reason for doing this will become clear in the next section.
 
As shown in \cite{Pestun:2009nn}, the insertion of a single scalar corresponds to the insertion of a dual field strength of the two-dimensional Yang-Mills theory, 
\beq
\tilde{\Phi}\quad \leftrightarrow\quad i\ast F_{\rm 2d}\comma
\eeq
which in turn is related to a small deformation of the (2d) Wilson loop. This correspondence allows us to relate the correlators of multiple $\tilde{\Phi}$'s to the area derivatives of the Wilson loop expectation value\fn{Owing to the relation to the 2d Yang-Mills theory, these correlators are also independent of the postions (namely $\tau_k$'s).},
\beq
\langle \mathcal{W}[\,\underbracket{\tilde{\Phi}\cdots \tilde{\Phi}}_{n}\,]\rangle=\frac{\del^{n}\langle \mathcal{W}\rangle}{(\del A)^{n}}\period
\eeq
As discussed in the previous work \cite{Giombi:2018qox}, it is also possible to relate the insertion of higher-charge operators $\mathcal{W}[\NO{\tilde{\Phi}^{L}}]$ to the area derivatives. This however requires the use of the Gram-Schmidt orthogonalization, which we will review and refine in the next section, and the results in general take a more complicated form (although they can computed systematically from the localization results).

Now, putting together these operator insertions with the operators discussed in the previous subsection, one can consider a variety of correlators of the form, 
\beq\label{eq:generalbd}
\begin{aligned}
 G_{L_1,\ldots, L_n|J_1,\ldots, J_m}\equiv \langle \mathcal{W}[\prod_{k=1}^{n}\NO{\tilde{\Phi}^{L_k}}]\prod_{k=1}^{m}\bb_{J_k} \rangle\period
\end{aligned}
\eeq
In what follows, we call these correlators {\it topological correlators} since they do not depend on the positions.
The main goal of this paper is to analyze these correlators\fn{Although we mostly focus on the simplest correlators $\langle \mathcal{W}[\NO{\tilde{\Phi}^{L}}]\bb_J\rangle$ in this paper, the general methodology that we develop is applicable also to more complicated correlators \eqref{eq:generalbd}.} by generalizing the arguments in \cite{Giombi:2018qox}. 
\subsection{Relation to the defect CFT data\label{subsec:cftdata}}
When the contour is a circle along the equator of $S^2$, the Wilson loop preserves higher amount of supersymmetries and becomes $1/2$-BPS:
\beq
\mathcal{W}_{{\rm 1/2-BPS}}=\frac{1}{N_c}{\rm tr}{\rm P}\exp \left[\oint_{{\rm equator}}\left(iA_j\dot{x}^j+\Phi^{3}|\dot{x}|\right)d\tau\right]\period
\eeq
An important feature of this Wilson loop is that it preserves the SL(2,R) conformal symmetry\fn{The full symmetry group preserved by the circular Wilson loop is OSP$(4^{\ast}|4)$.} and therefore can be regarded as a conformal defect. This in particular implies that one can extract the defect CFT data from the correlators in the supersymmetric subsector.

Before discussing how to do so, let us first introduce the {\it normalized correlators}, defined by
\beq
\llangle\mathcal{W}[ \cdots]\cdots\rrangle\equiv \frac{\langle \mathcal{W}[\cdots]\cdots \rangle}{\langle \mathcal{W}\rangle}\comma
\eeq
where $\cdots$ denote either the operators on the Wilson loop or the operators in the bulk depending on whether it is inside $\mathcal{W}[\,\,\,\,]$ or not.
Taking such a ratio renders the expectation value of the identity operator to be unity and make the correlators obey the standard defect CFT axioms. We then consider the following correlators of the protected operators,
\beq
\begin{aligned}
\mathcal{G}_{L_1,L_2}&\equiv\llangle\mathcal{W}[(u_1\cdot \vec{\Phi})^{L_1}(\tau_1)(u_2\cdot \vec{\Phi})^{L_2}(\tau_2)]\rrangle_{\rm circle}\comma\\
\mathcal{G}_{L_1,L_2,L_3}&\equiv \llangle\mathcal{W}[(u_1\cdot \vec{\Phi})^{L_1}(\tau_1)(u_2\cdot \vec{\Phi})^{L_2}(\tau_2)(u_3\cdot \vec{\Phi})^{L_3}(\tau_3)]\rrangle_{\rm circle}\comma \\
\mathcal{G}_{L|J}&\equiv\llangle \mathcal{W}[(u\cdot \vec{\Phi})^{L}(\tau)]\,\,{\rm tr}(U\cdot \vec{\Phi})^{J}(x^{\prime})\rrangle_{\rm circle}\period
\end{aligned}
\eeq
Here $\vec{\Phi}=\left(\Phi_1,\Phi_2,\Phi_3,\Phi_4,\Phi_5,\Phi_6 \right)$ and $U$'s and $u$'s are six-dimensional null vectors satisfying $U\cdot U=u\cdot u=0$. In addition, we require the third components of $u$'s to vanish in order to make the operators $(u_k\cdot \vec{\Phi})^{L_k}$ to have protected conformal dimensions (in other words, the null vector $u$ projects onto a symmetric traceless representation of the $SO(5)\subset SO(6)$ preserved by the 1/2-BPS Wilson loop).
Unlike the correlators in the supersymmetric subsector that we discussed in the previous subsections, these correlators depend on the positions of the operators. However, thanks to the conformal symmetry and the R-symmetry, the position dependence can be completely fixed to be
\beq
\begin{aligned}
\mathcal{G}_{L_1,L_2}&=n_{L_1} \times \frac{\delta_{L_1,L_2}(u_1\cdot u_2)^{L_1}}{(2\sin \frac{\tau_{12}}{2})^{2L_1}} \comma\\
\mathcal{G}_{L_1,L_2,L_3}&= c_{L_1,L_2,L_3}\times \frac{(u_1\cdot u_2)^{L_{12|3}}(u_2\cdot u_3)^{L_{23|1}}(u_3\cdot u_1)^{L_{31|2}}}{(2\sin \frac{\tau_{12}}{2})^{2L_{12|3}}(2\sin \frac{\tau_{23}}{2})^{2L_{23|1}}(2\sin \frac{\tau_{31}}{2})^{2L_{31|2}}}\comma\\
\mathcal{G}_{L|J}&=c_{L|J} \times\frac{(u\cdot U)^{L} (U_3)^{J-L} }{|x^{\prime}-x(\tau)|^{2L}|x^{\prime}_{\perp}|^{J-L}}\comma
\label{OPE-coeff}
\end{aligned}
\eeq
where $\tau_{ij}\equiv \tau_i-\tau_j$, $L_{ij|k}\equiv (L_i+L_j-L_k)/2$ and $|x^{\prime}_{\perp}|$ is given by
\beq
|x^{\prime}_{\perp}|=\frac{\sqrt{\left[1+(x_1^{\prime})^{2}+(x_2^{\prime})^{2}+(x_3^{\prime})^2+(x_4^{\prime})^2\right]^2-4\left[(x_1^{\prime})^2+(x_2^{\prime})^2\right]}}{2}\period
\eeq
The constants $n_{L_1}$, $C_{L_1,L_2,L_3}$ and $c_{L|J}$ are the defect CFT data which are nontrivial functions of the 't Hooft coupling and the rank of the gauge group.
One can also perform the conformal transformation to map the Wilson loop to a straight line. In that case, one simply needs to perform the following replacement for $\mathcal{G}_{L_1,L_2}$ and $\mathcal{G}_{L_1,L_2,L_3}$,
\beq
2\sin \frac{\tau_{ij}}{2} \quad \mapsto \quad |x(\tau_i)-x(\tau_j)|\comma
\eeq
while the expression for $\mathcal{G}_{L|J}$ still applies if we interpret $|x_{\perp}^{\prime}|$ as the distance in the direction perpendicular to the Wilson loop. 

These correlators reduce to the topological correlators \eqref{eq:generalbd} upon the following specification of the parameters
\beq
\begin{aligned}
u_i= (\cos \tau_i, \sin \tau_i,0,i,0,0)\comma\qquad U=(x^{\prime}_1,x^{\prime}_2, x^{\prime}_3,i,0,0)\comma \qquad x^{\prime}\in S^2\period
\end{aligned}
\eeq
The results read
\beq\label{eq:Gandsmallc}
\begin{aligned}
\frac{G_{L_1,L_2}}{\langle \mathcal{W}\rangle}&=\left(-\frac{1}{2}\right)^{L_1}\times\delta_{L_1,L_2}n_{L_1}\comma\qquad \frac{G_{L_1,L_2,L_3}}{\langle \mathcal{W}\rangle}=\left(-\frac{1}{2}\right)^{\frac{L_1+L_2+L_3}{2}}\times c_{L_1,L_2,L_3}\comma\\
\frac{G_{L|J}}{\langle \mathcal{W}\rangle}&=\left(-\frac{1}{2}\right)^{L}\times c_{L|J}\period
\end{aligned}
\eeq
This shows that the topological correlators coincide with the defect CFT data up to trivial overall factors. 

It is sometimes useful to unit-normalize the two-point functions on the Wilson loop. In such normalization, the structure constants  are given as follows:
\beq\label{eq:Gandbigc}
\begin{aligned}
C_{L_1,L_2,L_3}\equiv \frac{c_{L_1,L_2,L_3}}{\sqrt{n_{L_1}n_{L_2}n_{L_3}}}\comma\qquad
C_{L|J}\equiv \frac{c_{L|J}}{\sqrt{n_L}}\period
\end{aligned}
\eeq 
\section{Construction of higher-charge operators\label{sec:construction}}
In this section, we construct general operator insertions $\NO{\tilde{\Phi}^{L}}$ on the Wilson loop in the supersymmetric subsector  using the OPE  and the Gram-Schmidt process. The same idea was employed already in the previous work \cite{Giombi:2018qox}, but here we elucidate several important points which were not accounted for in \cite{Giombi:2018qox}. Although they are largely negligible at large $N_c$, they have important consequences on the bulk-defect correlators and the non-planar corrections, as will be shown in sections \ref{sec:largeN} and \ref{sec:nonplanar}.
\subsection{Basic idea\label{sec:basic}}
Before presenting a general construction, let us explain the basic idea of our approach using simple examples of correlators on the $1/2$-BPS circular loop. Along the way we clarify three important aspects ({\it bound state}, {\it degeneracy} and {\it mixing with multi-trace operators}) which were not discussed in our previous construction \cite{Giombi:2018qox}.

The basic strategy is to construct complicated operators from simpler operators using the OPE. The simplest operator (apart from the identity) on the $1/2$-BPS Wilson loop is the single-scalar insertion $\mathcal{W}[\NO{\tilde{\Phi}}] $. As discussed above, the correlator of this operator can be computed directly by taking area derivatives of the localization result.

\paragraph{Bound state}In addition to $\mathcal{W}[\NO{\tilde{\Phi}}] $, there is yet another operator on the Wilson loop with the same $R$-charge, $\NO{\mathcal{W}\, {\rm tr}[\tilde{\Phi}]}$,\footnote{In this paper we take the gauge group to be $U(N)$ for simplicity. In the case of $SU(N)$, this type of operator would appear first at charge 2, i.e. $\NO{\mathcal{W}\, {\rm tr}[\tilde{\Phi}^2]}$.}  namely a single-trace operator placed at a point on the Wilson loop. On the string-theory side, this operator, and its higher charge generalizations, correspond to a {\it bound state} of open and closed strings. Although it might not be so obvious why it must be included in the defect CFT spectrum, one can show that such operators are necessary for the consistency of the bulk-defect crossing equation as explained in Appendix \ref{ap:bulk2pt}.  
Unlike the insertion $\mathcal{W}[\NO{\tilde{\Phi}}] $, this operator is not related to the area derivatives of the Wilson loop expectation value. To construct this operator, we instead need to consider a correlator of the Wilson loop and a bulk single-trace operator. Since the correlator is independent of the positions, we can bring the bulk operator arbitrarily close to  a point on the loop without affecting its expectation value. After doing so, we perform the {\it bulk-defect OPE} to get
\beq\label{eq:constructionfrombulkope}
\left.{\rm tr}[\tilde{\Phi}]\right|_{\rm bulk} \mathcal{W}\quad \overset{\text{bulk-defect OPE}}{=}\quad \NO{\mathcal{W}\,{\rm tr}[\tilde{\Phi}]} + c_0 \mathcal{W}\period
\eeq
As shown above, in addition to $\NO{\mathcal{W}\,{\rm tr}[\tilde{\Phi}]}$, this process produces another operator, which is the identity operator on the Wilson loop (or, equivalently, the Wilson loop itself). The coefficient $c_0$ then corresponds to the bulk-defect structure constant between ${\rm tr}[\tilde{\Phi}]$ in the bulk and the identity operator on the Wilson loop, and is given by the expectation value
\beq
c_0 =\llangle \mathcal{W}[{\bf 1}]\,{\rm tr}[\tilde{\Phi}]\rrangle\quad\left(\equiv\langle \mathcal{W}\,{\rm tr}[\tilde{\Phi}]\rangle/\langle \mathcal{W}\rangle\right)\period
\eeq
Inverting the relation \eqref{eq:constructionfrombulkope}, we obtain
\beq
\NO{\mathcal{W}\,{\rm tr}[\tilde{\Phi}]} =\left.{\rm tr}[\tilde{\Phi}]\right|_{\rm bulk} \mathcal{W} -c_0 \mathcal{W}\period
\eeq
This expression allows us to relate the correlators of $\NO{\mathcal{W}\,{\rm tr}[\tilde{\Phi}]}$ to the correlators that are computable from supersymmetric localization.

\paragraph{Degeneracy and orthogonalization}Let us next consider the length-2 operators. In general, there are four different length-2 operators,
\beq
\mathcal{W}[\NO{\tilde{\Phi}^2}]\comma\qquad \NO{\mathcal{W}[\tilde{\Phi}]{\rm tr}[\Phi]}\comma\qquad \NO{\mathcal{W}{\rm tr}[\tilde{\Phi}^2]}\comma\qquad \NO{\mathcal{W}({\rm tr}[\tilde{\Phi}])^2}\period
\eeq
Among these four operators, the first operator $\mathcal{W}[\NO{\tilde{\Phi}^2}]$ can be constructed by taking the {\it defect OPE} of two scalar insertions,
\beq
\mathcal{W}[\tilde{\Phi}\tilde{\Phi}]\qquad \overset{\text{defect OPE}}{=}\qquad \mathcal{W}[\NO{\tilde{\Phi}^2}] +c_1\mathcal{W}\comma
\eeq
where the second term on the right hand side comes from the self-contraction of the two scalars and $c_1$ is the defect structure constant of two $\tilde{\Phi}$'s and the identity operator,
\beq
c_1 =\llangle\mathcal{W}[\tilde{\Phi}\tilde{\Phi}{\bf 1}] \rrangle\left(\equiv\langle \mathcal{W}[\tilde{\Phi}\tilde{\Phi}]\rangle/\langle \mathcal{W}\rangle\right)\period
\eeq
Then by subtracting this extra contribution, we can single out $\mathcal{W}[\NO{\tilde{\Phi}^2}]$. To compute other operators, we need to consider bulk-defect correlators as was the case for the operator $\mathcal{W}[\NO{\tilde{\Phi}}]$. For instance, the second operator $\NO{\mathcal{W}[\tilde{\Phi}]{\rm tr}[\Phi]}$ can be obtained by taking the combination of the defect OPE and the bulk-defect OPE of ${\rm tr}[\tilde{\Phi}]$ in the bulk and $\mathcal{W}[\tilde{\Phi}]$, namely
\beq
\left.{\rm tr}[\tilde{\Phi}]\right|_{\rm bulk}\mathcal{W}[\tilde{\Phi}]\qquad\overset{\rm OPE}{=}\qquad  \NO{\mathcal{W}[\tilde{\Phi}]{\rm tr}[\Phi]} +c_2\mathcal{W}[\NO{\tilde{\Phi}}]+c_3 \mathcal{W}\period
\eeq
Here the second term comes from the contraction of the bulk ${\rm tr}[\tilde{\Phi}]$ with the Wilson loop while the last term comes from the contraction of ${\rm tr}[\tilde{\Phi}]$ and the insertion on the loop $\tilde{\Phi}$. The OPE coefficients $c_2$ and $c_3$ are given by\fn{Note that, for $c_2$, one needs to divide by the norm of the resulting state $\mathcal{W}[\tilde{\Phi}]$ in order to have the correct OPE expansion. 
}
\beq
\begin{aligned}
&c_2=\frac{\llangle \mathcal{W}[\tilde{\Phi}\tilde{\Phi}]{\rm tr}[\tilde{\Phi}]\rrangle}{\llangle \mathcal{W}[\tilde{\Phi}\tilde{\Phi}]\rrangle}\comma\qquad
&c_3=\llangle \mathcal{W}[\tilde{\Phi}]{\rm tr}[\tilde{\Phi}]\rrangle\period
\end{aligned}
\eeq
Repeating the same procedures for the other two operators, one can define sets of normal-ordered operators. The resulting operators are free of admixiture of operators with different $R$-charges and they are all equally good defect primaries. However, they have one unsatisfactory feature that they are not necessarily orthogonal to each other; namely the two-point functions of different operators do not vanish in general. To make them orthogonal to each other, one has to take appropriate linear combinations of these four operators. The choice of the linear combinations is not unique since all these operator have the same quantum number and physically indistinguishable especially at finite $N_c$, where the distinction by the number of traces becomes obscured. In this paper, we make a choice that is most suited for studying the correlators at large $N_c$; namely we define $\mathcal{W}[\NO{\tilde{\Phi}^k}]$ without performing further subtraction while for the rest of the operators we take appropriate linear combinations so that they become orthogonal to each other and also to $\mathcal{W}[\NO{\tilde{\Phi}^k}]$. 

\paragraph{Mixing with ``multi-trace'' operators}
In the examples discussed so far, the self-contraction of insertions on the Wilson loop $\mathcal{W}[\tilde{\Phi}\cdots \tilde{\Phi}]$ only produced the operators of the same kind, $\mathcal{W}[\tilde{\Phi}\cdots \tilde{\Phi}]$. However, starting from length-$3$ operators, a new effect shows up which inevitably mixes the operators of different types, namely the operators involving multi-traces\fn{We call these operators multi-trace operators since the Wilson loop itself already contains a trace.} e.g.~$\mathcal{W}[\tilde{\Phi}\cdots \tilde{\Phi}]{\rm tr}[\tilde{\Phi}^{J}]$.

\begin{figure}
\centering
\begin{minipage}{0.49\hsize}
\centering
\includegraphics[clip,height=3cm]{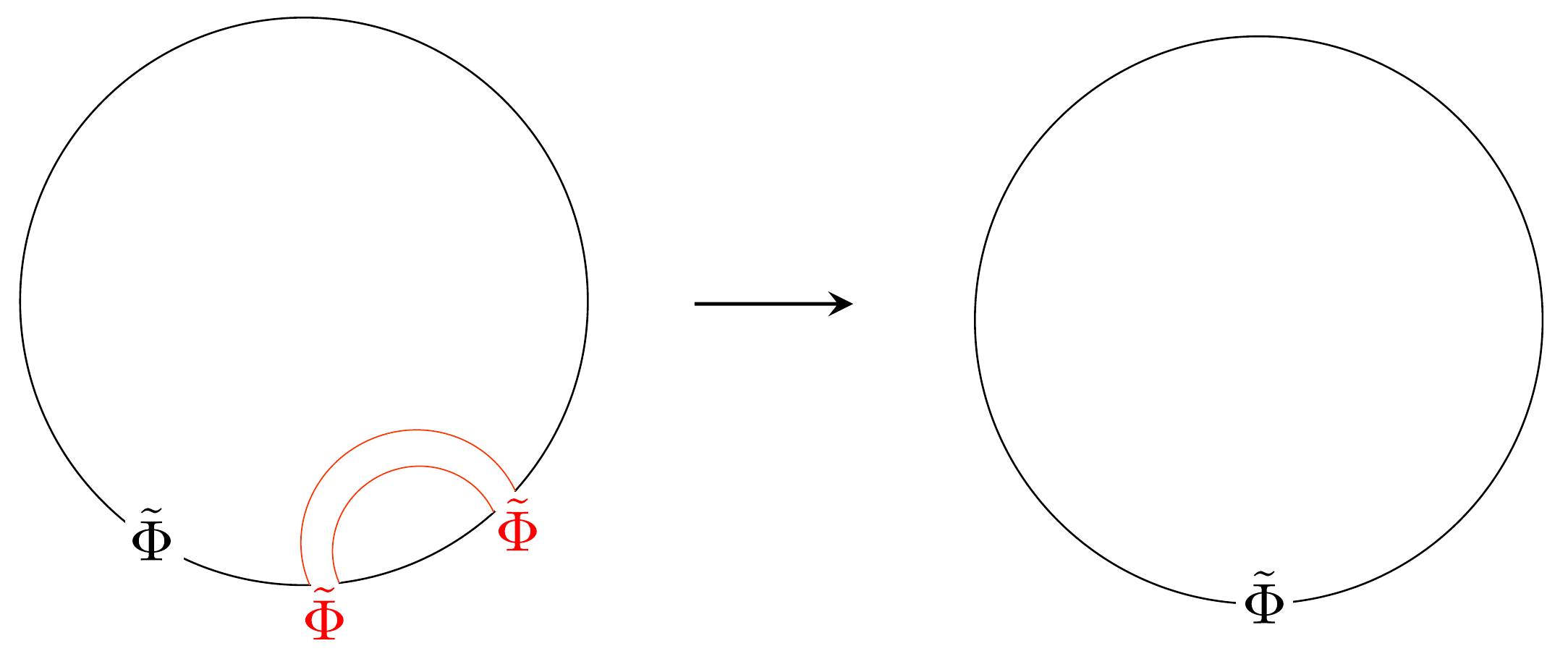}\\
(a)
\end{minipage}
\begin{minipage}{0.49\hsize}
\centering
\includegraphics[clip,height=3cm]{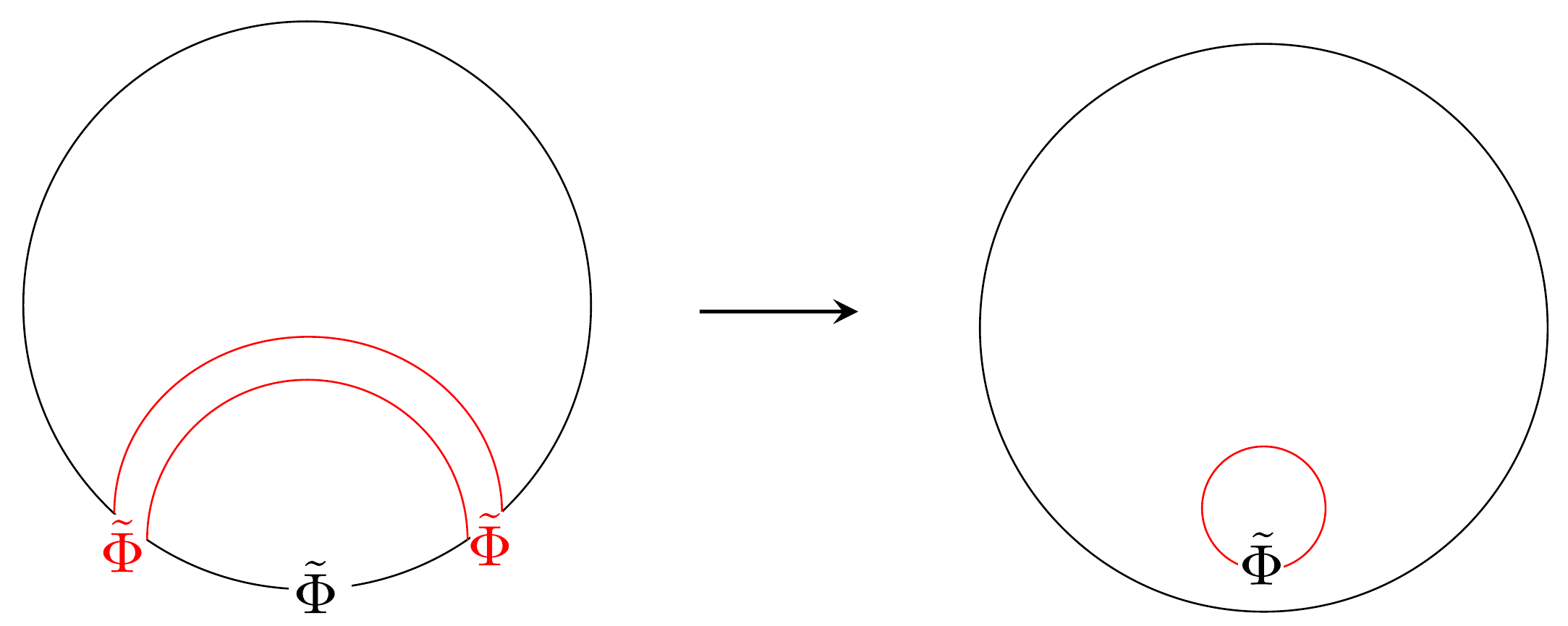}\\
(b)
\end{minipage}
\caption{Wick contractions of the insertions on the Wilson loop. (a) In the planar limit, the Wick contractions (denoted by red double lines) can only connect neighboring insertions. Therefore, the resulting operator is again an insertion on the Wilson loop. (b) On the other hand, at the non-planar level, one can connect non-neighboring insertions by a propagator. This produces a closed index loop for the remaining fields and converts them into a single-trace operator. It is therefore necessary to include the ``multi-trace'' operators in the spectrum on the Wilson loop.}
\label{fig:fig1}
\end{figure}
To see this explicitly, let us consider the OPE of three $\tilde{\Phi}$'s inserted on the Wilson loop. At weak coupling, the OPE simply amounts to performing the Wick contractions. At large $N_c$, due to planarity, the only allowed contraction is to connect two neighboring $\tilde{\Phi}$'s as shown in figure \ref{fig:fig1}-(a). As shown in the figure, we then get a single insertion $\tilde{\Phi}$ on the Wilson loop, namely $\mathcal{W}[\tilde{\Phi}]$. However, at the non-planar level, we can Wick-contract non-neighboring $\tilde{\Phi}$'s as shown in figure \ref{fig:fig1}-(b). This produces a closed color-index loop for the remaining $\tilde{\Phi}$ in the middle, and converts it into a single-trace operator ${\rm tr}[\tilde{\Phi}]$. Thus the OPE of three $\tilde{\Phi}$'s takes the following form:
\beq
\mathcal{W}[\tilde{\Phi}\tilde{\Phi}\tilde{\Phi}]=\mathcal{W}[\NO{\tilde{\Phi}^3}]+c_4 \mathcal{W}[\NO{\tilde{\Phi}}]+c_5 \mathcal{W}{\rm tr}[\tilde{\Phi}]\period
\eeq
This shows that, even if we are only interested in the operators of the form $\mathcal{W}[\NO{\tilde{\Phi}^k}]$, we cannot neglect other operators since they mix with each other through the OPE. (The only exception is at large $N_c$ which we will discuss in section \ref{subsec:simplification}.) Such a mixing was not discussed in our previoius work \cite{Giombi:2018qox}, but as we will see later it has important consequences at the non-planar level.

\subsection{Gram-Schmidt process\label{subsec:GS}}
By repeating the recursive procedure described in the previous subsection, one can construct arbitrary operators on the Wilson loop in the supersymmetric sector. This however is not useful for writing down general and/or closed-form expressions. In this subsection, we explain an alternative approach which is more algorithmic and applies straightforwardly also to the general $1/8$-BPS operators.

To explain the approach, let us first note that the normal-ordered operators constructed in the previous subsection share two important properties:
\begin{itemize}
\item They are given by a sum of bare (un-normal ordered) operators, $\mathcal{W}[\tilde{\Phi}\cdots \tilde{\Phi}] \prod_{k}{\rm tr}\tilde{\Phi}^{J_k}$.
\item They are orthogonal to each other i.e.~the two-point functions of different operators vanish. 
\end{itemize}
As was pointed out in \cite{Giombi:2018qox}, the operators with such properties can be systematically constructed by the application of the so-called Gram-Schmidt process.\fn{Recently, the Gram-Schmidt process has been used for the computation of various correlation functions in supersymmetric field theories. See \cite{Gerchkovitz:2016gxx,Baggio:2016skg,Rodriguez-Gomez:2016ijh,Rodriguez-Gomez:2016cem,Billo:2017glv,Bourget:2018obm,Bourget:2018fhe}.}

The Gram-Schmidt process is a recursive way of constructing the orthogonal vector basis from a given set of vectors. For instance, starting from a set of vectors $\{{\bf v}_1,\ldots,{\bf v}_n\}$  (to be called {\it bare vectors} in what follows) one can construct the following mutually-orthogonal vectors $\{{\bf u}_1,\ldots,{\bf u}_n\}$
\beq\label{eq:detrep}
\begin{aligned}
&{\bf u}_k =\frac{1}{m_{k-1}}\dmatrix{cccc}{({\bf v}_1,{\bf v}_1 )&({\bf v}_1,{\bf v}_2 )&\cdots&({\bf v}_1,{\bf v}_k )\\({\bf v}_2,{\bf v}_1 )&({\bf v}_2,{\bf v}_2 )&\cdots&({\bf v}_2,{\bf v}_k)\\\vdots&\vdots&\ddots&\vdots\\({\bf v}_{k-1},{\bf v}_1 )&({\bf v}_{k-1},{\bf v}_2)&\cdots&({\bf v}_{k-1},{\bf v}_k)\\{\bf v}_1&{\bf v}_2&\cdots&{\bf v}_k}\comma\\
&m_k= \dmatrix{cccc}{({\bf v}_1,{\bf v}_1 )&({\bf v}_1,{\bf v}_2)&\cdots&({\bf v}_1,{\bf v}_k )\\({\bf v}_2,{\bf v}_1 )&({\bf v}_2,{\bf v}_2 )&\cdots&({\bf v}_2,{\bf v}_k)\\\vdots&\vdots&\ddots&\vdots\\({\bf v}_{k},{\bf v}_1)&({\bf v}_{k},{\bf v}_2 )&\cdots&({\bf v}_{k},{\bf v}_k )}\comma
\end{aligned}
\eeq
where $( \ast,\ast)$ denotes the inner product of two vectors. Written more explicitly, the first two vectors are given by the following expressions:
\beq
{\bf u}_1={\bf v}_1\comma\qquad {\bf u}_2={\bf v}_2-\frac{({\bf v}_1,{\bf v}_2)}{({\bf v}_1,{\bf v}_1)}{\bf v}_1\period
\eeq

In our case, the role of the vectors ${\bf v}_k$ is played by the bare (un-normalized) operators while that of ${\bf u}_k$'s is played by the normal-ordered operators. The general bare operators are simply given by a collection of single-letter insertions on the Wilson loop and a multi-trace bulk local operator, namely
\beq\label{eq:defdgeneral}
{\bf v}_k:\qquad \dd_{L|J_1,J_2,\ldots, J_n}\,\,\equiv \,\,[\underbracket{\tilde{\Phi}\cdots \tilde{\Phi}}_{L}]\,\bb_{J_1,\ldots,J_n}\period
 \eeq
 Here the notation $[\ast]$ means that the fields in the square brackets will be inserted on the Wilson loop $\mathcal{W}$ when we evaluate the correlator. We chose this notation in order to avoid the confusion between the correlators of insertions on a {\it single Wilson loop} and the correlators involving {\it several different Wilson loops}. Throughout this paper, we only consider the former correlators: In other words, when we evaluate correlators of multiple $\dd$'s, we insert all the fields inside the square brackets into a {\it single} Wilson loop.  

The inner products of these bare operators $(\ast,\ast)$ are given by their two-point functions, namely
\beq
(\dd_{L|J_1,J_2,\ldots, J_n} \,,\,\dd_{L^{\prime}|J^{\prime}_1,J^{\prime}_2,\ldots, J^{\prime}_n})=\langle \mathcal{W}[\underbracket{\tilde{\Phi}\cdots \tilde{\Phi}}_{L+L^{\prime}}]\,\bb_{J_1,\ldots, J_n}\,\bb_{J^{\prime}_1,\ldots J^{\prime}_{n^{\prime}}}\rangle\period
\eeq
In what follows, we use the following shorthand notation for the quantities on the right hand side,
\beq
W_{L|J_1,\ldots,J_n\,;\,J^{\prime}_1\ldots, J^{\prime}_{n^{\prime}}}=\langle \mathcal{W}[\underbracket{\tilde{\Phi}\cdots \tilde{\Phi}}_{L}]\bb_{J_1,\ldots, J_n}\,\bb_{J^{\prime}_1,\ldots J^{\prime}_{n^{\prime}}}\rangle=\del_{A}^{L}\left[\langle \mathcal{W}\bb_{J_1,\ldots, J_n}\,\bb_{J^{\prime}_1,\ldots J^{\prime}_{n^{\prime}}}\rangle\right]\comma
\eeq
where in the last equality we used the relation between the single-letter insertion and the area derivative. Having understood what ${\bf v}_k$ and the inner product are, we can then use the general expression \eqref{eq:detrep} to write down the expressions for the normal ordered operators $\NO{\dd_{L|J_1,J_2,\ldots, J_n}}$. To see how it works in practice, below we construct explicitly operators with small $R$ charge.

\paragraph{L=0 :} Let us first consider the operators without R charge. There is only one operator with this property, which is the identity operator ${\bf 1}$ on the Wilson loop, or equivalently the Wilson loop itself.  Since it does not mix with any other operators, the bare operator ${\bf 1}$ is already normal-ordered. We thus have
\beq
\dd_0\equiv \NO{[{\bf 1}]}\left(=\mathcal{W}\right)\period
\eeq 
\paragraph{L=1 :} We next consider the operators with a unit R-charge. There are two bare operators with this charge in the supersymmetric subsector: 
\beq
\dd_{1}\equiv [\tilde{\Phi}] \comma\qquad \dd_{0|1}=[\,\,\,\,]\bb_1\,\,(= \mathcal{W}\,{\rm tr}[\tilde{\Phi}])\period
\eeq
As mentioned in the previous section, there is some ambiguity in choosing the basis of the normal-ordered operators because of the degeneracy, and in this paper we choose the basis in which the normal-ordered $\mathcal{W}[\NO{\tilde{\Phi}^k}]$ is defined without subtraction of the operators with the same charge. This can be achieved in the Gram-Schmidt process by bringing $\mathcal{W}[\tilde{\Phi}^{k}]$ first among the operators with the same charge.

In the case at hand, this amounts to performing the orthogonalization to the ordered set of vectors $\{\dd_{0},\dd_{1},\dd_{0|1}\}$, and as a result we get
\beq
\begin{aligned}
\NO{\dd_{1}}&=\frac{1}{W}\dmatrix{cc}{W&W_{1}\\\dd_0&\dd_1}\comma\\
\NO{\dd_{0|1}}&=\frac{1}{\dmatrix{cc}{W&W_{1}\\W_{1}&W_{2}}}\dmatrix{ccc}{W&W_{1}&W_{0|1}\\W_{1}&W_{2}&W_{1|1}\\\dd_{0}&\dd_{1}&\dd_{0|1}}\period
\end{aligned}
\eeq
\paragraph{L=2 :} For $L=2$, there are four different bare operators:
\beq
\dd_{2}=[\tilde{\Phi}^2]\comma\quad \dd_{1|1}=[\tilde{\Phi}]\bb_1\comma\quad \dd_{0|1}=[\,\,]\,\bb_2\comma\quad \dd_{0|1,1}=[\,\,]\,\bb_{1,1}\period
\eeq
By performing the Gram-Schmidt analysis, we obtain for instance
\beq
\begin{aligned}
\NO{\dd_{2}}=\frac{1}{\dmatrix{ccc}{W&W_{1}&W_{0|1}\\W_{1}&W_{2}&W_{1|1}\\W_{0|1}&W_{1|1}&W_{0|1\,;\,1}}}\dmatrix{cccc}{W&W_{1}&W_{0|1}&W_{2}\\W_{1}&W_{2}&W_{1|1}&W_{3}\\W_{0|1}&W_{1|1}&W_{0|1\,;\,1}&W_{2|1}\\\dd_0&\dd_1&\dd_{0|1}&\dd_{2}}\period
\end{aligned}
\eeq
We refrain from writing down the expressions for other operators since they are a bit lengthy. 

By repeating this procedure, one can express any operators in the supersymmetric subsector in terms of  bare operators $\dd_{L|J_1,\ldots J_n}$. Once we have such expressions, we can compute their correlators by decomposing them into the correlators of bare operators, and relate them from the results of localization as
\beq
\langle \prod_{k}\dd_{L_k|J_1^{(k)},\ldots, J_{n_k}^{(k)}}\rangle=(\del_A)^{\sum_k L_k}\left[\langle\mathcal{W}\prod_{k}\bb_{J_1^{(k)},\ldots,J_{n_k}^{(k)}}\rangle\right]\period
\eeq
The correlator inside the square bracket on the right hand side is given by a multi-matrix model as explained in section \ref{subsec:subsector} and in the references \cite{Giombi:2012ep}. 
\subsection{Simplification at large $N_c$\label{subsec:simplification}}
The method described so far in principle allows us to compute arbitrary correlation functions in the supersymmetric subsector. However, as is clear from the examples shown above, the number of operators participating in the Gram-Schmidt process proliferates as the charge increases, making the computation quite complicated in practice.

Below we show that, at first few orders in the $1/N_c$ expansion, one can truncate the operator spectrum and make the formalism more tractable. Such truncation was already implied by the analysis in our previous paper in which we neglected the mixing with the ``multi-trace'' operators but nevertheless reproduced the correct correlators at large $N_c$. The goal of this subsection is to show the existence of a similar truncation also at first few orders in the large $N_c$ expansion.
\paragraph{Truncation at large $N_c$}
Let us first analyze the strict large $N_c$ limit and see the decoupling of the multi-trace operators. 

As exemplified in the previous subsection  (see figure \ref{fig:fig1}), the planar self-contraction cannot produce the ``multi-trace'' operators. This implies that the operators made up purely of insertions on the Wilson loop $\mathcal{W}[\tilde{\Phi}^{L}]$ decouple from the multi-trace operators and form a closed subsector in the planar limit. 

\begin{figure}[t]
\centering
\includegraphics[clip,height=8cm]{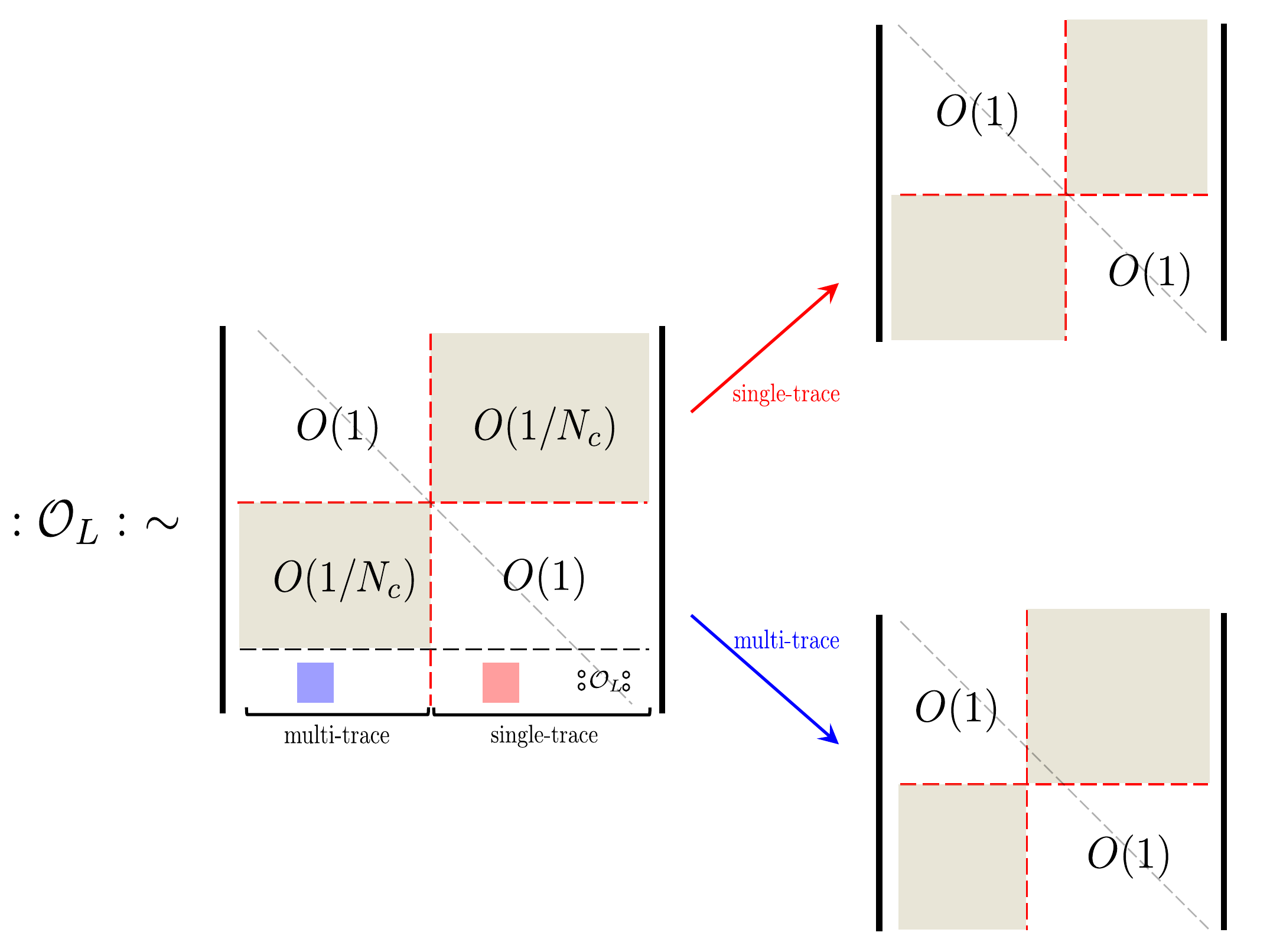}
\caption{The structure of the Gram-Schmidt determinant at large $N_c$. Owing to the large $N_c$ scaling of the two-point functions \eqref{eq:largeNestimate}, the Gram-Schmidt matrix for $\NO{\dd_{L}}$ consist of $O(1)$ diagonal blocks and $O(1/N_c)$ off-diagonal blocks. To faciliate seeing the structure of the matrix, we drew gray dashed lines which indicate the diagonal entries. To read off the contribution of each bare operator, one needs to compute cofactors (minors) of the elements in the last row. The cofactor of the single-trace operator, an example of which is denoted by a red square in the matrix on the left, is given by the determinant at the top-right corner. As is clear from the figure, its diagonal entries are all $O(1)$. On the other hand, the cofactor of the multi-trace operator, an example of which is denoted by a blue square in the matrix on the left, is given by the determinant at the lower-right corner and is $O(1/N_c)$.}
\label{fig:fig4}
\end{figure}
Let us now prove the decoupling more rigorously from the structure of the Gram-Schmidt determinant \eqref{eq:detrep}. For this purpose, we simply need to use the fact that the inner product between the ``single-trace'' operator\fn{Namely the operators made up purely of insertions on the Wilson loop.} $\dd_{L}$ and the ``multi-trace'' operator $\dd_{L^{\prime}|J_1,\ldots, J_n}$ is suppressed by $1/N_c$, which follows from the standard large $N_c$ counting:
\beq\label{eq:innerproductscalingnew}
(\dd_{L} \,,\,\dd_{L^{\prime}|J_1,\ldots, J_n})=\langle \mathcal{W}[\underbracket{\tilde{\Phi}\cdots \tilde{\Phi}}_{L+L^{\prime}}]\,\bb_{J_1,\ldots, J_n}\rangle \sim O(1/N_c)\period
\eeq
To actually prove the decoupling, we first rearrange the rows and the columns of the determinant \eqref{eq:detrep} and express the operator $\NO{\dd_{L}}$ as shown in figure \ref{fig:fig4}. Then, using the large $N_c$ scaling for the inner products \eqref{eq:innerproductscalingnew}, one can show that the off-diagonal blocks (shaded rectangular regions in figure \ref{fig:fig4}) are all of order $O(1/N_c)$ or higher while the diagonal blocks are of order $1$. Now, the coefficient multiplying each bare operator can be read off from the cofactors (also known as minors) of the elements in the last row. As shown in figure \ref{fig:fig4}, the cofactors of the insertions on the loop, $\NOB{\dd_{L^{\prime}}}$ are $O(1)$ since their diagonal entries are all $O(1)$. On the other hand, the cofactors of the multi-trace operators are at most $O(1/N_c)$ since one always needs to take at least one $O(1/N_c)$ element when computing the determinant. This shows that, in the strict large $N_c$ limit, the contribution from the multi-trace operators to the single-trace operators is negligible and one can therefore focus on the insertions on the Wilson loop.

Now, after the decoupling of all the multi-trace operators, the Gram-Schmidt analysis simplifies greatly since there is only one operator left for each $R$-charge. We thus obtain the following expression for the normal-ordered insertions at large $N_c$,
\beq\label{eq:defcircledot}
 \NOB{\dd_L}\equiv \frac{1}{D_L}\dmatrix{cccc}{W&W_1&\cdots&W_{L}\\W_{1}&W_{2}&\cdots&W_{L+1}\\ \vdots&\vdots&\ddots&\vdots\\ W_{L-1}&W_{L}&\cdots&W_{2L-1}\\ \dd_0&\dd_1&\cdots&\dd_L}\comma\qquad D_L\equiv \det{}_{i,j} W_{i+j-2}\quad (1\leq i,j\leq L)\comma
 \eeq
where $\dd_L$ is given by \eqref{eq:defdgeneral},  namely $\dd_L\equiv \mathcal{W}[\tilde{\Phi}^{L}]$, and the symbol $\NOB{\ast}$ denotes the normal ordering at large $N_c$. This is precisely the expression we derived in the previous paper \cite{Giombi:2018qox}, which correctly reproduces the perturbative results at weak and strong coupling.

For the analysis of the subleading corrections which we perform below, it is helpful to extend the definition of the large-$N_c$ normal ordering also to the multi-trace operators. The extension is simple; we simply define them as a product of $\NOB{\dd_{L}}$ and the bulk multi-trace operator:
\beq\label{eq:deflargeNcnormal}
\NOB{\dd_{L|J_1,\ldots,J_n}}\equiv \NOB{\dd_{L}}\,\,\bb_{J_1,\ldots,J_n}\period
\eeq
One can easily check that these operators are orthogonal to each other at large $N_c$, namely
 \beq\label{eq:largeNestimate}
 (\NOB{\dd_{L|{\bf J}}}\,,\,\NOB{\dd_{L^{\prime}|{\bf J}^{\prime}}}) \sim\begin{cases}O(1)\qquad &(L=L^{\prime} \text{ and }{\bf J}={\bf J}^{\prime})\\O(1/N_c)\text{ or higher}\qquad &(L\neq L^{\prime}\text{ or } {\bf J}\neq {\bf J}^{\prime})\end{cases}\comma
 \eeq
 where ${\bf J}$ and ${\bf J}^{\prime}$ denote sets of $J$'s, namely ${\bf J}=\{J_1,\ldots, J_n\}$ and ${\bf J}^{\prime}=\{J^{\prime}_{1},\ldots, {J}^{\prime}_{n^{\prime}}\}$.
In what follows, we use these operators as the basis for analyzing the subleading corrections.
 \paragraph{Correction at $O(1/N_c)$}
 At $O(1/N_c)$, the operators $\NOB{\dd_{L}}$'s can mix with the {\it double-trace} operators $\NOB{\dd_{\ell|j}}\simeq \mathcal{W}[\tilde{\Phi}^{\ell}]{\rm tr}[\tilde{\Phi}^{j}]$ since their two-point functions, or equivalently the inner products in the Gram-Schmidt language, are no longer orthogonal at $O(1/N_c)$:
 \beq
 \langle \NOB{\dd_L}\NOB{\dd_{\ell|j}}\rangle = W_{L+\ell|j}+\cdots \,\, \sim \,\,O(1/N_c)\period
 \eeq
  To restore the orthogonality, we need perform the following subtraction from $\NOB{\dd_L}$,
 \beq\label{eq:subtractnext}
 \NO{\dd_L}=\NOB{\dd_L}-S_1+O(1/N_c^2)\comma
 \eeq 
where $S_1$ is $O(1/N_c)$ and is given by
\beq
S_1=\sum_{\substack{\ell+j<L\\0\leq \ell,\,1\leq j}}\frac{\langle \NOB{\dd_{\ell}}\NOB{\dd_{\ell|j}}\rangle}{\langle \NOB{\dd_{\ell|j}}\NOB{\dd_{\ell|j}}\rangle}\NOB{\dd_{\ell|j}}\period
\eeq
Here the summation range is restricted to $\ell+j<L$ since the self-contraction of $L$ single-letter insertions only produces the operators with charge smaller than $L$ as was exemplified in section \ref{sec:basic}. At this order, there is no need for further subtraction since the two-point functions with higher-trace operators are $O(1/N_c^2)$ or higher\fn{This can be shown from the standard large $N_c$ counting.} and their mixing effects can be neglected.
As we will see in section \ref{sec:largeN}, the subtraction \eqref{eq:subtractnext} is necessary in order to reproduce the correct bulk-defect correlators. 
 \paragraph{Correction at $O(1/N_c^2)$}  
 Now at the next order, there are several sources of corrections. Firstly, the operator $\NOB{\dd_L}$ can now mix with the  {\it triple-trace} operators $\NOB{\dd_{\ell|\{j_1,j_2\}}}$. This mixing can be resolved by subtracting the following term from $\NOB{\dd_L}$:
 \beq
 S_2^{(1)}= \sum_{\substack{\ell + j_1+j_2<L\\0\leq \ell,\,1\leq  j_1,j_2}} \frac{\langle \NOB{\dd_L}\NOB{\dd_{\ell|j_1,j_2}}\rangle}{\langle \NOB{\dd_{\ell|j_1,j_2}}\NOB{\dd_{\ell|j_1,j_2}}\rangle}\NOB{\dd_{\ell|j_1,j_2}}\period
 \eeq
 
 Secondly, because of the subtraction added at $O(1/N_c)$ \eqref{eq:subtractnext}, different $\NOB{\dd_L}$'s are no longer orthogonal to each other. For instance the two-point function of $\NO{\dd_L}$ and $\NO{\dd_{\tilde{\ell}}}$ ($L\neq \tilde{\ell}$) is given by
 \beq
 \begin{aligned}
 \langle \left.\NO{\dd_L}\right|_{1/N_c}\left.\NO{\dd_{\tilde{\ell}}}\right|_{1/N_c}\rangle=&-\sum_{\substack{\ell+j<L\\0\leq \ell,\,1\leq j}}\frac{\langle\NOB{\dd_L}\NOB{\dd_{\ell|j}}\rangle\langle\NOB{\dd_{\tilde{\ell}}}\NOB{\dd_{\ell|j}}\rangle}{\langle\NOB{\dd_{\ell|j}}\NOB{\dd_{\ell|j}}\rangle}\\
 &-\sum_{\substack{\ell+j<\tilde{\ell}\\0\leq \ell,\,1\leq j}}\frac{\langle\NOB{\dd_L}\NOB{\dd_{\ell|j}}\rangle\langle\NOB{\dd_{\tilde{\ell}}}\NOB{\dd_{\ell|j}}\rangle}{\langle\NOB{\dd_{\ell|j}}\NOB{\dd_{\ell|j}}\rangle}\\
 &+\sum_{\substack{\ell+j<L,\tilde{\ell}\\0\leq \ell,\,1\leq j}}\frac{\langle\NOB{\dd_L}\NOB{\dd_{\ell|j}}\rangle\langle\NOB{\dd_{\tilde{\ell}}}\NOB{\dd_{\ell|j}}\rangle}{\langle\NOB{\dd_{\ell|j}}\NOB{\dd_{\ell|j}}\rangle}\period
\end{aligned}
 \eeq
 Here, the first and the second terms come from the $O(1/N_c)$ corrections to $\NOB{\dd_L}$ and $\NOB{\dd_{\tilde{\ell}}}$ while the third term comes from the corrections to both of the operators.

One can restore the orthogonality by further subtracting the following term from $\NOB{\dd_L}$:
 \beq
 S_2^{(2)}=-\sum_{\tilde{\ell}<L}\,\,\,\sum_{\ell+j<L}\frac{\langle \NOB{\dd_L},\NOB{\dd_{\ell|j}}\rangle\langle\NOB{\dd_{\tilde{\ell}}}\NOB{\dd_{\ell|j}}\rangle}{\langle\NOB{\dd_{\ell|j}}\NOB{\dd_{\ell|j}}\rangle\langle\NOB{\dd_{\tilde{\ell}}}\NOB{\dd_{\tilde{\ell}}}\rangle}\NOB{\dd_{\tilde{\ell}}}\period
 \eeq
 Again the summation range is restricted to $\tilde{\ell}<L$ due to the structure of the self-contraction.
 
 The last correction comes from the $O(1/N_c^2)$ correction to the expectation value of the Wilson loop and its derivatives $W_k$. This correction can be easily taken into account by taking the expression \eqref{eq:defcircledot}, and expand it to the next order.
 
 Therefore, everything combined, we obtain the following expression for $\NO{\dd_L}$ at $O(1/N_c^2)$:
 \beq
 \begin{aligned}
 \NO{\dd_L}=\left.\NOB{\dd_L}\right|_{1/N_c^2}-\underbrace{S_1}_{O(1/N_c)}-\underbrace{(S_2^{(1)}+S_2^{(2)})}_{O(1/N_c^2)}+O(1/N_c^3)\period
 \end{aligned}
 \eeq
 Here the notation $\left.\NOB{\dd_L}\right|_{1/N_c^2}$ denotes the determinant \eqref{eq:defcircledot} expanded up to $O(1/N_c^2)$. In section \ref{sec:nonplanar}, we will use this expression to compute the corrections to the correlators on the Wilson loop and check them against the direct perturbative computation.
 \section{Bulk-defect correlators at large $N_c$\label{sec:largeN}}
In this section, we compute the bulk-defect correlators
\beq\label{eq:whattocompute}
G_{L|J}\equiv \langle\mathcal{W}[\NO{\tilde{\Phi}^{L}}] \,\,\bb_J \rangle\comma
\eeq
at large $N_c$ using the normal-ordered operators that we constructed in the previous section. Generalization to two operator insertions on the Wilson loop, namely 
\beq\label{eq:onetwo}
G_{L_1,L_2|J}\equiv \langle\mathcal{W}[\NO{\tilde{\Phi}^{L_1}}\NO{\tilde{\Phi}^{L_2}}] \,\,\bb_J \rangle\comma
\eeq
 will be presented in Appendix \ref{ap:onetwo}.
\subsection{Generalities\label{subsec:general}}
To compute the result at large $N_c$, we simply need to substitute the large $N_c$ expansion of $\NO{\tilde{\Phi}^{L}}=\NO{\dd_L}$ \eqref{eq:subtractnext} to \eqref{eq:whattocompute}. The result is
\beq\label{eq:gljexpanded}
G_{L|J}=\langle \NOB{\dd_L}\bb_{J}\rangle-\sum_{\substack{L_1+J_1<L\\0\leq L_1,\,1\leq J_1}}\underbrace{\frac{\langle \NOB{\dd_L}\NOB{\dd_{L_1|J_1}}\rangle}{\langle \NOB{\dd_{L_1|J_1}}\NOB{\dd_{L_1|J_1}}\rangle}}_{O(1/N_c)}\underbrace{\langle\NOB{\dd_{L_1|J_1}}\bb_{J}\rangle}_{O(1)\text{ or }O(1/N_c)}+O(1/N_c^2)\period
\eeq
As indicated, the first factor in the sum is always $O(1/N_c)$ due to the large $N_c$ scaling given in \eqref{eq:largeNestimate}. On the other hand, the second factor in the sum is $O(1)$ if $J_1=J$ and $L_1=0$ while in all the other cases it is $O(1/N_c)$. Owing to the range of summation, the case with $J_1=J$ and $L_1=0$ is included in the sum only if $J<L$, and when it is included it just kills the leading term $\langle \NOB{\dd_L}\bb_{J}\rangle$. Therefore we arrive at the following result:
\beq
\begin{aligned}
G_{L|J}=\begin{cases}0&\qquad (J<L)\\\langle \NOB{\dd_L}\bb_J\rangle+O(1/N_c^2)&\qquad (J\geq L)\end{cases}\period
\end{aligned}
\eeq 
In fact $G_{L|J}$ for $J<L$ actually vanishes at any order of the $1/N_c$ expansion, not just at the leading order. This is easy to see in the case of the 1/2-BPS Wilson loop from $R$-symmetry selection rules, but it is also true in the general $1/8$-BPS case.\footnote{This can be seen to follow from the structure of the Gram-Schmidt process. Indeed,  by using the bulk-defect OPE, one can express the bulk operator $\hat O_J$ as a sum of defect operators with charges smaller than or equal to $J$. As a result, the bulk-defect correlator is given by a sum of the defect two-point functions. One can then use the orthogonality of the normal-ordered defect operators constructed in the Gram-Schmidt process to show that all these two-point functions vanish.}

To compute the leading expression $\langle \NOB{\dd_L}\bb_J\rangle$, it is convenient to consider the following polynomial \cite{Giombi:2018qox}:
\beq
F_{L}(X)=\frac{1}{D_L}\dmatrix{cccc}{W&W_1&\cdots&W_{L}\\W_{1}&W_{2}&\cdots&W_{L+1}\\ \vdots&\vdots&\ddots&\vdots\\ W_{L-1}&W_{L}&\cdots&W_{2L-1}\\ 1&X&\cdots&X^{L}}\period
\eeq
This reproduces the expression for $\NOB{\dd_L}$ \eqref{eq:defcircledot} once we replace $X^{k}$ by $\dd_k$. Now using the fact that the insertion of the bare operators $\dd_k(=\tilde{\Phi}^{k})$ can be traded for the area derivative $\del_A^{k}$, one can write down the following expression for $\langle\mathcal{W}[\NO{\tilde{\Phi}^{L}}]\bb_J\rangle$,
\beq\label{eq:leadingF}
\langle\mathcal{W}[\NO{\tilde{\Phi}^{L}}]\bb_J\rangle|_{N_c\to \infty}=\langle\NOB{\dd_L}\bb_J\rangle=F_{L}[\del_{A^{\prime}}]\left.\left[\langle\mathcal{W}\bb_J\rangle (A^{\prime})\right]\right|_{A^{\prime}=A}\period
\eeq
Let us make some clarification of this formula: In \eqref{eq:leadingF}, $A^{\prime}$ is the area of the region inside the Wilson loop in $\langle\mathcal{W}\bb_J \rangle$, and the area derivatives only act on $\langle \mathcal{W}O_J\rangle (A^{\prime})$ not on the coefficients of $F_{L}$.

One can also generalize this expression to the correlators involving several operator insertions on a single Wilson loop. The result for the generalization reads
\beq\label{eq:leadingFmulti}
\begin{aligned}
\langle \mathcal{W}[\prod_{k=1}^{n}\NO{\tilde{\Phi}^{L_k}}]\bb_J\rangle|_{N_c\to \infty}=\prod_{k=1}^{n}F_{L_k}[\del_{A^{\prime}}]\left.\left[\langle\mathcal{W}\bb_J\rangle (A^{\prime})\right]\right|_{A^{\prime}=A}\period
\end{aligned}
\eeq
For $n\geq 3$, this expression is valid when $J\geq L_i$ $(i=1,\ldots,n)$, and otherwise there are some corrections coming from the nonplanar corrections to $\NO{\tilde{\Phi}^{L_k}}$. This rule is slightly modified for $n=2$ as will be discussed in Appendix \ref{ap:onetwo}.
\subsection{Integral representation and Quantum Spectral Curve\label{subsec:integral}}
Let us now evaluate \eqref{eq:leadingF} more explicitly using the large $N_c$ expression for $\langle \mathcal{W}\bb_J\rangle$ given in \eqref{eq:WOWW}. For this purpose, it is useful to consider the following generating series\fn{This can be derived from the usual generating function for the modified Bessel function, $e^{\frac{\sqrt{\lambda^{\prime}}}{2}(y+\frac{1}{y})}=\sum_n I_n (\sqrt{\lambda^{\prime}})y^n$, after the change of variables $y=\sqrt{\frac{2\pi+a}{2\pi-a}}x$.}
\beq
\begin{aligned}
e^{2\pi g (x+\frac{1}{x})}e^{g a (x-\frac{1}{x})}=\sum_{n=-\infty}^{\infty}I_n (\sqrt{\lambda^{\prime}}) \left(\frac{2\pi +a}{2\pi -a}\right)^{n/2}x^n\comma
\end{aligned}
\eeq
where $g$ is related to the 't Hooft coupling $\lambda$ as
\beq
g\equiv \frac{\sqrt{\lambda}}{4\pi}\period
\eeq
 Using this generating series, $\langle \mathcal{W}\bb_J\rangle$ at large $N_c$ can be expressed as
 \beq
 \begin{aligned}
 \langle \mathcal{W}\bb_J\rangle=\frac{2^{-J/2}\sqrt{J}}{N_c}\oint \frac{dx \,x^{J-1}}{2\pi i}e^{2\pi g (x+\frac{1}{x})}e^{g a(x-\frac{1}{x})}\period
 \end{aligned}
 \eeq
 In this expression, the area-dependence only shows up in the exponent and we therefore have
 \beq
 (\del_A)^{k}\langle \mathcal{W}\bb_J\rangle=\frac{2^{-J/2}\sqrt{J}}{N_c}\oint \frac{dx \,x^{J-1}}{2\pi i}g^{k}\left(x-\frac{1}{x}\right)^{k}e^{2\pi g (x+\frac{1}{x})}e^{g a(x-\frac{1}{x})}\period
 \eeq
 We can then convert \eqref{eq:leadingF} into a simple integral expression
 \beq\label{eq:toexpandW}
 \langle \mathcal{W}[\NO{\tilde{\Phi}^{L}}]\bb_J\rangle =\frac{ 2^{-J/2}\sqrt{J}}{N_c}\oint d\mu \,\,B_J(x)Q_L(x)\comma 
 \eeq
 where $d\mu$ is the measure introduced in \cite{Giombi:2018qox},
 \beq\label{eq:measureprevious}
 d\mu =\frac{dx}{2\pi i}\frac{1+x^{-2}}{2} \frac{e^{2\pi g (x+\frac{1}{x})}e^{g a(x-\frac{1}{x})}}{2\pi g}\comma
 \eeq
 and $Q_L(x)$ and $B_J(x)$ are defined by
 \beq\label{eq:defofQLFLBJ}
 Q_L(x)\equiv F_{L}(g (x-x^{-1}))\comma\qquad B_J(x)=\frac{4\pi g x^{J+1}}{1+x^2}\period
 \eeq
 
 As shown in \cite{Giombi:2018qox}, the functions $Q_L(x)$'s are orthogonal to each other under the measure $\mu$ and their overlap integral gives the planar two-point function of the operators on the Wilson loop:
 \beq
 \langle\mathcal{W}[\NO{\tilde{\Phi}^{L}}\,\,\NO{\tilde{\Phi}^{L^{\prime}}}]\rangle|_{\rm planar} =\oint d\mu \,\, Q_{L}(x)Q_{L^{\prime}}(x)=\frac{D_{L+1}}{D_L}\delta_{L\,L^{\prime}}\period
 \eeq
 Furthermore, $Q_L(x)$ coincides with the so-called $Q$-function in the Quantum Spectral Curve formalism\fn{The same orthogonal polynomials appeared before in the integrability-based analysis of the operator spectrum on the Wilson loop \cite{Gromov:2012eu,Gromov:2013qga}.} \cite{Gromov:2015dfa,Gromov:2013pga}. The appearance of the Quantum Spectral Curve in \eqref{eq:toexpandW} is rather striking and indicates the existence of the integrability-based description for the bulk-defect correlators. We will come back to this point in the conclusion.
 
 One can also write down the result for multiple operator insertions on the Wilson loop. The result reads
 \beq\label{eq:generalizedintegralexpressions}
\langle \mathcal{W}[\prod_{k=1}^{n}\NO{\tilde{\Phi}^{L_k}}]\bb_J\rangle =\frac{ 2^{-J/2}\sqrt{J}}{N_c}\oint d\mu \,\,B_J(x)\prod_{k=1}^{n}Q_{L_k}(x)\period
 \eeq
 In what follows, we focus on the case of the $1/2$-BPS circular Wilson loop (namely we set $a=0$) and derive the weak- and the strong-coupling expansions of the integral \eqref{eq:toexpandW}.
\subsection{Weak coupling expansion\label{subsec:weak}}
To perform the weak-coupling expansion of the integral \eqref{eq:toexpandW}, we need to expand both the functions $Q_L$ and $B_J$, and the measure $\mu$. The expansion of $Q_L$ was computed in \cite{Giombi:2018qox} and it reads
\beq\label{eq:QLweakexpfirst}
Q_{L}=(-i g)^{L}\left(U_{L}(\cos \theta)+g^2\frac{2\pi ^2}{3}U_{L-2}(\cos\theta)+\cdots\right)\comma
\eeq
where $U_L$ is the Chebyshev polynomial of the second kind and $\theta$ is related to $x$ by
\beq
x=-i e^{i\theta}\period
\eeq
In terms of the variable $\theta$, the measure $d\mu$, $B_{L}$ and their product read\fn{
Note that we already set $a=0$ since we are studying the $1/2$-BPS Wilson loop.}
\beq\label{eq:weakvariousquantities}
\begin{aligned}
&d\mu =\frac{d\theta \sin \theta}{4\pi^2 g} e^{4\pi g\sin \theta}\comma\qquad B_J=\frac{2\pi g}{\sin \theta} \frac{e^{iJ\theta}}{i^{J}}\comma\qquad d\mu \,\,B_J= \frac{d\theta\,e^{4\pi g\sin \theta}}{2\pi }\frac{e^{iJ \theta}}{i^{J}}\comma
\end{aligned}
\eeq
and the weak-coupling expansion corresponds to expanding the exponential $e^{2g\sin \theta}$. Since $U_L$ is an order $L$ polynomial, one has to expand $e^{2g\sin \theta}$ up to the $O(g^{J-L})$ term in order to compensate the factor $e^{iJ\theta}$ and get a non-zero integral:
\beq
d\mu \,\,B_J=\frac{d\theta}{2\pi }\frac{e^{iJ \theta}}{i^{J}}\left(1+\cdots + g^{J-L}\frac{2^{J-L}(2\pi)^{J-L}}{(J-L)!}\sin^{J-L}\theta+\cdots \right)\period
\eeq
Plugging in that term in the integral, we get
\beq
\begin{aligned}
\langle \mathcal{W}[\NO{\tilde{\Phi}^{L}}]\bb_J\rangle|_{\rm leading} &= \frac{2^{-J/2}\sqrt{J}}{N_c}(-i)^{J+L}\frac{(4\pi)^{J-L}g^{J}}{(J-L)!}\int \frac{d\theta}{2\pi}e^{iJ \theta}\sin^{J-L}\theta\,\,
U_L(\cos\theta)\\
&=\frac{(-1)^{L}2^{-J/2}\sqrt{J}g^{J}}{N_c }\frac{(2\pi)^{J-L}}{(J-L)!}\comma
\label{bd-loc-weak}
\end{aligned}
\eeq
where we used in the second equality,
\beq\label{eq:chebyshev}
U_{L}(\cos \theta)=\frac{\sin (L+1) \theta}{\sin \theta}\period
\eeq

One can also compute the higher orders in the expansion. At the order $O(g^{J+1})$, one can show by explicit computation that the contribution vanishes. This is of course in line with the perturbation theory in which the corrections always come with even powers of $g$ (recall that $g\propto \sqrt{\lambda}$). Now, at the order $O(g^{J+2})$, the contribution only comes from the correction to $d\mu \,\,B_J$, not from the correction to $Q_L$\fn{The correction to $Q_L$ is of $O(g^{L+2})$ and proportional to $U_{L-2}$. In order to make the integral involving $U_{L-2}$ nonvanishing, we need to take the $O(g^{J-L+2})$ term in the expansion of $d\mu\,\, B_J$. In total, it is of order $O(g^{J+4})$.}. Thus the result reads
\beq\label{eq:integralexpnext}
\begin{aligned}
\langle \mathcal{W}[\NO{\tilde{\Phi}^{L}}]\bb_J\rangle|_{\rm next} &= \frac{2^{-J/2}\sqrt{J}}{N_c}(-i)^{J+L}\frac{(4\pi)^{J-L+2}g^{J+2}}{(J-L+2)!}\int \frac{d\theta}{2\pi}e^{iJ \theta}\sin^{J-L+2}\theta\,\,
U_L(\cos\theta)\\
&=\frac{(-1)^{L}2^{-J/2}\sqrt{J}g^{J+2}}{N_c (J-L)!}\frac{(2\pi)^{J-L+2}}{J-L+2}\period
\end{aligned}
\eeq
Here in the second equality we assumed $L$ is strictly larger than $0$. We will comment on this point later in this section. 

By dividing these results by the expectation of the Wilson loop, 
\beq
\langle \mathcal{W}\rangle=1+2\pi^2 g^2+\cdots\comma
\eeq
we obtain the normalized correlators
\beq
\llangle \mathcal{W}[\NO{\tilde{\Phi}^{L}}]\bb_J\rrangle=\frac{(-1)^{L}2^{-J/2}\sqrt{J}g^{J}}{N_c }\frac{(2\pi)^{J-L}}{(J-L)!}\left[1-2\pi^2g^2\frac{J-L}{J-L+2}+O(g^4) \right]\period
\eeq
We can then read off the defect CFT data using the relations \eqref{eq:Gandsmallc} and  \eqref{eq:Gandbigc} as
\beq
\begin{aligned}
c_{L|J}&=\frac{2^{L-\frac{J}{2}}\sqrt{J}g^{J}}{N_c }\frac{(2\pi)^{J-L}}{(J-L)!}\left[1-2\pi^2g^2\frac{J-L}{J-L+2}+O(g^4) \right]\comma\\
C_{L|J}&=\frac{2^{L-\frac{J}{2}}(-i)^{L}\sqrt{J}g^{J-L}}{N_c }\frac{(2\pi)^{J-L}}{(J-L)!}\left[1-2\pi^2 g^2\left(\frac{J-L}{J-L+2}-\frac{1}{3}\right)+O(g^4) \right]\comma
\end{aligned}
\eeq
where we used
\beq
n_L=(-g^{2})^{L}\left[1-\frac{2\pi^2 g^2}{3}+O(g^4)\right]\comma
\eeq
which was computed in \cite{Giombi:2018qox}. As we will see in the next section, these results are in perfect agreement with the direct perturbative computation.

\begin{figure}
\centering
\includegraphics[clip,height=3cm]{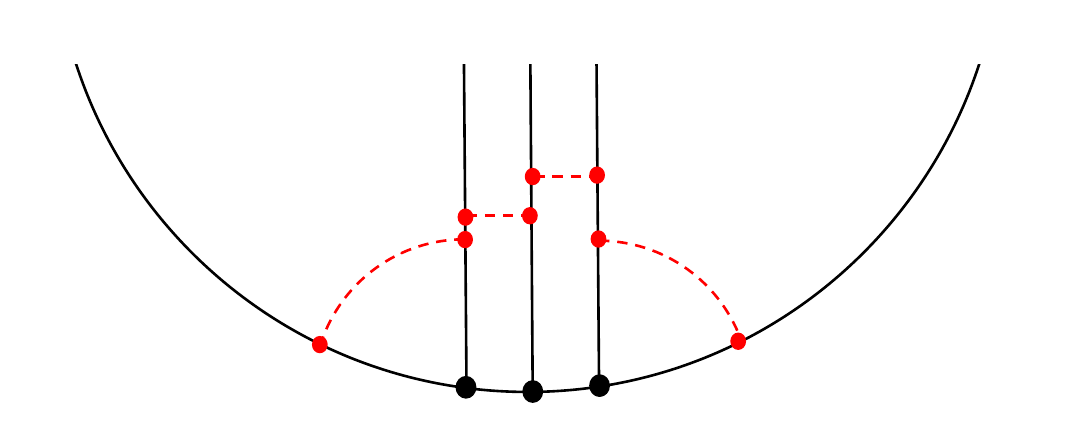}
\caption{An example of the wrapping diagram for the bulk-defect correlator. The black dots denote the field insertions $\tilde{\Phi}$ and the black straight lines emanating from them are the propagators. When the loop order $\ell$ is larger than $L+1$, one can connect the two edges of the Wilson loop (denoted by a black semi-circle) using a collection of gluon propagators (denoted by red dashed lines).}
\label{fig:fig2}
\end{figure}
Before closing this section, let us make one remark: In the computation of the next-leading correction \eqref{eq:integralexpnext}, we assumed that $L$ is strictly larger than zero. In that case one can replace $U_L = \sin (L+1)\theta/\sin \theta$ in the integrand with $-e^{-i(L+1)\theta}/(2i\sin\theta)$ since the integral of the other exponential $e^{+i(L+1)\theta}/(2i\sin\theta)$ vanishes. On the other hand, for $L=0$ the other exponential does contribute and modifies the result to
\beq
\begin{aligned}
\langle \mathcal{W}\bb_J\rangle|_{\rm next} &= \frac{2^{-J/2}\sqrt{J}}{N_c}(-i)^{J}\frac{(4\pi)^{J+2}g^{J+2}}{(J+2)!}\int \frac{d\theta}{2\pi}e^{iJ \theta}\sin^{J+2}\theta\,\,
U_0(\cos\theta)\\
&=\frac{2^{-J/2}\sqrt{J}(2\pi g)^{J+2}}{N_c (J+1)!}\period
\end{aligned}
\eeq 
Because of this, one cannot simply analytically continue the $L>0$ result to get the result for $L=0$. More generally, at $\ell$ loops, the other exponential starts to contribute when the length $L$ is equal to  $\ell-1$ and the result therefore exhibits non-analyticity beyond that point. Although it might appear peculiar at first sight, such non-analyticity is not unheard of, and shows up also in the study of the anomalous dimensions\fn{An analogous phenomenon was found also for the structure constants of the single-trace operators \cite{Basso:2015zoa}.} of the single-trace operators \cite{Beisert:2004hm,Sieg:2005kd}; in that context, it is known that when the loop order exceeds the length of the operator a new type of planar diagrams which wrap around the operator show up and produce non-analyticity in the length $L$. Such a phenomenon is called the wrapping correction.\fn{From the integrability point of view, these wrapping corrections correspond to the finite-size corrections which are produced by the virtual particles going around the spin chain \cite{Ambjorn:2005wa}.} This seems to be true also in our case since precisely at $L=\ell-1$ one can draw a new diagram which passes through the defect operator and connects the two edges of the Wilson loop (see figure \ref{fig:fig2}). It would be interesting to make this connection more precise by directly analyzing the Feynman diagrams.
\subsection{Strong coupling expansion\label{subsec:strong}}
To perform the strong-coupling expansion, it is convenient to rewrite the integral in terms of
\beq
y\equiv \frac{i(x-x^{-1})}{2}\period
\eeq
The result after rewriting reads
\begin{align}\label{eq:strongstarting}
\begin{aligned}
\oint d\mu \, B_J =\int^{1}_{-1}\frac{dy}{2\pi \sqrt{1-y^2}}&\left[(\sqrt{1-y^2}-iy)^{J}e^{4\pi g \sqrt{1-y^2}}\right.\\
&\left.+(-\sqrt{1-y^2}-iy)^{J}e^{-4\pi g \sqrt{1-y^2}}\right]\period
\end{aligned}
\end{align}
Here we converted the $x$-integral to the integral of $y$ around the cut $[-1,1]$ and then further rewrote it as an integral along the cut by summing the contributions from the contour above and below the cut.

At strong coupling, the integral can be approximated by its saddle point. In particular, if $J$ and $L$ remain finite, the saddle point equation is determined purely by the exponential factors $e^{\pm 4\pi g \sqrt{1-y^2}}$ and it sets the saddle point to be at $y=0$. In this case, one just needs to keep the first term in \eqref{eq:strongstarting} since the second term is always exponentially suppressed:
\beq
\oint d\mu \,\,B_J\sim\int^{1}_{-1}\frac{dy}{2\pi \sqrt{1-y^2}}(\sqrt{1-y^2}-iy)^{J}e^{4\pi g \sqrt{1-y^2}}\qquad (g\gg 1)\period
\eeq
To study the fluctuation around the saddle point, it is convenient to rewrite it in terms of
\beq
t=\sqrt{2\pi g}y\comma
\eeq
as was done in \cite{Giombi:2018qox}. The result reads
\begin{align}
\oint d\mu &=\int^{\infty}_{-\infty}\frac{dt\, e^{4\pi g}e^{-t^2}}{(2\pi)^{5/2}g^{3/2}}\left[1-\frac{t^4}{8\pi g}+\cdots\right]\comma\\
B_{J}&=2\pi g \sum_{k=0}^{\infty}\frac{(-1)^{k}}{k!}\left(i t\sqrt{\frac{2}{\pi g}}\right)^{k}\frac{\Gamma[\frac{1+J+k}{2}]}{\Gamma[\frac{1+J-k}{2}]}\period\label{eq:Bjexpandedstrong}
\end{align}

Now, as discussed in \cite{Giombi:2018qox}, the functions $Q_L$ form a set of orthogonal polynomials in $t$ whose highest power is given by\fn{Note that this is true at finite $g$, not just in the strong coupling limit.}
\beq
Q_L (t)=(-i)^{L} \left(\frac{2g}{\pi}\right)^{L/2}t^{L}+\cdots\period
\eeq 
This in particular means that $Q_L(t)$ is orthogonal to any polynomials whose degree is less than $L$ under the measure $d\mu$. Therefore, the leading strong coupling answer comes from the $O(t^{L})$ term in the expansion of $B_J$ \eqref{eq:Bjexpandedstrong}, which leads to the following integral:
\beq\label{eq:tQintegral}
\begin{aligned}
&\int d\mu\,\,t^{L}Q_{L}(t)=i^{L}\left(\frac{\pi}{2g}\right)^{L/2}\int d\mu\,\,Q_{L}(t)Q_{L}(t)\\
&=\left(-i\sqrt{\frac{g}{2\pi}}\right)^{L}\frac{e^{4\pi g}L!}{2(2 g)^{3/2}\pi^2}\left[1-\frac{3}{32\pi g}(2L^2+2L+1)+O(g^{-2})\right]\comma
\end{aligned}
\eeq
Here in the last equality we used the strong-coupling expansion of the overlap integral of $Q_L$'s computed in \cite{Giombi:2018qox}:
\beq\label{eq:QLoverlap}
\int d\mu\,Q_L (t)Q_L(t)=(-1)^{L}\frac{e^{4\pi g}L!}{2(2 g)^{3/2}\pi^2}\left(\frac{g}{\pi}\right)^{L}\left[1-\frac{3}{32\pi g}(2L^2+2L+1)+O(g^{-2})\right]\period
\eeq
 Then the leading answer at strong coupling can be evaluated as
\beq
\label{bd-leading}
\langle \mathcal{W}[\NO{\tilde{\Phi}^{L}}]\bb_J\rangle|_{\rm leading} = \frac{2^{-J/2}\sqrt{J}}{N_c}\frac{e^{4\pi g}}{2\sqrt{2g}}\left(\frac{-1}{\pi}\right)^{L}\frac{\Gamma[\frac{1+J+L}{2}]}{\Gamma[\frac{1+J-L}{2}]}\period
\eeq

At the next order, there are two sources of corrections: The first one comes from the subleading term in \eqref{eq:tQintegral} while the second one comes from the $t^{L+2}$ term in the expansion of $B_J$ \eqref{eq:Bjexpandedstrong}, which gives the integral $\int d\mu \,t^{L+2} Q_L$. To evaluate this integral, we use the strong coupling result for $Q_L(t)$ \cite{Giombi:2018qox}:
\beq
Q_L(t)=(-i)^{L}\left(\frac{g}{2\pi}\right)^{L/2}H_L (t)\period
\eeq
Here $H_L$ is the $L$-th Hermite polynomial. We can then evaluate the integral using the property of the Hermite polynomial as
\beq
\begin{aligned}
\int d\mu\,\, t^{L+2}Q_{L}(t)&=\left(i\sqrt{\frac{\pi}{2g}}\right)^{L+2}\int d\mu \,\, \left(Q_{L+2}(t)-\frac{g}{2\pi}\frac{(L+2)!}{L!}Q_L(t)+\cdots \right) Q_L (t)\comma\\
&=\frac{1}{4}\left(-i\sqrt{\frac{g}{2\pi}}\right)^{L}\frac{e^{4\pi g}(L+2)!}{2(2 g)^{3/2}\pi^2}\left[1 +O(g^{-1})\right]\period
\end{aligned}
\eeq
Here we used the orthogonality of $Q_L$'s and the overlap integral \eqref{eq:QLoverlap}.

Combining the two contributions, we obtain the following expression for the next-leading correction:
\beq
\begin{aligned}
\frac{\langle \mathcal{W}[\NO{\tilde{\Phi}^{L}}]\bb_J\rangle|_{\rm next}}{\langle \mathcal{W}[\NO{\tilde{\Phi}^{L}}]\bb_J\rangle|_{\rm leading}}&=-\frac{3(2L^2+2L+1)}{32\pi g}-\frac{(J-L-1)(J+L+1)}{8\pi g}\\
&=\frac{1-4J^2 +2L-2L^2}{32\pi g}\period
\end{aligned}
\eeq

Now, by dividing these results by the expectation value of the Wilson loop,
\beq\label{eq:expectationvaluewilsonstrong}
\langle \mathcal{W}\rangle=\frac{e^{4\pi g}}{2(2 g)^{3/2}\pi^2}\left[1-\frac{3}{32\pi g}+\cdots\right]\comma
\eeq
we can compute the normalized correlator as
\beq
\label{loc-strong}
\llangle\mathcal{W} [\NO{\tilde{\Phi}_{L}}]\bb_J\rrangle =\frac{2^{-J/2}\sqrt{J}}{N_c}\frac{2\pi g}{(-\pi)^{L}}\frac{\Gamma[\frac{1+J+L}{2}]}{\Gamma[\frac{1+J-L}{2}]}\left[1+\frac{2-2J^2+L-L^2}{16\pi g}+O(g^{-2})\right]\period
\eeq
We can also read off the defect CFT data using \eqref{eq:Gandsmallc} and  \eqref{eq:Gandbigc}:
\beq
\begin{aligned}
c_{L|J}&=\frac{2^{L-\frac{J}{2}}\sqrt{J}}{N_c}\frac{2\pi g}{\pi^{L}}\frac{\Gamma[\frac{1+J+L}{2}]}{\Gamma[\frac{1+J-L}{2}]}\left[1+\frac{2-2J^2+L-L^2}{16\pi g}+O(g^{-2})\right]\comma\\
C_{L|J}&=\frac{2^{L-\frac{J}{2}}(-i)^{L}\sqrt{J}}{N_c}\frac{2\pi g}{(\pi g)^{L/2}\sqrt{L!}}\frac{\Gamma[\frac{1+J+L}{2}]}{\Gamma[\frac{1+J-L}{2}]}\left[1+\frac{4-4J^2+5L+L^2}{32\pi g}+O(g^{-2})\right]\period
\end{aligned}
\eeq 
Here we used the strong-coupling expansion of $n_L$ computed in \cite{Giombi:2018qox}:
\beq
n_L=\left(-\frac{g}{\pi}\right)^{L}L! \left[1-\frac{3}{32\pi g}(2L^2+2L)+O(g^{-2})\right]\period
\eeq
As we will see in the next section, the leading strong-coupling results computed above match perfectly with the direct string theory computation. 

In this section, we focused on the limit where the lengths of operators $L$ and $J$ remain finite. There are also other interesting limits in which $L$ and/or $J$ become $O(\sqrt{\lambda})$ at strong coupling. It would be interesting to perform the computation in such limits and compare the results with the holographic computation based on nontrivial classical string solutions \cite{Zarembo:2002ph}.
\section{Comparison with the perturbative analyses\label{sec:comparison}}

\subsection{$\mathcal{N}=4$ SYM at weak coupling\label{subsec:compareweak}}
\begin{figure}
\centering
\includegraphics[clip,height=9cm]{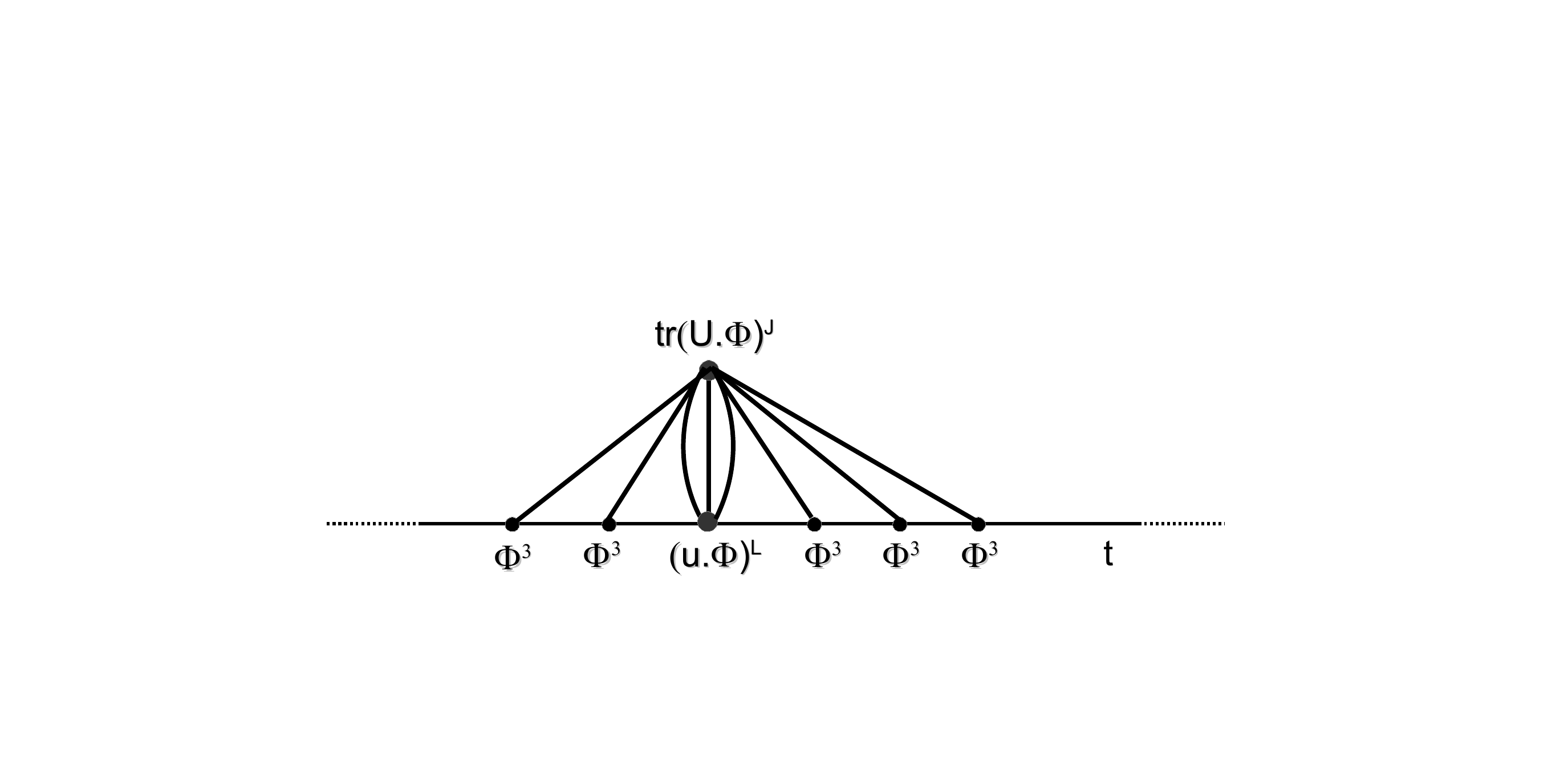}
\vskip -1.7cm
\caption{Leading weak coupling contribution to the correlator of one defect insertion and a bulk operator. There are $(J-L)$ insertions of $\Phi_3$ integrated along the line, which come from expanding the Wilson loop exponential factor.}
\label{fig:bulk-def-weak}
\end{figure}  
In this section we compute the leading weak coupling contribution to the bulk-defect correlator by a direct perturbative computation in the gauge theory. Let us carry out the calculation assumning the infinite straight line Wilson loop for simplicity. It is not difficult to see that the leading contribution to the bulk-defect correlator $\llangle \mathcal{W}[(u\cdot \vec{\Phi})^{L}(t_0)]\,\,{\rm tr}(U\cdot \vec{\Phi})^{J}(x^{\prime}) \rrangle$ is obtained by expanding the Wilson loop exponential factor to $(J-L)$-th order (picking the $\Phi_3$ scalar contribution only), and performing the resulting free Wick contraction between the bulk operator and the insertions on the Wilson line, as shown in Figure \ref{fig:bulk-def-weak}. Focusing on the planar limit, there are $J$ possible cyclic contractions that contribute. Furthermore, since the integrand obtained by contracting the propagators is symmetric under permuations of the insertion points, path-ordering is not important and we can extend the integrations over the whole line. Recalling that in our conventions the free propagator is 
\begin{equation}
\langle (\Phi_A)^i_{\ j}(x_1) (\Phi_B)^k_{\ l}(x_2)\rangle = \frac{g^2_{\rm YM}}{8\pi^2} \frac{\delta^i_l \delta^k_j}{(x_1-x_2)^2}\delta_{AB}\,,
\end{equation} 
we find 
\begin{equation}
\begin{aligned}
\llangle \mathcal{W}[(u\cdot \vec{\Phi})^{L}(t_0)]\,\,{\cal N}_J {\rm tr}(U\cdot \vec{\Phi})^{J}(x^{\prime}) \rrangle 
&= \frac{1}{N_c}{\cal N}_J  \left(\frac{g^2_{\rm YM} N}{8\pi^2}\right)^J \frac{J}{(J-L)!}\times \\
&\times \frac{1}{((t_0-t')^2+(x_{\perp}^{\prime})^2)^L}
\left(\int_{-\infty}^{\infty} dt \frac{1}{(t-t')^2+(x_{\perp}^{\prime})^2}\right)^{J-L}
\end{aligned}
\end{equation}
where ${\cal N}_J$ is the chiral primary normalization factor defined in (\ref{eq:Nj}). After evaluating the integral, we obtain
\begin{equation}
\llangle \mathcal{W}[(u\cdot \vec{\Phi})^{L}(t_0)]\,\,{\cal N}_J {\rm tr}(U\cdot \vec{\Phi})^{J}(x^{\prime}) \rrangle 
= \frac{2^{-\frac{3 J}{2}} \sqrt{J} \lambda^{J/2}}{\pi ^L N_c (J-L)!}\, \frac{ (u\cdot U)^L (U_3)^{J-L}}{(x_{\perp}^{\prime})^{J-L}\left((t_0-t')^2+(x_{\perp}^{\prime})^2\right)^L}\,.
\end{equation}
This gives the leading weak coupling contribution to the bulk-defect OPE coefficient defined in (\ref{OPE-coeff}). Going to the topological configuration by using (\ref{eq:Gandsmallc}), we then find
\begin{equation}
\llangle\mathcal{W} [\NO{\tilde{\Phi}_{L}}]\bb_J\rrangle_{\rm circle} =(-1)^L \frac{2^{-\frac{3 J}{2}-L} \sqrt{J} \lambda^{J/2}}{\pi ^L N_c (J-L)!}\left(1+O(\lambda)\right)
\end{equation}
which is in agreement with the leading weak coupling term of the localization prediction, eq. (\ref{bd-loc-weak}). 

\subsection{Strong coupling from AdS$_2$ string worldsheet\label{subsec:comparestrong}}
At strong coupling, the 1/2-BPS Wilson loop is dual to an AdS$_2$ minimal surface embedded in AdS$_5$ and sitting at a point on $S^5$ (the choice of direction on the five-sphere corresponds to the choice of scalar field that 
couples to the Wilson loop operator). Using the Poincare coordinates in Euclidean AdS$_5$
\begin{equation}
ds^2 =\frac{1}{z^2}\left(dz^2+(dx^0)^2+dx^{i}dx^{i}\right)\equiv  \frac{1}{z^2}dx^{\mu}dx^{\mu} 
\end{equation}
where $i=1,2,3$, and for later convenience we have introduced the notation $x^{\mu} \equiv (x^0,x^i,z)$, the string embedding dual to the 1/2-BPS straight line is given by $x^0=t$, $z=s$ (with $(t,s)$ being the worldsheet coordinates). We will carry out the calculations below using this straight-line geometry for simplicity (the circular Wilson loop can be obtained by a conformal transformation). 

Fluctuations of the open string coordinates around this minimal surface are in correspondence with defect operator insertions on the Wilson loop (see \cite{Giombi:2017cqn, Giombi:2018qox} for a review and more details). On the other hand, the single-trace ``bulk" operators $\tr (U\cdot \Phi)^J$ inserted away from the Wilson loop are dual to closed string modes that can be described at strong coupling as fluctuations of certain light supergravity fields in the graviton supermultiplet. They couple to the AdS$_2$ string worldsheet via a vertex operator $V_J$. 

The bulk-defect correlators studied in this paper can be then computed from a mixed ``open-closed" string amplitude given by the correlation function of the vertex operator and the 
(products of) fluctuations inserted at a point $t_0$ the AdS$_2$ boundary:
\begin{equation}
\llangle \mathcal{W}[(u\cdot \vec{\Phi})^{L}(t_0)]\,\,{\cal N}_J{\rm tr}(U\cdot \vec{\Phi})^{J}(x^{\prime})\rrangle_{\rm SYM} = \langle (u\cdot y(t_0))^L V_J(x^{\prime};U)\,\rangle_{{\rm AdS}_2} 
\label{bulk-defect-strong}
\end{equation}   
Here $y(t_0)$ denotes (the boundary limit of) the five massless fluctuations in the $S^5$ directions that arise in the expansion of the string action around the minimal surface: they are dual to the protected insertions of the five $\Delta=1$ scalars that do not couple to the Wilson loop\fn{For the relation between fluctuations on the worldsheet and the operator insertions, see \cite{Giombi:2017cqn,Sakaguchi:2007ea,Sakaguchi:2007zsa}.}. As explained in Section \ref{subsec:cftdata}, $u$ and $U$ are auxiliary null vectors that project onto the symmetric traceless representations of $SO(5)$ and $SO(6)$ respectively. The bosonic part\footnote{The full vertex operator will also include terms involving fermions, but we will not need them for the leading order calculation carried out here.} of the vertex operator has the following explicit form \cite{Lee:1998bxa,Berenstein:1998ij} (for a summary and review of the derivation, see also \cite{Giombi:2009ds}, whose conventions and notations we follow here) 
\begin{eqnarray}
&&V_J(x^{\prime};U) =\frac{\sqrt{\lambda}}{4\pi}{\cal C}_J \int d^2\sigma\left(V_J^{{\rm AdS}_5}+V_J^{{\rm S}^5}\right)\cr
&&V_J^{{\rm AdS}_5}=\left(2 \frac{J(J-1)}{J+1}K_J(x^{\mu};x^{\prime})\frac{1}{z^2}\partial_{\alpha}x^{\mu}\partial_{\alpha}x^{\mu}
+\frac{4}{J+1}\partial_{\alpha}x^{\mu} \partial_{\alpha} x^{\nu} \nabla_{\mu}\partial_{\nu} K_J(x^{\mu};x^{\prime})\right)Y^J(\Theta;U)\cr 
&&V_J^{{\rm S}^5} = 2J \partial_{\alpha}\Theta^A\partial_{\alpha}\Theta^A \,K_J(x^{\mu};x^{\prime}) Y^J(\Theta;U)\,.
\end{eqnarray}
Here $K_J$ is the AdS$_5$ bulk-to-boundary propagator
\begin{equation}
K_J(x^{\mu};x^{\prime}) = \left(\frac{z}{z^2+(x-x^{\prime})^2}\right)^J
\end{equation}
which describes the propagation of the supergravity mode dual to $\tr(U\cdot \Phi)^J$ from the string worldsheet to the insertion point $x^{\prime}$ at the boundary (essentially this is the analog of the plane-wave factor in the more familiar flat space vertex operators); $Y^J$ is the $SO(5)$ spherical harmonic that comes from the KK reduction on $S^5$
\begin{equation}
Y^J(\Theta;U) = \left(U\cdot \Theta\right)^J\,,\qquad U^2=0\,,
\end{equation}
where $\Theta^A$ are the six embedding coordinates of $S^5$, which in our parametrization of the fluctuations can be taken to be
\begin{equation}
\Theta^A = \frac{1}{1+\frac{1}{4} y^a y^a}\left(y^1,y^2,1-\frac{1}{4}y^ay^a,y^3,y^4,y^5\right)\,,
\end{equation} 
where $y^a$, $a=1,\ldots, 5$ are the massless fluctuations in the fundamental of the $SO(5)$ preserved by the Wilson loop.\footnote{The particular choice of parametrization above is such that $y=0$ corresponds to the Wilson loop that couples to $\Phi^3$.} Finally, the constant ${\cal C}_J$ is a normalization factor that corresponds to the gauge theory normalization in (\ref{eq:Nj}), and is given by
\begin{equation}
{\cal C}_J = 2^{J/2-2} \frac{J+1}{N_c\sqrt{J}}\,.
\end{equation}
Note that due this factor, the bulk-defect correlator goes as $\sim 1/N_c$ in the large-$N$ limit, as expected. 

\begin{figure}
\centering
\includegraphics[clip,height=9cm]{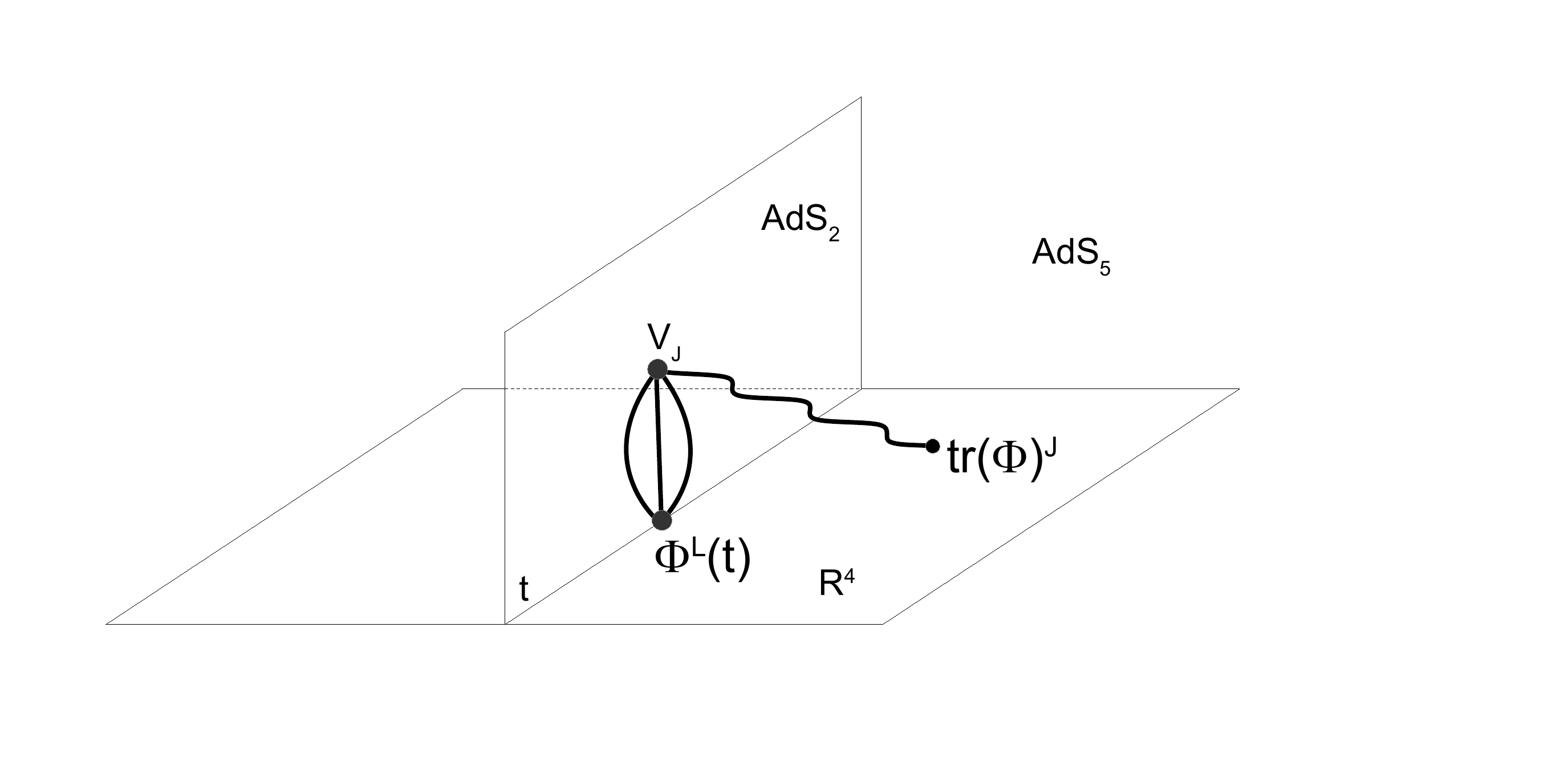}
\vskip -1cm
\caption{Depiction of the leading order contribution to the bulk-defect correlator at strong coupling.}
\label{fig:bulk-def-pic}
\end{figure}

To leading order at large $\lambda$, the correlator (\ref{bulk-defect-strong}) is given by simply expanding the vertex operator to $L$-th order in the $y^a$ fluctuations and performing the corresponding Wick contractions with the AdS$_2$ bulk-to-boundary propagator for $y$:
\begin{equation}
\langle y^a(t,s) y^b(t',s) \rangle_{{\rm AdS}_2} = \frac{1}{\pi} \frac{s}{s^2+(t-t')^2}\delta^{ab}\,.
\end{equation}
A schematic depiction of the calculation is given in Figure \ref{fig:bulk-def-pic}.
It is not difficult to see that only $V_J^{{\rm AdS}_5}$ contributes to this order, because expansion of $V_J^{{\rm S}_5}$ to $L$-th order in $y^a$ will always produce terms that vanish upon contraction with the $(u\cdot y)^L$ boundary insertion, due to $u^2=0$. Then, the relevant part of the vertex operator expanded to $L$-th power in $y$ is found to be, after a bit of algebra:
\begin{equation}
V_J(x^{\prime};U) = \frac{4J {\cal C}_J \sqrt{\lambda}}{\pi}\int \frac{dt ds}{s^2} \left[(x_{\perp}^{\prime})^2 \left(\frac{s}{s^2+(t-t')^2+(x_{\perp}^{\prime})^2}\right)^{J+2}  (U\cdot y)^L (U_3)^{J-L}\frac{J!}{L!(J-L)!}+\ldots\right]
\end{equation} 
Plugging into (\ref{bulk-defect-strong}) and performing the $L!$ free Wick contractions, we find 
\begin{equation}
\begin{aligned}
\langle (u\cdot y(t_0))^L V_J(x^{\prime};U)\,\rangle_{{\rm AdS}_2}  &= \frac{4J {\cal C}_J \sqrt{\lambda}}{\pi}  (u\cdot U)^L (U_3)^{J-L}\frac{J!}{(J-L)!}(x_{\perp}^{\prime})^2\times \\ 
& \times \int_{{\rm AdS}_2} \frac{dt ds}{s^2}\left(\frac{s}{s^2+(t-t')^2+(x_{\perp}^{\prime})^2}\right)^{J+2} \left(\frac{s}{\pi(s^2+(t-t_0)^2)}\right)^L
\end{aligned}
\end{equation}
Evaluating the integral, for instance by introducing Schwinger parameters, yields
\begin{equation}
\begin{aligned}
\label{KK-integ}
&\int_{{\rm AdS}_2} \frac{dt ds}{s^2}\left(\frac{s}{s^2+(t-t')^2+(x_{\perp}^{\prime})^2}\right)^{J+2} \left(\frac{s}{s^2+(t-t_0)^2}\right)^L =\\
& \frac{\sqrt{\pi}\Gamma \left(\frac{1}{2} (J-L+2)\right) \Gamma \left(\frac{1}{2} (J+L+1)\right)}{2 \Gamma (J+2)}\, \frac{(x_{\perp}^{\prime})^{L-J-2}}{\left((t_0-t')^2+(x_{\perp}^{\prime})^2\right)^L}\,.
\end{aligned}
\end{equation}
Putting everything together, we arrive at the final result
\begin{equation}
\langle (u\cdot y(t_0))^L V_J(x^{\prime};U)\,\rangle_{{\rm AdS}_2} 
= \frac{2^{-\frac{J}{2}+L-1} \sqrt{J \lambda} \Gamma \left(\frac{1}{2} (J+L+1)\right)}{N_c \pi^L \Gamma (L+1) \Gamma \left(\frac{1}{2} (J-L+1)\right)}
\frac{ (u\cdot U)^L (U_3)^{J-L}}{(x_{\perp}^{\prime})^{J-L}\left((t_0-t')^2+(x_{\perp}^{\prime})^2\right)^L}
\end{equation}
Through (\ref{bulk-defect-strong}) and (\ref{OPE-coeff}), from this we can read off the bulk-defect OPE coefficient $c_{L|J}$ at strong coupling. 
Now going to the topological configuration on the sphere by using (\ref{eq:Gandsmallc}), we find
\begin{equation}
\llangle\mathcal{W} [\NO{\tilde{\Phi}_{L}}]\bb_J\rrangle_{\rm circle} =(-1)^L \frac{2^{-\frac{J}{2}-1} \sqrt{J \lambda} \Gamma \left(\frac{1}{2} (J+L+1)\right)}{N_c \pi^L \Gamma (L+1) \Gamma \left(\frac{1}{2} (J-L+1)\right)}
\end{equation}
which precisely agrees with the localization prediction (\ref{loc-strong}).
\begin{figure}
\centering
\includegraphics[clip,height=9cm]{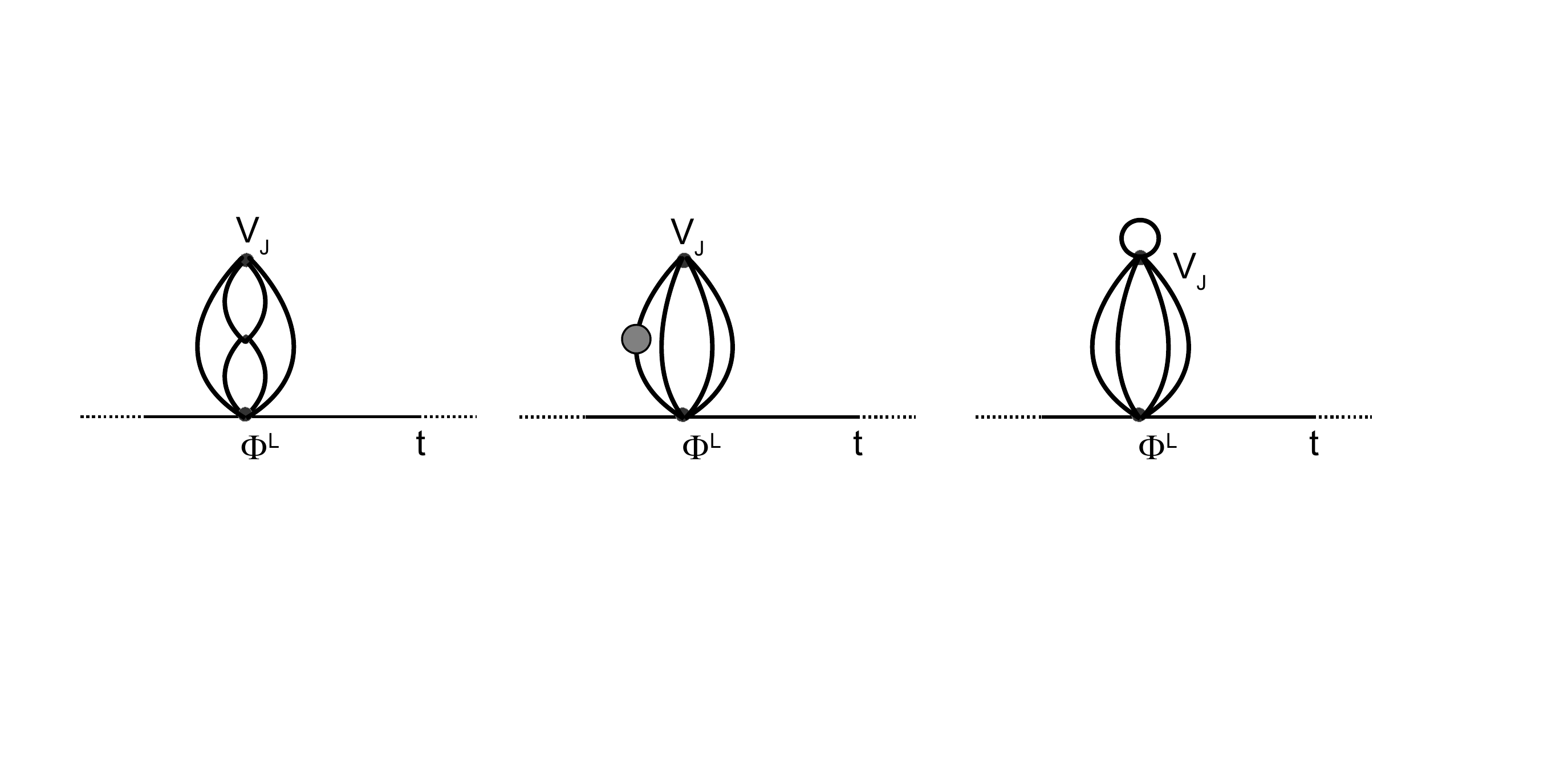}
\vskip -2.5cm
\caption{Diagrams contributing to the next-to-leading order at strong coupling. The first two pictures from the left correspond to bringing down the interaction vertices from the string action while still expanding the vertex operator to $L$-th power in fluctuations. The third diagram arises from expanding the vertex operator to $(L+2)$-th order and self-contracting two of the fluctuations.}
\label{fig:bulk-def-subleading}
\end{figure}  
It would be interesting to compute the subleading strong coupling corrections to this result and compare to localization. To the first subleading order, there are two type of contributions: those that involve the interaction vertices of fluctuations coming from the expansion of the string action, depicted in the first two pictures in Figure \ref{fig:bulk-def-subleading}, and those that involve expanding the vertex operator $V_J$ to $(L+2)$-th order in fluctuations and self-contracting two of those fluctuations, as shown in the third picture in Figure \ref{fig:bulk-def-subleading}. We leave the explicit calculation of these corrections to the future. It is interesting to notice that a precisely analogous structure of subleading corrections also arises in the localization result, see the discussion below eq. (\ref{bd-leading}). This suggests that the insertion of $B_J$ in the contour integral (\ref{eq:generalizedintegralexpressions}) essentially plays the role of the vertex operator $V_J$ on the string theory side.

\subsubsection{Two defect insertions and one bulk operator}
\label{Two-def-strong}
\begin{figure}
\centering
\includegraphics[clip,height=9cm]{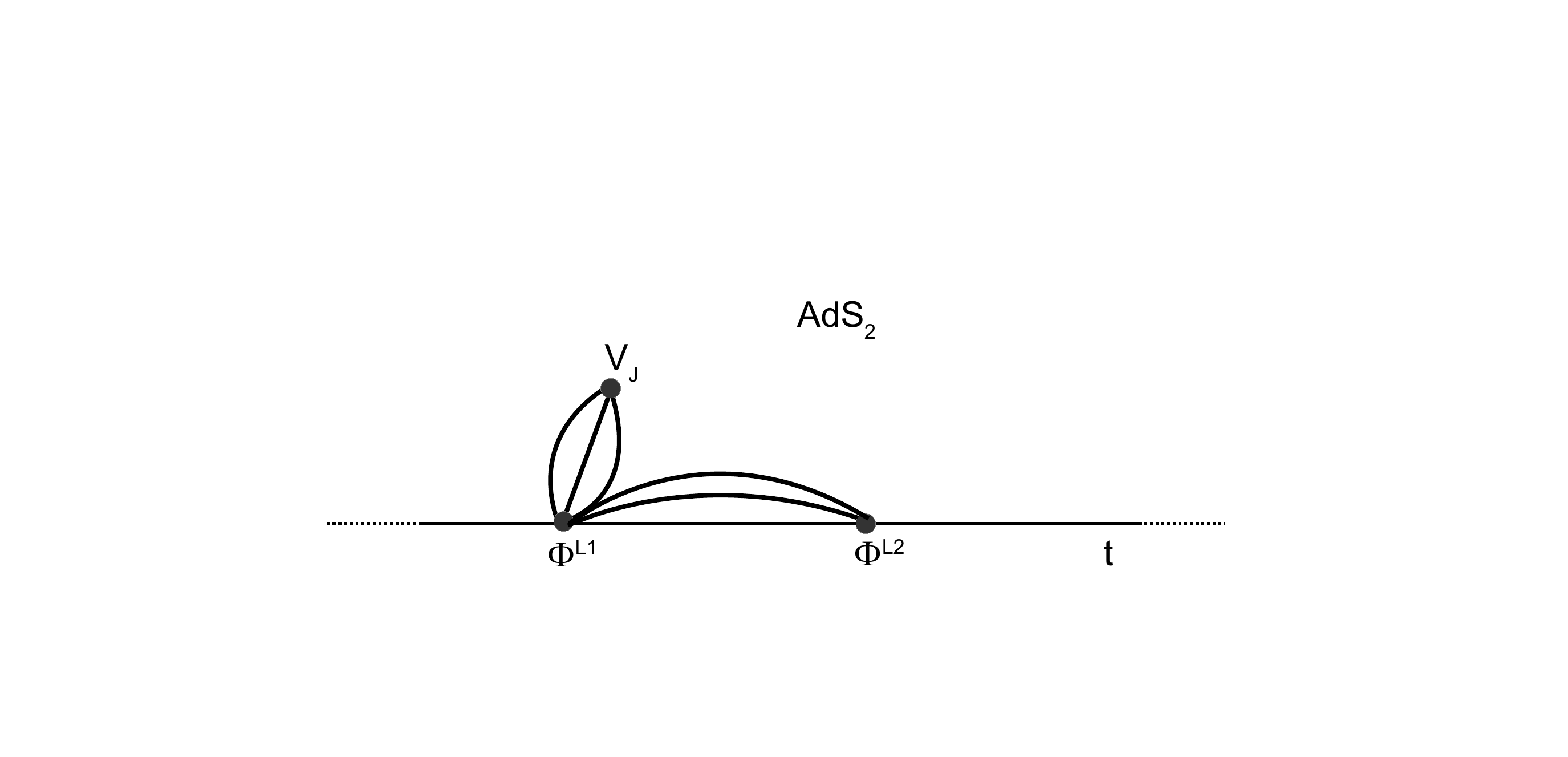}
\vskip -1.4cm
\caption{Leading strong coupling contribution to the correlator of two defect insertions $\Phi^{L_1}$, $\Phi^{L_2}$ and one bulk insertion $\tr \Phi^J$.}
\label{fig:bulk-two-def}
\end{figure}  
More general correlators involving several defect and bulk insertions can be computed following a similar procedure. As an example, let us consider the case of two defect insertions and one bulk operator: 
\begin{equation}
\llangle \mathcal{W}[(u_1\cdot \vec{\Phi})^{L_1}(t_1)(u_2\cdot \vec{\Phi})^{L_2}(t_2)]\,\,{\rm tr}(U\cdot \vec{\Phi})^{J}(x^{\prime})\rrangle_{\rm SYM} = \langle (u_1\cdot y(t_1))^{L_1} (u_2\cdot y(t_2))^{L_2} V_J(x^{\prime};U)\,\rangle_{{\rm AdS}_2} 
\label{bulk-two-defect-strong}
\end{equation}  
Let us assume without loss of generality that $L_1\ge L_2$. 
The leading order contribution at strong coupling comes from expanding the vertex operator to the smallest possible order in fluctuations, namely to the $(L_1-L_2)$-th order, contracting those fluctuations to the boundary insertion $(u\cdot y)^{L_1}$, and then connecting the remaining fluctuations with $L_2$ ``boundary-to-boundary" propagators between the defect insertions. This is depicted in Figure \ref{fig:bulk-two-def}. Evaluating this diagram, we find
\begin{equation}
\begin{aligned}
&\langle (u_1\cdot y(t_1))^{L_1} (u_2\cdot y(t_2))^{L_2} V_J(x^{\prime};U)\,\rangle_{{\rm AdS}_2}   = \\
&=\frac{4J {\cal C}_J \sqrt{\lambda}}{\pi}  (u_1\cdot U)^{L_1-L_2} (U_3)^{J-L_1+L_2} \frac{J! L_1!}{(J-L_1+L_2)!(L_1-L_2)!}(x_{\perp}^{\prime})^2\times \\ 
& \times \left(\frac{\sqrt{\lambda}}{2\pi^2}\frac{u_1\cdot u_2}{(t_1-t_2)^2}\right)^{L_2}\int_{{\rm AdS}_2} \frac{dt ds}{s^2}\left(\frac{s}{s^2+(t-t')^2+(x_{\perp}^{\prime})^2}\right)^{J+2} \left(\frac{s}{\pi(s^2+(t-t_1)^2)}\right)^{L_1-L_2}\,.
\end{aligned}
\end{equation}
Using the integral (\ref{KK-integ}), the final result is then
\begin{eqnarray}
&&\langle (u_1\cdot y(t_1))^{L_1} (u_2\cdot y(t_2))^{L_2} V_J(x^{\prime};U)\,\rangle_{{\rm AdS}_2} =c_{L_1,L_2|J}\frac{  (u_1\cdot U)^{L_1-L_2} (U_3)^{J-L_1+L_2}(u_1\cdot u_2)^{L_2}}{(x_{\perp}^{\prime})^{J-L_1+L_2}\left((t_1-t')^2+(x_{\perp}^{\prime})^2\right)^{L_1-L_2}(t_1-t_2)^{2L_2}}\cr
&&c_{L_1,L_2|J}=\frac{\sqrt{J}  2^{-\frac{J}{2}+L_1-2 L_2-1} \lambda ^{\frac{1}{2} \left(L_2+1\right)} L_1! \Gamma \left(\frac{1}{2} \left(J+L_1-L_2+1\right)\right)}{N_c \pi ^{L_1+L_2}  \Gamma \left(L_1-L_2+1\right) \Gamma \left(\frac{1}{2} \left(J-L_1+L_2+1\right)\right)}
\end{eqnarray}
This gives the correlator on the straight Wilson line. Conformally mapping to the circular loop, and choosing the null polarization vectors corresponding 
to the topological operators, we obtain the position independent result
\begin{equation}
\llangle\mathcal{W}[\NO{\tilde{\Phi}^{L_1}}\NO{\tilde{\Phi}^{L_2}}] \,\,\bb_J \rrangle_{\rm circle}=
(-1)^{L_1}\frac{\sqrt{J}  2^{-\frac{J}{2}-2 L_2-1} \lambda ^{\frac{1}{2} \left(L_2+1\right)} L_1! \Gamma \left(\frac{1}{2} \left(J+L_1-L_2+1\right)\right)}{N_c \pi ^{L_1+L_2}  \Gamma \left(L_1-L_2+1\right) \Gamma \left(\frac{1}{2} \left(J-L_1+L_2+1\right)\right)}
\end{equation}
This is again in precise agreement with the strong coupling limit of the localization result computed in the Appendix, see eq. (\ref{loc-twodef-strong}).

\section{Non-planar corrections to the defect correlators\label{sec:nonplanar}}
As another application of the results in section \ref{subsec:simplification}, in this section we study the $1/N_c^2$ correction to the correlation functions on the Wilson loop. As will be clear in the analysis below, one has to take into account the mixing with the multi-trace operators in order to reproduce the correct answer.
For simplicity, we will focus on the two-point functions,
\beq
G_{L,L}=\langle \mathcal{W}[\NO{\tilde{\Phi}^{L}}\NO{\tilde{\Phi}^{L}}]\rangle \comma
\eeq
but the generalization to the higher-point functions is straightforward (although it becomes more complicated).
\subsection{Generalities\label{subsec:nonplanargeneral}}
The nonplanar corrections to $G_{L,L}$ and $G_{L_1,L_2,L_3}$ can be studied by using the expansion of the operator   determined in section \ref{subsec:simplification}. For convenience, let us display the formulae again:
\beq
\begin{aligned}
&\NO{\dd_L}=\left.\NOB{\dd_L}\right|_{1/N_c^2}-S_1-(S_2^{(1)}+S_2^{(2)})+O(1/N_c^2)\comma\\
&S_1=\sum_{\substack{\ell+j<L\\0\leq \ell,\,1\leq j}}\frac{\langle \NOB{\dd_{\ell}}\NOB{\dd_{\ell|j}}\rangle}{\langle \NOB{\dd_{\ell|j}}\NOB{\dd_{\ell|j}}\rangle}\NOB{\dd_{\ell|j}}\comma\\
& S_2^{(1)}= \sum_{\substack{\ell + j_1+j_2<L\\0\leq \ell,\,1\leq  j_1,j_2}} \frac{\langle \NOB{\dd_L}\NOB{\dd_{\ell|j_1,j_2}}\rangle}{\langle \NOB{\dd_{\ell|j_1,j_2}}\NOB{\dd_{\ell|j_1,j_2}}\rangle}\NOB{\dd_{\ell|j_1,j_2}}\comma\\
& S_2^{(2)}=-\sum_{\tilde{\ell}<L}\,\,\,\sum_{\ell+j<L}\frac{\langle \NOB{\dd_L},\NOB{\dd_{\ell|j}}\rangle\langle\NOB{\dd_{\tilde{\ell}}}\NOB{\dd_{\ell|j}}\rangle}{\langle\NOB{\dd_{\ell|j}}\NOB{\dd_{\ell|j}}\rangle\langle\NOB{\dd_{\tilde{\ell}}}\NOB{\dd_{\tilde{\ell}}}\rangle}\NOB{\dd_{\tilde{\ell}}}\period
\end{aligned}
\eeq
By inspecting these expressions, we conclude that there are four sources of corrections to the correlators at $1/N_c^2$: 
\begin{itemize}
\item The first correction (to be denoted by ${\tt first}$) comes from the terms involving a single $S_1$. The correction $S_1$ itself is $O(1/N_c)$, but since the correlator involving a single $\NOB{\dd_{L|J}}$ is $O(1/N_c)$, in total it gives $O(1/N_c^2)$ corrections. 
\item The second correction (to be denoted by ${\tt second}$) comes from the terms involving two $S_1$'s. They can contribute at $O(1/N_c^2)$ since the correlators involving two identical $\NOB{\dd_{L|J}}$ can be $O(1)$ (cf.~\eqref{eq:largeNestimate}). 
\item The third correction (to be denoted by ${\tt third}$) comes from the terms involving a single $S_{2}^{(2)}$. This is by itself $O(1/N_c^2)$.
\item In addition to the aforementioned corrections which are induced by the mixing with the multi-trace operators, there are also corrections coming from the $1/N_c^2$ correction to the expectation value of the Wilson loop.  Such corrections are already discussed in the previous paper \cite{Giombi:2018qox}, and they can be computed by first expressing the planar result in terms of the planar expectation value of the Wilson loop $\langle \mathcal{W}\rangle$, and replacing $\langle \mathcal{W}\rangle$ with its $O(1/N_c^2)$ result given in \eqref{eq:WOWW}. We will denote this correction by adding the subscript $ |_{1/N_c^2}$ to the planar expression.
\end{itemize}
Note that the terms involving $S_2^{(1)}$ never contribute at this order since the correlator involving $\NOB{\dd_{L|J_1,J_2}}$ is $O(1/N_c^2)$, making the whole contribution $O(1/N_c^4)$ or higher.

Let us now write down the results explicitly.  For this purpose it is convenient to use the large $N_c$ factorization and simplify the correlator as\fn{Here we used $\langle \bb_J \bb_J\rangle=\left(-\frac{1}{2}\right)^{J}$, which follows from our normalization of $\bb_J$ \eqref{eq:Nj}.}
\beq
\begin{aligned}
\langle\NOB{\dd_{L_1|J}}\NOB{\dd_{L_2|J}}\NOB{\dd_{L_3}}\cdots \NOB{\dd_{L_n}}\rangle&=\langle\NOB{\dd_{L_1}}\NOB{\dd_{L_2}}\NOB{\dd_{L_3}}\cdots \NOB{\dd_{L_n}}\rangle \langle\bb_J \bb_J\rangle\\
&=\left(-2\right)^{-J}\langle\NOB{\dd_{L_1}}\NOB{\dd_{L_2}}\NOB{\dd_{L_3}}\cdots \NOB{\dd_{L_n}}\rangle \period
\end{aligned}
\eeq
We can then evaluate the corrections to the two-point functions as follows:
\beq
\begin{aligned}
G_{L,L}|_{\tt first}&=-2 \sum_{\substack{\ell+j<L\\0\leq \ell,\,1\leq j}}(-2)^{j}\frac{\langle\NOB{\dd_{L}}\NOB{\dd_{\ell|j}} \rangle^2}{\langle \NOB{\dd_{\ell}}\NOB{\dd_{\ell}}\rangle}\comma\\
G_{L,L}|_{\tt second}&=\sum_{\substack{\ell+j<L\\0\leq \ell,\,1\leq j}}(-2)^{j}\frac{\langle\NOB{\dd_{L}}\NOB{\dd_{\ell|j}} \rangle^2}{\langle \NOB{\dd_{\ell}}\NOB{\dd_{\ell}}\rangle}\comma\\
G_{L,L}|_{\tt third}&=0\period
\end{aligned}
\eeq
Note that the third term is zero owing to the restriction on the range of summation $L_0<L$.
Summing them up, we obtain the following expression for $1/N_c^2$ corrections to $G_{L,L}$:
\beq\label{eq:GLLexplicit}
G_{L,L}=\langle \NOB{\dd_L}\NOB{\dd_L}\rangle|_{1/N_c^2}-\sum_{\substack{\ell+j<L\\0\leq \ell,\,1\leq j}}(-2)^{j}\frac{\langle\NOB{\dd_{L}}\NOB{\dd_{\ell|j}} \rangle^2}{\langle \NOB{\dd_{\ell}}\NOB{\dd_{\ell}}\rangle}+O(1/N_c^4)\period
\eeq
As mentioned above, the first term $\langle \NOB{\dd_L}\NOB{\dd_L}\rangle|_{1/N_c^2}$ can be computed by taking the planar expression and replacing the planar expectation value $\langle\mathcal{W}\rangle$ with its non-planar counterpart. 
Similarly, one can also obtain the expressions for the three-point function
\beq\label{eq:threepntdisplay}
G_{L_1,L_2,L_3}=\langle \mathcal{W}[\NO{\tilde{\Phi}^{L_1}}\NO{\tilde{\Phi}^{L_2}}\NO{\tilde{\Phi}^{L_3}}]\rangle\period
\eeq
The results are summarized in Appendix \ref{ap:3pt}.

In the following subsections, we evaluate the expression \eqref{eq:GLLexplicit} explicitly at weak and strong couplings.
\subsection{Weak coupling expansion\label{subsec:nonplanarweak}}
Here we evaluate the non-planar corrections for the two-point functions at tree level and compare them with the perturbative answers.
\paragraph{Computation of $\NOB{\dd_{L}}|_{1/N_c^2}$} Before performing the weak-coupling expansion of the correlator, let us first determine the non-planar correction to the operator $\NOB{\dd_{L}}|_{1/N_c^2}$, which comes purely from the non-planar corrections to the expectation value of the Wilson loop $\langle \mathcal{W}\rangle$. Note that, as we will see later, such corrections do not contribute to the two-point functions. Nevertheless here we evaluate them explicitly since the results can be used for the computation of higher-point functions. 

As shown in the previous paper \cite{Giombi:2018qox}, the corrections coming from the non-planar correction to $\langle \mathcal{W}\rangle$ can be incorporated by adding the following term\fn{Note that we already set $a=0$ since we are analyzing the $1/2$-BPS loops in this section.} to the measure for the integral representation \eqref{eq:measureprevious}:
\beq\label{eq:nonplanarmeasure}
\begin{aligned}
d\mu_{1/N_c^2}&=\frac{dx}{2\pi i x}e^{2\pi g (x+\frac{1}{x})} f(2\pi g (x+1/x))\comma\\
f(z)&=\frac{(2\pi g)^4}{N_c^2}\frac{z^2-3z+3}{3z^4}\period
\end{aligned}
\eeq
At the leading order in the weak coupling expansion, we obtain
\beq
\left.d\mu_{1/N_c^2}\right|_{\rm weak}=\frac{1}{N_c^2}\frac{dx}{2\pi i x}\frac{1}{(x+\frac{1}{x})^4}\period
\eeq
Because of this correction to the measure, the planar normal-ordered operators $\NOB{\dd_L}$'s are no longer orthogonal to each other. For instance, the two-point function of $\NOB{\dd_L}$ and $\NOB{\dd_M}$ is given by
\beq\label{eq:integralLMweak}
\begin{aligned}
&\langle\NOB{\dd_L}  \NOB{\dd_M}\rangle|_{1/N_c^2,\,\, {\rm weak}}=\\
&\frac{(-ig)^{L+M}}{N_c^2}\oint\frac{dx}{2\pi i x}\frac{\left(i^{L}x^{L+1}+i^{-L}x^{-(L+1)}\right)\left(i^{M}x^{M+1}+i^{-M}x^{-(M+1)}\right)}{(x+\frac{1}{x})^6}\comma
\end{aligned}
\eeq 
where we used the weak-coupling expression for the function $Q_L$ (see also \eqref{eq:QLweakexpfirst} and \eqref{eq:chebyshev}):
\beq
\begin{aligned}
Q_L(x)&=(-i g)^{L} \left[U_L\left(\frac{i (x-x^{-1})}{2}\right) +O(g^2)\right]\\
&=(-ig)^{L}\left[\frac{i^{L}x^{L+1}+i^{-L}x^{-(L+1)}}{x+\frac{1}{x}}+O(g^2)\right]\period
\end{aligned}
\eeq
The integral in \eqref{eq:integralLMweak} can be computed by using the expansion
\beq\label{eq:usefulexp}
\frac{1}{\left(x+\frac{1}{x}\right)^{2k}}=(-1)^{k}\sum_{j=0}^{\infty}(-1)^{j}\frac{(j+k-1)!}{(2k-1)!(j-k)! }x^{2j}\qquad (k\in \mathbb{Z}_{>0})\comma
\eeq
as
\beq\label{eq:LMweaknonplanar}
\langle\NOB{\dd_L}  \NOB{\dd_M}\rangle|_{1/N_c^2,\,\, {\rm weak}}=\begin{cases}-\frac{(-i g)^{L+M}}{120N_c^2}\left[\frac{\left(\frac{L-M}{2}+2\right)!}{\left(\frac{L-M}{2}-3\right)!}-\frac{\left(\frac{L+M}{2}+3\right)!}{\left(\frac{L+M}{2}-2\right)!}\right]&\quad(L-M:\text{ even})\\0&\quad (L-M:\text{ odd})
\end{cases}\comma
\eeq
where we assumed $L>M$. To restore the orthogonality, one simply needs to subtract $\NOB{\dd_{M}}$ from $\NOB{\dd_{L}}$ as
\beq
\NOB{\dd_{L}}\quad \to\quad \NOB{\dd_{L}}-\sum_{0\leq M<L}\,\,\frac{\langle\NOB{\dd_L}  \NOB{\dd_M}\rangle|_{1/N_c^2,\,\text{weak}}}{\langle\NOB{\dd_M}  \NOB{\dd_M}\rangle}\times \NOB{\dd_{M}}\period
\eeq
Subtracting all the operators with smaller lengths, we arrive at the following weak-coupling expression for  $\NOB{\dd_{L}}|_{1/N_c^2}$:
\beq\label{eq:operatormodifyweak}
\NOB{\dd_{L}}|_{1/N_c^2,\,\, {\rm weak}}=\NOB{\dd_{L}} +\sum_{0\leq M<L}\frac{(-i g)^{L-M}}{120N_c^2}\left[\frac{\left(\frac{L-M}{2}+2\right)!}{\left(\frac{L-M}{2}-3\right)!}-\frac{\left(\frac{L+M}{2}+3\right)!}{\left(\frac{L+M}{2}-2\right)!}\right] \NOB{\dd_{M}}\period
\eeq
Here we used the weak-coupling expansion of the planar two-point function $\langle\NOB{\dd_M}  \NOB{\dd_M}\rangle =(-g^2)^{M}+\cdots$.
\paragraph{Two-point functions}
Let us now compute the two-point functions at weak coupling by evaluating each term in \eqref{eq:GLLexplicit}. The first term $\langle \NOB{\dd_{L}}\NOB{\dd_L}\rangle_{1/N_c^2}$ can be obtained by setting $M=L$ in \eqref{eq:LMweaknonplanar} and the result is
\beq\label{eq:firsttermweak}
\langle \NOB{\dd_{L}}\NOB{\dd_L}\rangle|_{1/N_c^2} =\frac{(-g^2)^{L}}{120N_c^2}\frac{(L+3)!}{(L-2)!}[1+O(g^2)]\period
\eeq
Note that the correction to the operator \eqref{eq:operatormodifyweak} does not contribute\fn{On the other hand, it does contribute for higher-point functions, but we will not perform such computation explicitly in this paper.} to the two-point function due to the orthogonality of the planar normal-ordered operators; $\langle\NOB{\dd_L}\NOB{\dd_M}\rangle=0$ for $L\neq M$.

The second term is expressed in terms of planar correlators which can be computed by using the integral expression,
\beq\label{eq:integralddb}
\langle\NOB{\dd_{L}}\NOB{\dd_{\ell|j}} \rangle=\frac{2^{-j/2}\sqrt{j}}{N_c}\int d\mu \,\,B_j(x) Q_L(x) Q_{\ell}(x)\period
\eeq 
Using the weak-coupling expansion of each object, we can compute it as\fn{To compute the integral, we used the expansion \eqref{eq:usefulexp}.}
\beq
\begin{aligned}
&\langle\NOB{\dd_{L}}\NOB{\dd_{\ell|j}} \rangle|_{\rm weak}\\
&=\frac{(-ig)^{L+\ell}2^{-\frac{j}{2}}\sqrt{j}}{N_c}\int \frac{dx\,\,x^{j}}{2\pi i x}\frac{\left(i^{L}x^{L+1}+i^{-L}x^{-(L+1)}\right)\left(i^{\ell}x^{\ell+1}+i^{-\ell}x^{-(\ell+1)}\right)}{(x+\frac{1}{x})^2}\\
&=\begin{cases}\frac{g^{L+\ell}2^{-\frac{j}{2}}\sqrt{j}}{N_c}(-1)^{\frac{L+\ell+j}{2}}(\ell+1)&\quad (L-\ell-j:\text{ even})\\0&\quad (L-\ell-j:\text{ odd})\end{cases}
\end{aligned}
\eeq
We can thus obtain the following expression for the second term:
\beq\label{eq:secondtermweak}
\begin{aligned}
-\sum_{\substack{\ell+j<L\\0\leq \ell,\,1\leq j}}(-2)^{j}\frac{\langle\NOB{\dd_{L}}\NOB{\dd_{\ell|j}} \rangle^2}{\langle \NOB{\dd_{\ell}}\NOB{\dd_{\ell}}\rangle}&=-\frac{(-g^2)^{L}}{N_c^2}\sum_{\substack{\ell+j<L \\0\leq \ell,\,1\leq j}}j(\ell+1)^2\frac{(-1)^{L-\ell-j}+1}{2}\\
&=-\frac{(-g^2)^{L}}{120N_c^2}\frac{(L+2)!}{(L-3)!}\period
\end{aligned}
\eeq
Adding \eqref{eq:firsttermweak} and \eqref{eq:secondtermweak} we obtain
\beq\label{eq:weakfinalGLL}
G_{L,L}|_{1/N_c^2,\,\,{\rm weak}}=\frac{1}{24 N_c^2}\frac{(L+2)!}{(L-2)!} =\frac{1}{N_c^2}\left(\begin{array}{c}L+2\\4\end{array}\right)\period
\eeq
\begin{figure}
\centering
\includegraphics[clip,height=6cm]{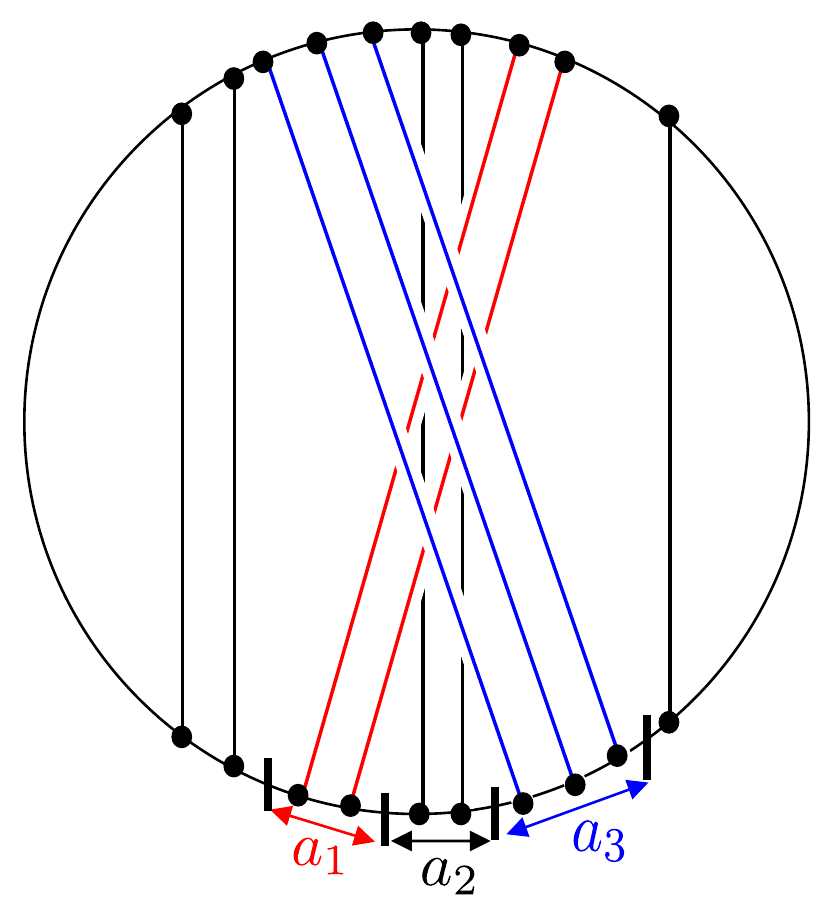}
\caption{The general structure of $O(1/N_c^2)$ Wick contractions for the two-point functions. The $O(1/N_c^2)$ Wick contractions can be obtained by choosing three consecutive segments (whose lengths are denoted by $a_{1,2,3}$) inside the operator and reconnect them as shown in the figure. In order to make the diagram truly non-planar, one further needs to assume that two of the three segments have nonzero lengths. One can count the number of such contractions by counting the number of ways to distribute the four end-points of the segments (denoted by short vertical black lines) inside the operator. }
\label{fig:fig3}
\end{figure}

Let us now compare the result \eqref{eq:weakfinalGLL} with the (non-planar) Wick contractions in $\mathcal{N}=4$ SYM.  The general structure of the $O(1/N_c^2)$ Wick contractions is depicted in figure \ref{fig:fig3}. As shown there, they are characterized by three non-negative integers $a_1$, $a_2$ and $a_3$, which are the lengths of three consecutive segments in the figure. Alternatively, they are specified by the four end-points of these segments. If it were not for any restrictions on the lengths of the segments, the number of ways to specify the four end-points is given by 
\beq
({\tt combinatorics})|_{\rm naive}=\left(\begin{array}{c}L+4\\4\end{array}\right)\period
\eeq 
This however is not correct since if two of the three lengths $a_{1,2,3}$ are zero, the Wick contraction reduces to the planar Wick contraction. To avoid counting such planar contractions, we need to impose the condition $a_{1}> 0$ and $a_2> 0$. After imposing this condition, the number of Wick contractions becomes
\beq
({\tt combinatorics})|_{\rm correct}=\left(\begin{array}{c}L+2\\4\end{array}\right)\comma
\eeq  
which precisely matches the result we computed \eqref{eq:weakfinalGLL}.

Before ending this subsection, let us emphasize again that if we did not take into account the mixing with multi-trace operators we would obtain \eqref{eq:firsttermweak}, which does not match with the result of the Wick contraction. The analysis in this subsection therefore illustrates the importance of the mixing effect and also provides the evidence for our Gram-Schmit analysis.
\subsection{Strong coupling expansion\label{subsec:nonplanarstrong}}
We next compute the non-planar two-point functions at strong coupling.
 \paragraph{Computation of $\NOB{\dd_{L}}|_{1/N_c^2}$} As in the previous subsection, let us first determine the correction to the operator $\NOB{\dd_{L}}|_{1/N_c^2}$. For this purpose, we again use the non-planar correction to the measure \eqref{eq:nonplanarmeasure}. At strong coupling, the integral is dominated by the saddle point $x=0$ and by expanding the measure around the saddle-point in terms of the coordinate used in \cite{Giombi:2018qox},
 \beq
 x=\frac{\sqrt{2\pi g -t^2}-i t}{\sqrt{2\pi g}}=1-\frac{it}{\sqrt{2\pi g}}+O(g^{-1})\comma
 \eeq 
 we obtain
 \beq
 \oint d\mu_{1/N_c^2}=\frac{e^{4\pi g}g^{3/2}}{6N_c^2}\sqrt{\frac{\pi}{2}}\int_{-\infty}^{\infty} dt \,\,e^{-t^2}\left[1-\frac{6-6t^2+t^4}{8\pi g}+O(g^{-2})\right]\period
 \eeq
It is convenient to express it in terms of the strong-coupling expansion of the planar measure\fn{See \cite{Giombi:2018qox} for derivation.}
 \beq
  d\mu= \frac{e^{4\pi g}}{(2\pi)^{5/2}g^{3/2}}dt \,\,e^{-t^2}\left[1-\frac{t^4}{8\pi g}+O(g^{-2})\right] \comma
 \eeq
 as
 \beq\label{eq:dmunpdmu}
  d\mu_{1/N_c^2}= \frac{2g^3\pi^3}{3N_c^2} d\mu \left[1+\frac{3(t^2-1)}{4\pi g}+O(g^{-2})\right]\period
 \eeq
 
 As was computed in the previous paper, the functions $Q_L(x)$ can be expanded at strong coupling as
 \beq
 Q_L(x)=\frac{1}{i^{L}}\left(\frac{g}{2\pi}\right)^{\frac{L}{2}} \left[H_L(t)+\frac{1}{8\pi g}\left(\frac{L! (2L-1)}{(L-2)!}H_{L-2}(t)+\frac{L!}{(L-4)!}H_{L-4}(t)\right)+O(g^{-2})\right]\comma\nn
 \eeq
 and are orthogonal to each other under the measure $d\mu$. They are however not orthogonal any more at the non-planar level owing to the correction term in \eqref{eq:dmunpdmu}; $3(t^2-1)/(4\pi g)$. For instance, the two-point functions of $\NOB{\dd_L}$ and $\NOB{\dd_M}$ can be computed as
\beq\label{eq:2ptstrongnon}
\begin{aligned}
&\langle\NOB{\dd_L}  \NOB{\dd_M}\rangle|_{1/N_c^2,\,\, {\rm strong}}=\\
&\frac{2g^3\pi^3}{3N_c^2}({\tt planar})+\frac{e^{4\pi g}}{8 N_c^2}\frac{1}{i^{L+M}}\left(\frac{ g}{2\pi}\right)^{\frac{L+M+1}{2}}\int_{-\infty}^{\infty}dt\, e^{-t^2}(t^2-1)H_{L}(t)H_M(t) \comma
\end{aligned}
\eeq
where $({\tt planar})$ is the planar result computed in \cite{Giombi:2018qox},
\beq
({\tt planar})=\frac{e^{4\pi g}}{2(2 g)^{3/2}\pi^2}\left(-\frac{g}{\pi}\right)^{L}L! \left[1-\frac{3}{32\pi g}(2L^2+2L+1)\right]\delta_{L,M}\period
\eeq
 
 The integral in \eqref{eq:2ptstrongnon} can be computed by using the following properties of the Hermite polynomials:
 \beq
 \begin{aligned}
 &\int_{-\infty}^{\infty}dt \, e^{-t^2} H_L(t)H_M(t)=2^{L}L! \sqrt{\pi}\delta_{L,M}\comma\\
 &(t^2-1)H_M(t)=\frac{H_{M+2}(t)}{4}+\frac{2M-1}{2}H_M (t)+M(M-1)H_{M-2} (t)\period
 \end{aligned}
 \eeq
 The result reads
\beq\label{eq:strong2ptresult}
\begin{aligned}
&\langle\NOB{\dd_L}  \NOB{\dd_M}\rangle|_{1/N_c^2,\,\, {\rm strong}}=\frac{2g^3\pi^3}{3N_c^2}({\tt planar})+\\
&\frac{e^{4\pi g}}{8 N_c^2}\sqrt{\frac{g}{2}}\left(-\frac{g}{\pi}\right)^{L}L!\left[-\frac{\pi}{2g}\delta_{L-2,M}+\frac{(2L-1)}{2}\delta_{L,M}-\frac{g}{2\pi}(L+2)(L+1)\delta_{L+2,M}\right] \period
\end{aligned}
\eeq
 As in the weak-coupling analysis, we can restore the orthogonality by performing the following subtraction:
 \beq
 \begin{aligned}
\NOB{\dd_{L}}|_{1/N_c^2} &= \NOB{\dd_{L}}-\,\,\frac{\langle\NOB{\dd_L}  \NOB{\dd_{L-2}}\rangle|_{1/N_c^2,\,\,\text{strong}}}{\langle\NOB{\dd_{L-2}}  \NOB{\dd_{L-2}}\rangle}\times \NOB{\dd_{L-2}}\\
&=\NOB{\dd_{L}}+\frac{\pi^{5/2}g}{4N_c^2}\NOB{\dd_{L-2}}\period
\end{aligned}
 \eeq
 Here we used the planar two-point functions at strong coupling \cite{Giombi:2018qox}
 \beq\label{eq:planartwoptstrongsub}
 \langle \NOB{\dd_L}\NOB{\dd_L}\rangle=\frac{e^{4\pi g}}{2 (2 g)^{3/2}\pi^2}\left(-\frac{g}{\pi}\right)^{L}L!\left[1-\frac{3}{32\pi g}+O(g^{-2})\right]\period
 \eeq
 \paragraph{Two-point functions} We now proceed to the computation of the non-planar correction to the two-point function. The first term in \eqref{eq:GLLexplicit} can be computed by setting $M=L$ in \eqref{eq:strong2ptresult},
 \beq\label{eq:strongtwoptnon}
 \langle \NOB{\dd_L}\NOB{\dd_L}\rangle|_{1/N_c^2,\,{\rm strong}}=\frac{e^{4\pi g}g^{3/2}\pi}{6\sqrt{2}N_c^2}\left(-\frac{g}{\pi}\right)^{L}L!\left[1-\frac{3(2L^2-6L+5)}{32\pi g}\right]\period
 \eeq
 
 On the other hand, the second term, which comes from the mixing with the multi-trace operators, is given in terms of planar correlators involving both the bulk and the defect operators. They can be computed by taking the strong coupling limit of the integral expression \eqref{eq:integralddb} as was done in section \ref{subsec:strong}. However, it turns out that such an integral always scales as
 \beq
 \langle \NOB{\dd_{L}}\NOB{\dd_{\ell|j}}\rangle \sim e^{4\pi g} g^{\ell-1/2}\comma
 \eeq
 and once divided by the two-point function it yields
 \beq
 \frac{\langle \NOB{\dd_{L}}\NOB{\dd_{\ell|j}}\rangle^2}{\langle \NOB{\dd_{\ell}}\NOB{\dd_{\ell}}\rangle}\sim e^{4\pi g}g^{-1}\period
 \eeq
 This is always subleading as compared to the first term \eqref{eq:strongtwoptnon}. Thus, at the first two orders at strong coupling, one can completely neglect the effect of the mixing with the multi-trace operators and the non-planar correction to the two-point function is simply given by
 \beq
 G_{L,L}|_{1/N_c^2,\,{\rm strong}}=\frac{e^{4\pi g}g^{3/2}\pi}{6\sqrt{2}N_c^2}\left(-\frac{g}{\pi}\right)^{L}L!\left[1-\frac{3(2L^2-6L+5)}{32\pi g}\right]\period
 \eeq
 
 Using the result for the planar two-point functions \eqref{eq:planartwoptstrongsub} and the strong-coupling expansion of the Wilson loop expectation value
 \beq
 \langle\mathcal{W}\rangle=\frac{e^{4\pi g}}{2(2g)^{3/2}\pi^2}\left[\left(1-\frac{3}{32\pi g}+O(g^{-2})\right) +\frac{2(\pi g)^3}{3N_c^2}\left(1-\frac{15}{32 \pi g}+O(g^{-2})\right)+O(1/N_c^4)\right]\comma\nn
 \eeq
 we obtain the following expression for the normalized correlator:
 \beq
 \begin{aligned}
 &\llangle\NO{\dd_L}\NO{\dd_L}\rrangle=\\
 &\left(-\frac{g}{\pi}\right)^{L}L!\left[\left(1-\frac{3(L+L^2)}{16\pi g}+O(g^{-2})\right)+\frac{(\pi g)^2 L}{2N_c^2}\Big(1+O(g^{-1})\Big)+O(1/N_c^{4})\right]\period
 \end{aligned}
 \eeq
 As shown here, the leading non-planar correction at strong coupling is linear in $L$. It would be an interesting future direction to reproduce this result from the string worldsheet by extending the analysis in \cite{Giombi:2017cqn} to the non-planar surface. 
 \subsection{Integral representation and Quantum Spectral Curve\label{subsec:qscnonplanar}}
As we already saw in the previous two subsections, the integral representation for the planar correlators is useful for computing the weak and strong coupling expansions of the non-planar two-point functions. At least formally, one can perform the same analysis also at finite 't Hooft coupling and write down an integral representation of the full non-planar two-point function using the Quantum Spectral Curve.

In particular, it is straightforward to express the first term $\langle \NOB{\dd_L}\NOB{\dd_L}\rangle|_{1/N_c^2}$ of \eqref{eq:GLLexplicit} in terms of an integral since this correction comes from the nonplanar correction to the expectation value of the Wilson loop, which admits an integral representation. Namely, we have\fn{Note that the function $Q_L(x)$ used in this expression is the {\it planar} Q-functions given in \eqref{eq:defofQLFLBJ}.}
\beq
\langle \NOB{\dd_L}\NOB{\dd_L}\rangle|_{1/N_c^2}=\int d\mu_{1/N_c^2}\,\, Q_{L} (x) Q_{L}(x)\comma
\eeq
where $d\mu_{1/N_c^2}$ is the non-planar correction to the measure given in \eqref{eq:nonplanarmeasure}. 

On the other hand, the second term is given by a sum of ratios of planar correlators. Although the expression might seem quite complicated at first sight, one can simplify it to some extent using the properties of the orthogonal polynomials. To see this, let us first express the summand using the integral expression for the planar correlators:
\beq
\begin{aligned}
(-2)^{j}\frac{\langle\NOB{\dd_{L}}\NOB{\dd_{\ell|j}} \rangle^2}{\langle \NOB{\dd_{\ell}}\NOB{\dd_{\ell}}\rangle} =&(-2)^{j} \int d\mu (x) d\mu (y) B_j (x) Q_L(x) B_j (y)Q_L(y)\frac{Q_{\ell}(x)Q_{\ell}(y)}{\int d\mu \,Q_{\ell} Q_{\ell}} \period
\end{aligned}
\eeq 
Since the function $Q_L(x)$ is an orthogonal polynomial of $X=g(x-x^{-1})$, one can perform the summation over $\ell$ using the Christoffel-Darboux theorem (see for instance \cite{simon2008christoffel}),
\beq
\sum_{\ell=0}^{L-j-1}\frac{Q_{\ell}(x)Q_{\ell}(y)}{\int d\mu \,Q_{\ell} Q_{\ell}}=K_{L-j-1}(x,y)\comma
\eeq 
where $K_{L}(x,y)$ is the Christoffel-Darboux kernel
\beq
K_{L}(x,y)=\frac{1}{\int d\mu \,Q_{L} Q_{L}}\frac{Q_{L+1}(x)Q_{L}(y)-Q_{L}(x)Q_{L+1}(y)}{g(x-y)(1+\frac{1}{xy})}\period
\eeq

Using the Christoffel-Darboux kernel, one can express the sum as
\beq\label{eq:CDkernel}
\sum_{\substack{\ell+j<L\\0\leq \ell,\,1\leq j}}(-2)^{j}\frac{\langle\NOB{\dd_{L}}\NOB{\dd_{\ell|j}} \rangle^2}{\langle \NOB{\dd_{\ell}}\NOB{\dd_{\ell}}\rangle}=\int d\mu(x)d\mu(y)\mathcal{K}_{L}(x,y)Q_L(x)Q_L(y)\comma
\eeq
where the kernel $\mathcal{K}_L$ is given by
\beq
\begin{aligned}
\mathcal{K}_L (x,y)=&\sum_{j=0}^{L-2}(-2)^{L-j-1}B_{L-j-1}(x)B_{L-j-1}(y)K_{j}(x,y)\\
=&\frac{(4\pi g)^2 (-2)^{L-1}x^{L}y^{L}}{(1+x^2)(1+y^2)}\sum_{j=0}^{L-2}\frac{K_{j}(x,y)}{(-2 x y)^{j}}\period
\end{aligned}
\eeq
Note that, although we explicitly performed one of the two sums, the expression \eqref{eq:CDkernel} is slightly formal since the kernel still contains the sum over $j$. It would be interesting to see if we can further simplify the expression.

Now, combining the two terms, we finally get the following integral expression for the non-planar two-point function, which consists of single and double integrals:
\beq\label{eq:integralnonplanarQSC}
G_{L,L}|_{1/N_c^2}=\int d\mu_{1/N_c^2}  Q_L(x)Q_L(x)-\int d\mu(x)d\mu(y)\mathcal{K}_{L}(x,y)Q_L(x)Q_L(y)\period
\eeq
Recently, the integrability-based methods were applied to the non-planar corrections \cite{Bargheer:2017nne,Bargheer:2018jvq,Ben-Israel:2018ckc,Eden:2017ozn}. It would be interesting to compare our formula with such methods and clarify the relation between them.
\section{Conclusion\label{sec:conclusion}}
In this paper, we studied the correlation functions of the bulk single-trace operator and the insertions on the Wilson loop using the results from localization. The results are exact and depend nontrivially on the 't Hooft coupling constant $\lambda$. At weak coupling, we checked the results against the perturbation theory of $\mathcal{N}=4$ SYM. At strong coupling, we evaluated the correlation functions of fluctuations of the string worldsheet and the vertex operator, where the former corresponds to the insertion on the Wilson loop and the latter describes the single-trace operator. In both cases, the results are in perfect agreement with what we obtained from localization. We also showed that the results at large $N_c$ can be simply expressed in terms of the $Q$-functions of the quantum spectral curve.

Along the way, we also clarified several aspects of the Gram-Schmidt analysis which were not addressed in the previous paper \cite{Giombi:2018qox}. Most importantly, we pointed out the necessity of including the multi-trace operators in the operator spectrum for the analysis of non-planar corrections. We then confirmed it by computing the two-point function and comparing it with the perturbation theory.

One possible future direction is to apply our techniques to other theories, such as $\mathcal{N}=2$ superconformal field theories in four dimensions and ABJM theory in three dimensions. Supersymmetric Wilson loops in these theories were studied from the defect CFT point of view in \cite{Bianchi:2017ozk,Bianchi:2017svd,Bianchi:2018scb,Fiol:2015spa,Bianchi:2018zpb}. In particular, the correlators of a Wilson loop and chiral local operators in the bulk were analyzed recently in \cite{Billo:2018oog} using the matrix model and the Gram-Schmidt orthogonalization. It would be interesting to compute more general correlators involving operator insertions on the Wilson loop by generalizing the analysis in this paper.
  
There are also several other interesting future directions. Firstly, both in this paper and in the previous paper \cite{Giombi:2018qox}, we focused on the Wilson loop in the fundamental representation, and it would be interesting to generalize the analysis to higher-dimensional  representations; In particular, when the size of the representation is of order $N_c$, the AdS dual is given by a probe D-brane \cite{Drukker:2005kx,Yamaguchi:2006tq,Gomis:2006sb}. One can therefore try to test the results from localization at strong coupling by computing the correlation functions of fluctuations on the D-branes. Work in that direction is in progress \cite{InProgress}. Also interesting would be to consider the representation with $O(N_c^2)$ size which corresponds to a nontrivial bubbling geometry \cite{Yamaguchi:2006te,Lunin:2006xr,DHoker:2007mci}.

Secondly, it was recently shown that the algebra of operators on the Wilson loop in a certain higher-dimensional representation in $2d$ BF theory is isomorphic to $\mathfrak{gl}(K)$ Yangian \cite{Ishtiaque:2018str}. Their setup seems to be intimately related to our setup since the BF theory is a zero-coupling limit of $2d$ Yang-Mills. It would therefore be interesting to compute the operator algebra in our setup and see if there is an analogous structure.

Thirdly, it would be worth performing a more thorough analysis on the non-planar corrections to the correlators on the Wilson loop and understand their underlying structure. For the expectation value of a single Wilson loop \cite{Okuyama:2017feo} and the correlator of Wilson loops \cite{Okuyama:2018aij}, it was recently shown that one can apply the topological recursion \cite{Eynard:2007kz} and systematically compute the non-planar corrections. It would be interesting if such a method can be generalized to the observables that we studied in this paper.
 
 Yet another direction is to study these correlators using the conformal bootstrap as was initiated in \cite{Liendo:2016ymz,Liendo:2018ukf}. In supersymmetric theories with exactly marginal directions, one needs to specify the values of the marginal parameters using some additional inputs in order to pin down the theory one wants to bootstrap. The bulk-defect correlators and the defect structure constants which we computed would provide such inputs and can be used as a starting point for the numerical analysis of the defect bootstrap equation \cite{Billo:2016cpy,Gadde:2016fbj}. 
 
Finally, it would be interesting to try to reproduce the results in this paper by using the integrability machinery. Thanks to the recent developments, we now have integrability-based frameworks to study the correlation functions of single-trace operators \cite{Fleury:2016ykk,Eden:2016xvg} and the correlation functions of operators on the Wilson loop \cite{Kim:2017phs,KiryuToAppear}. By contrast, the correct framework to analyze the bulk-defect correlators is not known at present time. It would be interesting to try to figure out such a framework by performing explicit computations at weak and strong coupling. Also, the appearance of Q-functions in our final results suggests the utility of the quantum spectral curve for the study of correlation functions. So far there have been three data points (including this paper) \cite{Giombi:2018qox,Cavaglia:2018lxi} which point to such a direction. It would be interesting to connect these dots and try to come up with a fully non-perturbative, integrability-based framework for studying the correlation functions.
  
\section*{Acknowledgement}
SK thanks K.~Costello, N.~Ishtiaque, F.~Moosavian, A.~Polychronakos, M.~Rapcak and Y.~Zhou for discussions on related topics.
The work of SG is supported in part by the US NSF under Grant No. PHY-1620542. The work of SK is supported by DOE grant number DE-SC0009988.
\newpage
\appendix
\section{Bulk two-point functions and multi-trace operators\label{ap:bulk2pt}}
In this appendix, we provide a brief argument on why the multi-trace operators must be included in the defect CFT spectrum using the defect OPE expansion of the bulk two-point function.

In general defect CFTs, the two-point functions of the bulk operators can be expressed as a sum of products of bulk-defect two-point functions,
\beq
\langle O_{\rm bulk} O_{\rm bulk}\rangle \sim \sum_{O_{\rm defect}}\langle O_{\rm bulk} O_{\rm defect}\rangle\langle O_{\rm defect} O_{\rm bulk}\rangle\period
\eeq
This defect OPE expansion applies to physical correlators which depend on the cross ratios. In supersymmetric theories, one can also write down a truncated version of it which is closed in a supersymmetric subsector (called the ``micro-bootrstrap equation'') as was shown in \cite{Liendo:2016ymz}. 

Let us now analyze the consequence of such an expansion on our bulk two-point function. In the large $N_c$ limit, the bulk two-point function is given by a sum of the disconnected term \eqref{eq:bulktwoptdisconnected}, and the connected term
\beq
\begin{aligned}
\langle \mathcal{W} \hat{O}_{J_1} \hat{O}_{J_2}\rangle_{\rm conn}=&\frac{2^{-(J_1+J_2)/2}\sqrt{J_1 J_2}}{N_c^2}\left(\frac{2\pi-a}{2\pi+a}\right)^{(J_1+J_2)/2}\times\\
&\left[\sum_{k=1}^{{\rm min}(J_1,J_2)}(-1)^{k}\left(\frac{2\pi+a}{2\pi-a}\right)^{k}(J_1+J_2-2k)I_{J_1+J_2-2k}(\sqrt{\lambda^{\prime}})\right.\\
&\left.+\sum_{k=1}^{\infty}(J_1+J_2+2k-2)I_{J_1+J_2+2k-2}(\sqrt{\lambda^{\prime}})\right]+O(1/N_c^4)\period
\end{aligned}
\eeq
To interpret them in terms of the defect CFT, one has to divide them by the expectation value of the Wilson loop as was done in section \ref{subsec:cftdata}. After doing so, the disconnected term becomes
\beq\label{eq:leadingdisc}
\llangle \bb_{J_1}\bb_{J_2}\rrangle_{\rm disc}=\left(-\frac{1}{2}\right)^{J_1} \delta_{J_1,J_2}\period
\eeq 
Now, from the fact that this quantity is $O(1)$ (rather than $O(1/N_c^2)$), it immediately follows that one has to include operators other than $\mathcal{W}[\NO{\tilde{\Phi}^{L}}]$ in the defect CFT spectrum: The bulk-defect two-point functions of $\mathcal{W}[\NO{\tilde{\Phi}^{L}}]$ is $O(1/N_c)$ as was shown \ref{sec:largeN} and we would not be able to reproduce the $O(1)$ contribution if we only had these operators. In fact, one can interpret \eqref{eq:leadingdisc} as a result of exchanging $\dd_{0|J_1}=\mathcal{W} \bb_{J_1} $. This explains the necessity of including the ``multi-trace'' operators in the defect CFT spectrum. 

It would be interesting to perform a more thorough analysis of the OPE expansion and the bootstrap equation in the supersymmetric subsector using the results computed in this paper.
\section{Correlators with two operator insertions on the Wilson loop\label{ap:onetwo}}
Let us now generalize the computation to the correlation functions of a single bulk operator and two operator insertions on the Wilson loop,
\beq
G_{L_1,L_2|J}\equiv \langle\mathcal{W}[\NO{\tilde{\Phi}^{L_1}}\NO{\tilde{\Phi}^{L_2}}] \,\,\bb_J \rangle\period
\eeq
Using the large $N_c$ expansion of $\NO{\tilde{\Phi}^{L_1}}=\NO{\dd_{L_1}}$ and $\NO{\tilde{\Phi}^{L_2}}=\NO{\dd_{L_2}}$, one obtains
\beq\label{eq:onetwoexplicitexpression}
\begin{aligned}
G_{L_1,L_2|J}=&\langle\NOB{\dd_{L_1}}\NOB{\dd_{L_2}}\bb_J\rangle-\sum_{\substack{\ell_1+J_1<L_1\\0\leq \ell_1,1\leq J_1}}\frac{\langle \NOB{\dd_{L_1}}\NOB{\dd_{\ell_1|J_1}}\rangle}{\langle\NOB{\dd_{\ell_1|J_1}}\NOB{\dd_{\ell_1|J_1}} \rangle}\langle\NOB{\dd_{\ell_1|J_1}}\NOB{\dd_{L_2}}\bb_{J} \rangle\\
&-\sum_{\substack{\ell_2+J_2<L_2\\0\leq \ell_2,1\leq J_2}}\frac{\langle \NOB{\dd_{L_2}}\NOB{\dd_{\ell_2|J_2}}\rangle}{\langle\NOB{\dd_{\ell_2|J_2}}\NOB{\dd_{\ell_2|J_2}} \rangle}\langle\NOB{\dd_{L_1}}\NOB{\dd_{\ell_2|J_2}}\bb_{J} \rangle+O(1/N_c^2)\period
\end{aligned}
\eeq

The second and the third terms are generically suppressed by $1/N_c$. The only cases where these terms are not suppressed are 1.~$\ell_1=L_2$ and $J_1=J$, and 2.~$\ell_2=L_1$ and $J_2=J$.
In the first case, the three-point function $\langle\NOB{\dd_{\ell_1|J_1}}\NOB{\dd_{L_2}}\bb_{J} \rangle$ becomes $O(1)$ and therefore is not suppressed by $1/N_c$ while in the second case the same thing happens for $\langle\NOB{\dd_{L_1}}\NOB{\dd_{\ell_2|J_2}}\bb_{J} \rangle$. Owing to the restriction on the range of the summation over $\ell_k$ and $J_k$ in \eqref{eq:onetwoexplicitexpression}, these additional contributions exist only for $J<|L_1-L_2|$.

These additional contributions, when they exist, cancel precisely the first term in \eqref{eq:onetwoexplicitexpression}. Therefore we arrive at the following selection rule for the correlator $G_{L_1,L_2|J}$:
\beq
G_{L_1,L_2|J}=\begin{cases}0\qquad&(J<|L_1-L_2|)\\\langle\NOB{\dd_{L_1}}\NOB{\dd_{L_2}}\bb_J\rangle\qquad &(J\geq |L_1-L_2|)\end{cases}\period
\eeq
In what follows, we compute the leading large $N_c$ results at weak and strong couplings\fn{For simplicity, here we only present the leading weak- and strong-coupling results. } using the integral expression \eqref{eq:generalizedintegralexpressions} assuming $J\geq |L_1-L_2|$.
\subsection{Weak coupling expansion}
At the leading order at weak coupling, the function $Q_{L_i}$ is given by (see section \ref{subsec:weak})
\beq
Q_{L_i} =(-i g)^{L_i}U_{L_i} (\cos \theta)=(-i g)^{L_i}\frac{\sin (L_i+1)\theta}{\sin \theta}\qquad (x=-ie^{i\theta})\comma
\eeq
while the expressions for the measure $d\mu$ and $B_J$ are given in \eqref{eq:weakvariousquantities}. Let us now first consider the case where the single-trace operator is sufficiently long, or more precisely the case in which $J\geq L_1+L_2$ is satisfied. In this case, the analysis is similar to the one for $G_{L|J}$ performed in section \ref{subsec:weak}: Namely in order to compensate the factor $e^{iJ\theta}$ contained in $B_J$, one has to expand $e^{2g\sin\theta}$ factor up to $O(g^{J-L_1-L_2})$,
\beq
d\mu\,\,B_J=\frac{d\theta}{2\pi}\frac{e^{iJ\theta}}{i^{J}}\left(1+\cdots +g^{J-L_1-L_2}\frac{2^{J-L_1-L_2}(2\pi)^{J-L_1-L_2}}{(J-L_1-L_2)!}\sin^{J-L_1-L_2}\theta+\cdots\right)\period
\eeq
Substituting this term into the integral, we get
\begin{align}
\left.G_{L_1,L_2|J}\right|_{\rm weak}&=\frac{2^{-J/2}\sqrt{J}}{N_c}(-i)^{J+L_1+L_2}\frac{(4\pi)^{J-L_1-L_2}g^{J}}{(J-L_1-L_2)!}\int \frac{d\theta}{2\pi}e^{iJ\theta}\sin^{J-L_1-L_2}\theta \, U_{L_1}(\cos\theta)U_{L_2}(\cos\theta)\,\nn\\
&=\frac{(-1)^{L_1+L_2}2^{-J/2}\sqrt{J}g^{J}}{N_c}\frac{(2\pi)^{J-L_1-L_2}}{(J-L_1-L_2)!}\period
\end{align}

On the other hand, when $J$ satisfies 
\beq
|L_1-L_2|\leq J<L_1+L_2 \comma\qquad L_1+L_2-J:{\rm even}\comma
\eeq
we do not need to expand $e^{2g \sin \theta}$ to compensate $e^{iJ \theta}$. Therefore the integral is given by 
\beq
\begin{aligned}
\left.G_{L_1,L_2|J}\right|_{\rm weak}&=\frac{2^{-J/2}\sqrt{J}}{N_c}(-i)^{J+L_1+L_2}g^{L_1+L_2}\int \frac{d\theta}{2\pi}e^{iJ\theta} \, U_{L_1}(\cos\theta)U_{L_2}(\cos\theta)\,\\
&=\frac{2^{-J/2}\sqrt{J}}{N_c}\frac{(-i)^{J+L_1+L_2}g^{L_1+L_2}(L_1+L_2-J+2)}{2}\period
\end{aligned}
\eeq
When $J$ satisfies
\beq
|L_1-L_2|\leq J<L_1+L_2 \comma\qquad L_1+L_2-J:{\rm odd}\comma
\eeq
we need to take the $O(g)$ term in $e^{2g\sin \theta}$, and the result reads
\begin{align}
\left.G_{L_1,L_2|J}\right|_{\rm weak}&=\frac{(4\pi)2^{-J/2}\sqrt{J}}{N_c}(-i)^{J+L_1+L_2}g^{L_1+L_2+1}\int \frac{d\theta}{2\pi}e^{iJ\theta} \sin \theta\, U_{L_1}(\cos\theta)U_{L_2}(\cos\theta)\,\nn\\
&=\frac{\pi 2^{1-J/2}\sqrt{J}}{N_c}(-i)^{J+L_1+L_2-1}g^{L_1+L_2+1}\period
\end{align}

In summary, the results at weak coupling are given by the following expressions:
\beq\nn
\begin{aligned}
&G_{L_1,L_2|J}=\\
&\begin{cases}0&J<|L_1-L_2|\\\frac{2^{-J/2}\sqrt{J}}{N_c}\frac{(-i)^{J+L_1+L_2}g^{L_1+L_2}(L_1+L_2-J+2)}{2}&|L_1-L_2|\leq J<L_1+L_2\,, \quad L_1+L_2-J:{\rm even}\\\frac{\pi 2^{1-J/2}\sqrt{J}}{N_c}(-i)^{J+L_1+L_2-1}g^{L_1+L_2+1}&|L_1-L_2|\leq J<L_1+L_2\,, \quad L_1+L_2-J:{\rm odd}\\\frac{(-1)^{L_1+L_2}2^{-J/2}\sqrt{J}g^{J}}{N_c}\frac{(2\pi)^{J-L_1-L_2}}{(J-L_1-L_2)!}&L_1+L_2\leq J \period
\end{cases}
\end{aligned}
\eeq
\subsection{Strong coupling expansion}
We now study the strong coupling limit. As shown in section \ref{subsec:strong}, the strong-coupling limit of the relevant quantities is given by
\beq
\begin{aligned}
&\oint d\mu =\int^{\infty}_{-\infty}\frac{dt\, e^{4\pi g}e^{-t^2}}{(2\pi)^{5/2}g^{3/2}}\comma\quad
B_{J}=2\pi g \sum_{k=0}^{\infty}\frac{(-1)^{k}}{k!}\left(i t\sqrt{\frac{2}{\pi g}}\right)^{k}\frac{\Gamma[\frac{1+J+k}{2}]}{\Gamma[\frac{1+J-k}{2}]}\comma\\
&Q_{L}(t)=(-i)^{L}\left(\frac{g}{2\pi}\right)^{L/2}H_{L}(t)\period
\end{aligned}
\eeq
To get a non-zero answer, one needs to take $O(t^{|L_1-L_2|})$ term in the expansion of $B_J$. Then the integral becomes
\beq
\begin{aligned}
\left.G_{L_1,L_2|J}\right|_{\rm strong}=&\frac{e^{4\pi g}2^{-J/2}\sqrt{J}}{N_c(2\pi)^{3/2}g^{1/2}\,\,|L_1-L_2|!}\left(-\frac{1}{\pi}\right)^{L_{\rm max}}\left(\frac{g}{2}\right)^{L_{\rm min}}\frac{\Gamma[\frac{1+J+|L_1-L_2|}{2}]}{\Gamma[\frac{1+J-|L_1-L_2|}{2}]}\\
&\times \int^{\infty}_{-\infty}\, dt \, e^{-t^2} t^{|L_1-L_2|}H_{L_1}(t)H_{L_2}(t)\comma\\
=&\frac{e^{4\pi g}2^{-1-J/2}\sqrt{J}L_{\rm max}!}{N_c\pi\sqrt{2g}\,\,|L_1-L_2|!}\left(-\frac{1}{\pi}\right)^{L_{\rm max}}g^{L_{\rm min}}\frac{\Gamma[\frac{1+J+|L_1-L_2|}{2}]}{\Gamma[\frac{1+J-|L_1-L_2|}{2}]}\comma
\end{aligned}
\eeq
where $L_{\rm max}\equiv {\rm max}(L_1,L_2)$ and $L_{\rm min}\equiv {\rm min}(L_1,L_2)$.

Let us also consider the normalized correlator by dividing the correlator by the expectation value of the Wilson loop \eqref{eq:expectationvaluewilsonstrong}. We then obtain
\beq
\left.\frac{G_{L_1,L_2|J}}{\langle\mathcal{W}\rangle}\right|_{\rm strong}=\frac{(-1)^{L_{\rm max}}2^{1-\frac{J}{2}}g^{1+L_{\rm min}}\sqrt{J}L_{\rm max}!}{\pi^{L_{\rm max}-1}N_c \,\,|L_1-L_2|!}\frac{\Gamma[\frac{1+J+|L_1-L_2|}{2}]}{\Gamma[\frac{1+J-|L_1-L_2|}{2}]}\period
\label{loc-twodef-strong}
\eeq
This result coincides with the direct perturbative string theory analysis performed in section \ref{Two-def-strong}.
\section{Non-planar correction to the three-point functions\label{ap:3pt}}
Here we show the result for the non-planar correction to the three-point functions on the Wilson loop \eqref{eq:threepntdisplay}. By extending the argument in section \ref{subsec:nonplanargeneral}, one can straightforwardly compute the corrections as
\beq
\begin{aligned}
G_{L_1,L_2,L_3}|_{\tt first}=&- \!\!\!\sum_{\substack{\{s,t,u\}\\=\{1,2,3\}}}\sum_{\substack{\ell_s+j_s<L_s\\0\leq \ell_s,\,1\leq j_s}}(-2)^{j_s}\frac{\langle\NOB{\dd_{L_s}}\NOB{\dd_{\ell_s|j_s}} \rangle\langle\NOB{\dd_{\ell_s|j_s}}\NOB{\dd_{L_t}}\NOB{\dd_{L_u}}\rangle}{\langle \NOB{\dd_{\ell_s}}\NOB{\dd_{\ell_s}}\rangle}\comma\\
G_{L_1,L_2,L_3}|_{\tt second}=& \!\!\!\sum_{\substack{\{s,t,u\}\\=\{1,2,3\}}}\sum_{1\leq j}\sum_{\substack{\ell_s+j<L_s\\0\leq \ell_s}}\sum_{\substack{\ell_t+j<L_t\\0\leq \ell_t}}(-2)^{j}\frac{\langle\NOB{\dd_{L_s}}\NOB{\dd_{\ell_s|j}} \rangle\langle\NOB{\dd_{L_t}}\NOB{\dd_{\ell_t|j}} \rangle}{\langle \NOB{\dd_{\ell_s}}\NOB{\dd_{\ell_s}}\rangle\langle \NOB{\dd_{\ell_t}}\NOB{\dd_{\ell_t}}\rangle}\\
&\quad \times \langle\NOB{\dd_{\ell_s}}\NOB{\dd_{\ell_t}}\NOB{\dd_{L_u}}\rangle\\
G_{L_1,L_2,L_3}|_{\tt third}=&\sum_{\substack{\{s,t,u\}\\=\{1,2,3\}}}\sum_{\tilde{\ell}_s<L_s}\,\,\,\sum_{\ell_s+j_s<L_s}(-2)^{j_s}\frac{\langle\NOB{\dd_{L_s}}\NOB{\dd_{\ell_s|j_s}}\rangle\langle\NOB{\dd_{\tilde{\ell}_s}}\NOB{\dd_{\ell_s|j_s}}\rangle}{\langle\NOB{\dd_{\ell_s}}\NOB{\dd_{\ell_s}}\rangle\langle\NOB{\dd_{\tilde{\ell}_s}}\NOB{\dd_{\tilde{\ell}_s}}\rangle}\\
&\quad \times \langle\NOB{\dd_{\tilde{\ell}_s}}\NOB{\dd_{L_t}}\NOB{\dd_{L_u}}\rangle\period
\end{aligned}
\eeq
Here the first sum $\{s,t,u\}=\{1,2,3\}$ denotes the sum over cyclic permutations; namely  $\{s,t,u\}=\{1,2,3\},\{2,3,1\},\{3,1,2\}$. Thus, in total we obtain the following expression for the correlator:
\beq\label{eq:GL123explicit}
\begin{aligned}
&G_{L_1,L_2,L_3}=\\
&\langle\NOB{\mathcal{O}_{L_1}}\NOB{\mathcal{O}_{L_2}}\NOB{\mathcal{O}_{L_3}} \rangle|_{1/N_c^2}+G_{L_1,L_2,L_3}|_{\tt first}+G_{L_1,L_2,L_3}|_{\tt second}+G_{L_1,L_2,L_3}|_{\tt third}\period
\end{aligned}
\eeq

Using \eqref{eq:GL123explicit}, one can also compute the normalized three-point function at $O(1/N_c^2)$,
\beq
\llangle\NO{\tilde{\Phi}^{L_1}}\NO{\tilde{\Phi}^{L_2}}\NO{\tilde{\Phi}^{L_3}}\rrangle\equiv \frac{G_{L_1,L_2,L_3}}{\langle \mathcal{W}\rangle}\period
\eeq
Here we display some of the explicit results for the first two leading orders at weak coupling\fn{The results are correct up to the order $O(1/N_c^2)$.}:

{\footnotesize
\beq
\begin{aligned}\label{eq:3ptvarious}
\llangle\NO{\tilde{\Phi}^{2}}\NO{\tilde{\Phi}^{1}}\NO{\tilde{\Phi}^{1}}\rrangle=&g^4\left(1+\frac{1}{N_c^2}\right)+\frac{2\pi^2g^{6} }{3}\left(-1+\frac{1}{N_c^2}\right)\comma\\
\llangle\NO{\tilde{\Phi}^{2}}\NO{\tilde{\Phi}^{2}}\NO{\tilde{\Phi}^{2}}\rrangle=&-g^6\left(1+\frac{7}{N_c^2}\right)+\frac{4\pi^2g^{8} }{N_c^2}\comma\\
\llangle\NO{\tilde{\Phi}^{3}}\NO{\tilde{\Phi}^{2}}\NO{\tilde{\Phi}^{1}}\rrangle=&-g^6\left(1+\frac{5}{N_c^2}\right)+\frac{2\pi^2g^{8} }{3}\left(1+\frac{2}{N_c^2}\right)\comma\\
\llangle\NO{\tilde{\Phi}^{3}}\NO{\tilde{\Phi}^{3}}\NO{\tilde{\Phi}^{2}}\rrangle=&g^8\left(1+\frac{22}{N_c^2}\right)+\frac{38\pi^2g^{10} }{3N_c^2}\comma\\
\llangle\NO{\tilde{\Phi}^{4}}\NO{\tilde{\Phi}^{2}}\NO{\tilde{\Phi}^{2}}\rrangle=&g^8\left(1+\frac{15}{N_c^2}\right)-\frac{2\pi^2g^{8} }{3}\left(1+\frac{11}{N_c^2}\right)\comma\\
\llangle\NO{\tilde{\Phi}^{4}}\NO{\tilde{\Phi}^{3}}\NO{\tilde{\Phi}^{1}}\rrangle=&g^8\left(1+\frac{15}{N_c^2}\right)-\frac{2\pi^2g^{8} }{3}\left(1+\frac{11}{N_c^2}\right)\comma\\
\llangle\NO{\tilde{\Phi}^{5}}\NO{\tilde{\Phi}^{3}}\NO{\tilde{\Phi}^{2}}\rrangle=&-g^{10}\left(1+\frac{35}{N_c^2}\right)+\frac{2\pi^2g^{12} }{3}\left(1+\frac{10}{N_c^2}\right)\comma\\
\llangle\NO{\tilde{\Phi}^{4}}\NO{\tilde{\Phi}^{4}}\NO{\tilde{\Phi}^{4}}\rrangle=&g^{12}\left(1+\frac{135}{N_c^2}\right)-\frac{74\pi^2g^{14} }{N_c^2}\period
\end{aligned}
\eeq
}

\noindent The leading weak coupling terms are in precise agreement with the direct (planar and non-planar) Wick contractions.

It would be interesting to perform more systematic analysis and also to compute the correlators at strong coupling.
\section{Non-planar two-point functions at subleading order}
Here we test our results for the non-planar two-point functions \eqref{eq:GLLexplicit} at the next leading order at weak coupling. 

Using the expression \eqref{eq:GLLexplicit}, one can in principle compute the correlators at weak coupling straightforwardly. However, at the next leading order, deriving a formula for general lengths $L$ becomes quite complicated in practice. Instead, here we display some explicit results for small-length operators at first two leading orders\fn{As in \eqref{eq:3ptvarious}, the results are truncated at $O(1/N_c^2)$.}:

{\footnotesize
\begin{align}
\llangle \NO{\tilde{\Phi}}\NO{\tilde{\Phi}}\rrangle=&-g^2+\frac{2\pi^2g^{4}}{3}\left(1-\frac{1}{N_c^2}\right)\comma\label{eq:nonplanar2ptlength1}\\
\llangle \NO{\tilde{\Phi}^2}\NO{\tilde{\Phi}^2}\rrangle=&g^4\left(1+\frac{1}{N_c^2}\right)-\frac{2\pi^2g^{6}}{3}\left(1-\frac{1}{N_c^2}\right)\comma\label{eq:nonplanar2ptlength2}\\
\llangle \NO{\tilde{\Phi}^3}\NO{\tilde{\Phi}^3}\rrangle=&-g^6\left(1+\frac{5}{N_c^2}\right)+\frac{2\pi^2g^{8}}{3}\left(1+\frac{2}{N_c^2}\right)\comma\label{eq:nonplanar2ptlength3}\\
\llangle \NO{\tilde{\Phi}^4}\NO{\tilde{\Phi}^4}\rrangle=&g^8\left(1+\frac{15}{N_c^2}\right)-\frac{2\pi^2g^{10}}{3}\left(1+\frac{11}{N_c^2}\right)\comma\\
\llangle \NO{\tilde{\Phi}^5}\NO{\tilde{\Phi}^5}\rrangle=&-g^{10}\left(1+\frac{35}{N_c^2}\right)+\frac{2\pi^2g^{12}}{3}\left(1+\frac{30}{N_c^2}\right)\comma\\
\llangle \NO{\tilde{\Phi}^6}\NO{\tilde{\Phi}^6}\rrangle=&g^{12}\left(1+\frac{70}{N_c^2}\right)-\frac{2\pi^2g^{14}}{3}\left(1+\frac{64}{N_c^2}\right)\comma
\end{align}
 }

\noindent More generally, we find the following pattern for the first two orders at weak coupling:
\beq\label{eq:generalplanarnonplanar}
\llangle \NO{\tilde{\Phi}^L}\NO{\tilde{\Phi}^L}\rrangle=(-g^2)^{L}\left[\left(1+\frac{\left(\begin{array}{c}L+2\\4\end{array}\right)}{N_c^2}\right)-\frac{2\pi^2g^2}{3} \left(1+\frac{\left(\begin{array}{c}L+2\\4\end{array}\right)-L}{N_c^2}\right)\right]\period
\eeq
As shown in the previous paper \cite{Giombi:2018qox} and in section \ref{subsec:nonplanarweak} of this paper, both the planar results and the tree-level non-planar correction can be derived analytically and they reproduce the results from the perturbation theory. The goal of this appendix is to show that the non-planar correction at one loop, in particular the combinatorial number
$\left(\begin{array}{c}L+2\\4\end{array}\right)-L$,
is also consistent with the perturbation theory.
\begin{figure}[t]
\centering
\includegraphics[clip,height=4cm]{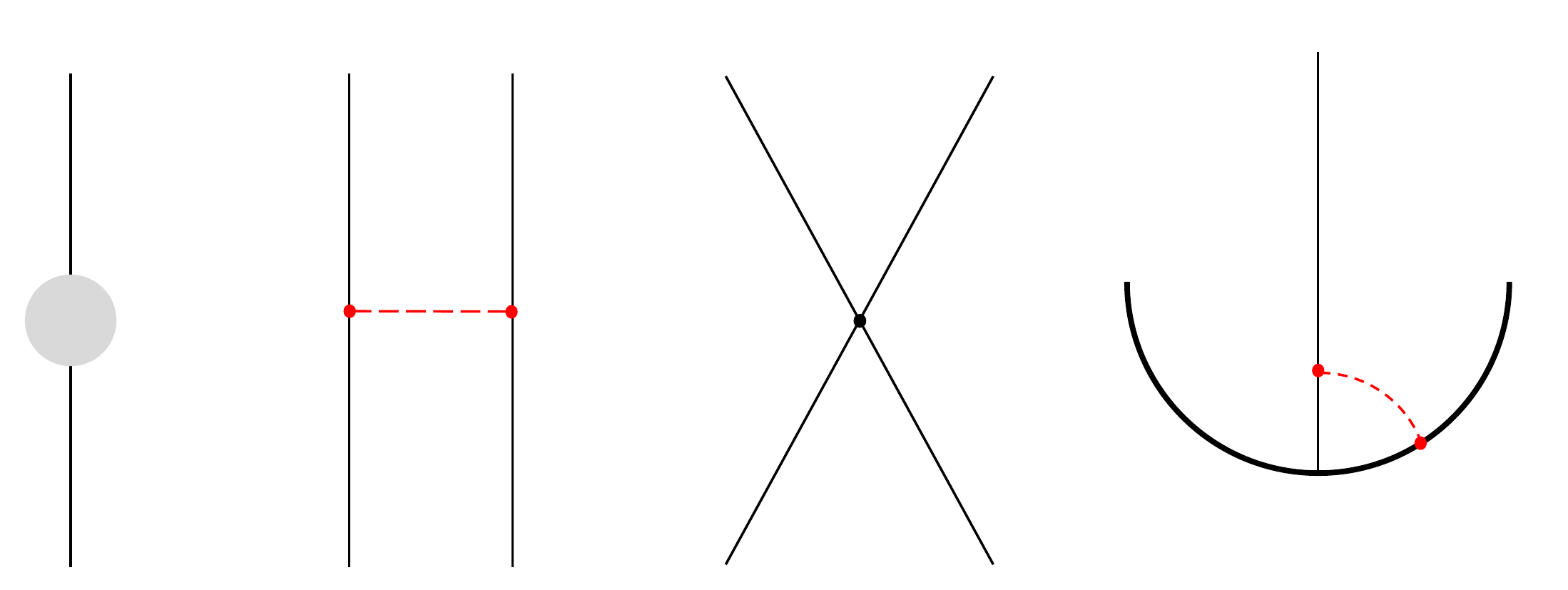}
\caption{The diagrams that show up at one loop; the self-energy of a propagator (denoted by a black line), the gluon exchange (denoted by a red dashed line) between two propagators, the scalar quartic interaction and the gluon exchange between a propagator and the Wilson loop (denoted by a thick black curve).}
\label{fig:figd1}
\end{figure}
\begin{figure}[t]
\centering
\includegraphics[clip,height=4cm]{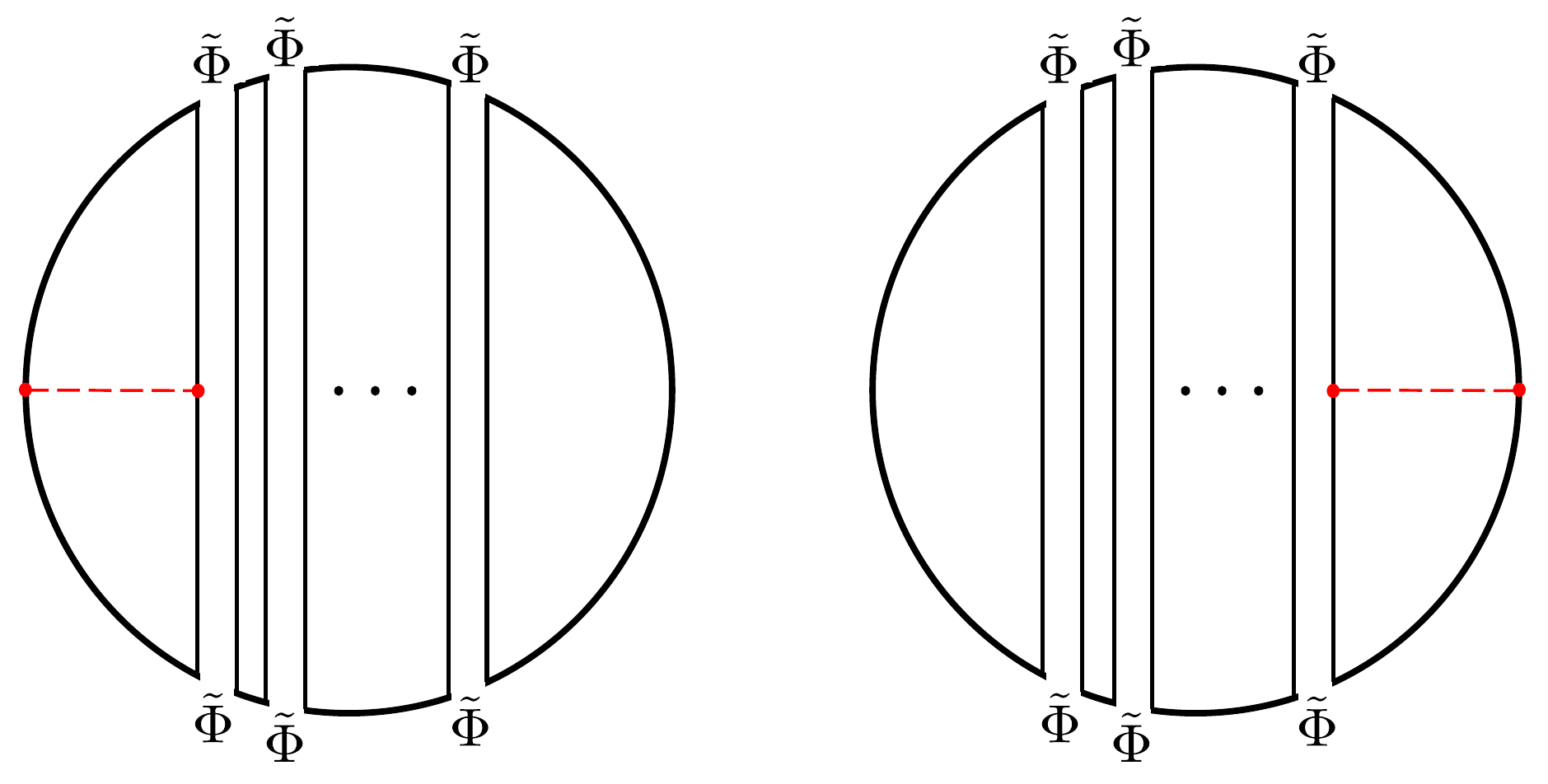}
\caption{The gluon exchange diagrams for the planar two-point functions. Here the propagators of the scalar fields are denoted in the double-line notation. The thick black circle denotes the Wilson loop and the red dashed lines are the gluon propagators. Regardless of the lengths of the operators, there are always two such diagrams for the planar two-point functions.}
\label{fig:figd2}
\end{figure}

To see this, let us first give a brief summary of the planar results at one loop computed in \cite{KiryuToAppear}. At one loop, there are basically four different diagrams (see figure \ref{fig:figd1}) that can contribute to the final answer; the self-energy of a propagator, the gluon exchange between two propagators, the scalar quartic vertex, and the gluon exchange between the Wilson loop and a propagator. Although all the four diagrams are separately divergent, the divergence cancel out when we sum up these diagrams. After the summation, one finds that the finite piece comes solely from the gluon exchanges between the Wilson loop and a propagator. For the two-point functions, there are always two such diagrams (see figure \ref{fig:figd2}) regardless of the lengths of the operators, and the sum of the two gives
\beq
({\tt planar})=-\frac{2\pi^2g^2}{3}\period
\eeq
Multiplying this number to the planar tree-level answer, we reproduce the one-loop planar answer in \eqref{eq:generalplanarnonplanar}.

\begin{figure}[t]
\centering
\includegraphics[clip,height=4.5cm]{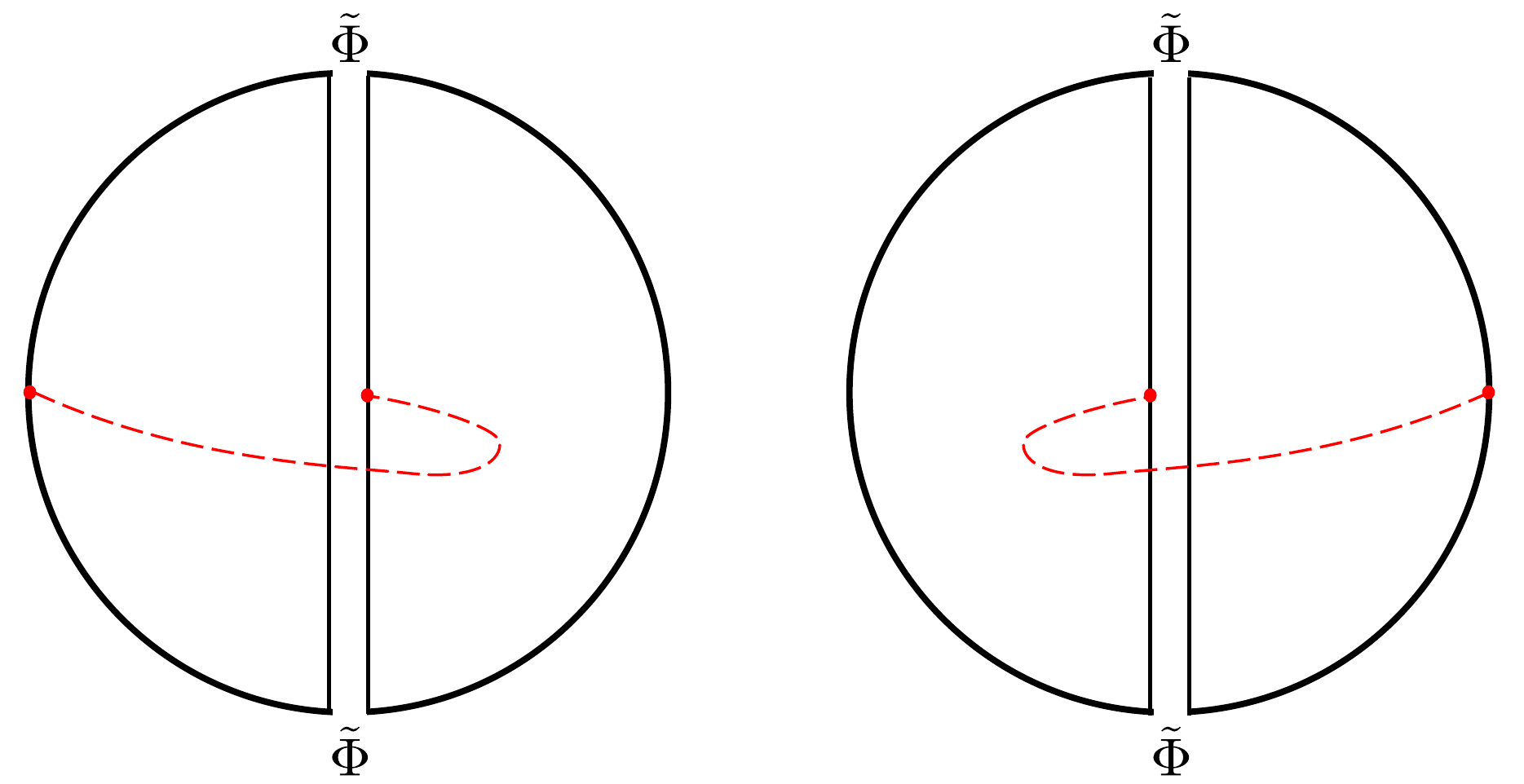}
\caption{The non-planar correction to the two-point function of length $1$ operators at one loop. As compared to the planar correlator, the gluon is attached to the propagator from the opposite side and this results in an extra minus sign.}
\label{fig:figd3}
\end{figure}
We now turn to the non-planar correction. Also at the non-planar level, the same four diagrams show up at one loop and we expect\fn{We checked this for a few cases. Showing this in general requires more careful analysis which we leave for furture investigation. Instead here we proceed assuming that it is true. } that the finite piece comes purely from the gluon exchanges between the Wilson loop and a propagator. The main difference from the planar case is that at the non-planar level the number of such diagrams depends on the length of the operator.

Let us now examine a few cases explicitly. For the length $1$ operator, there are two such diagrams as shown in figure \ref{fig:figd3}. As compared to the planar diagrams, the gluon is attached to the propagator from the opposite side and this produces an extra minus sign owing to the anti-symmetry of the structure constant of the gauge group, $f_{abc}$. We therefore conclude that the leading non-planar correction at one loop for the length 1 operator is given by
\beq
({\tt nonplanar})|_{L=1}=-({\tt planar})=+\frac{2\pi^2g^2}{3}\period
\eeq
Multiplying this number to the planar tree-level answer, we reproduce the result in \eqref{eq:nonplanar2ptlength1}.

\begin{figure}[t]
\centering
\includegraphics[clip,height=7cm]{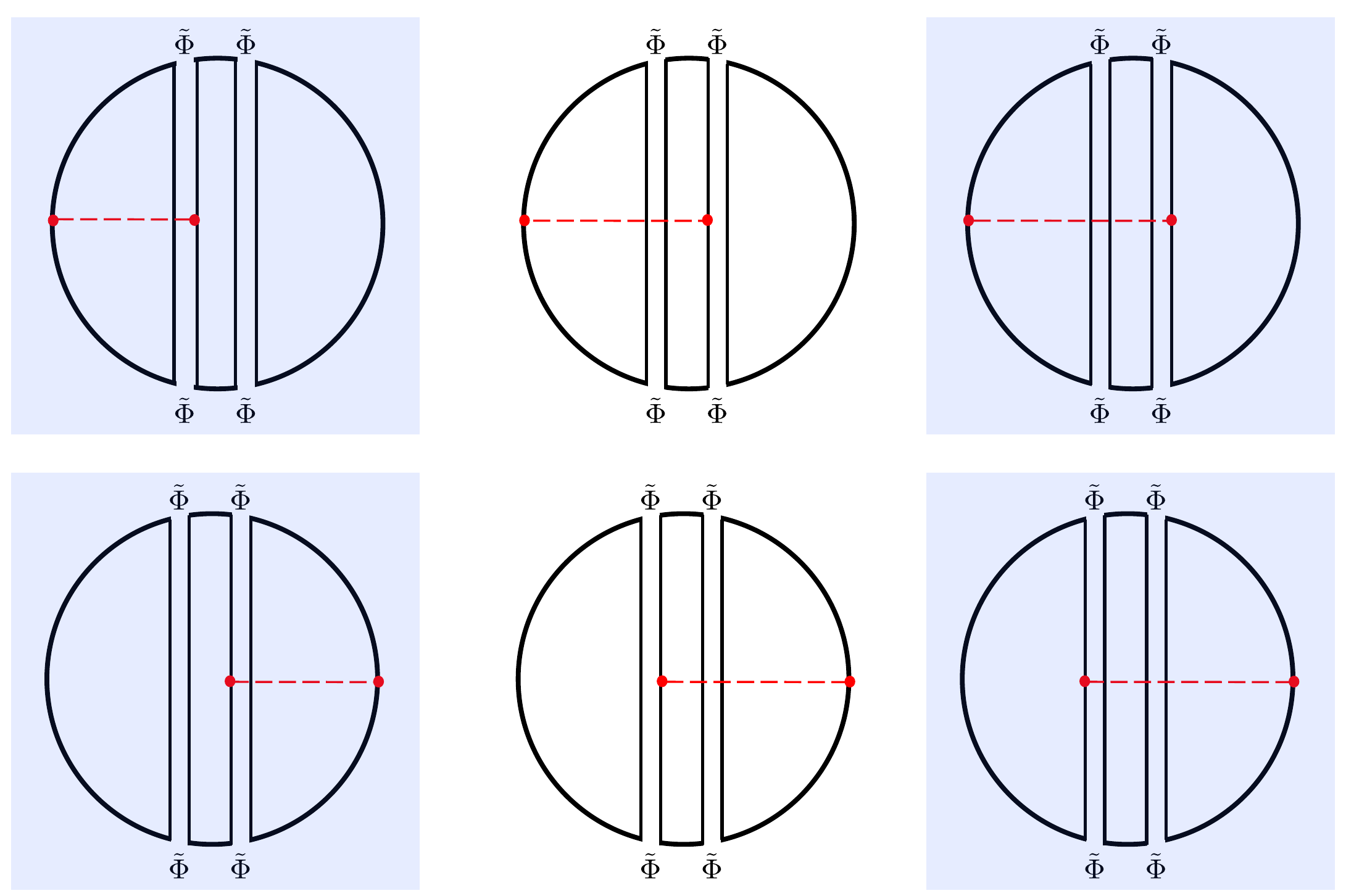}
\caption{The non-planar correction to the planar Wick contractions for the length-2 operators. The diagrams depicted in the shaded blue regions come with extra minus signs (as compared to the planar contribution). The result sums up to $-({\tt planar})$}
\label{fig:figd4}
\end{figure}
\begin{figure}[h]
\centering
\includegraphics[clip,height=5cm]{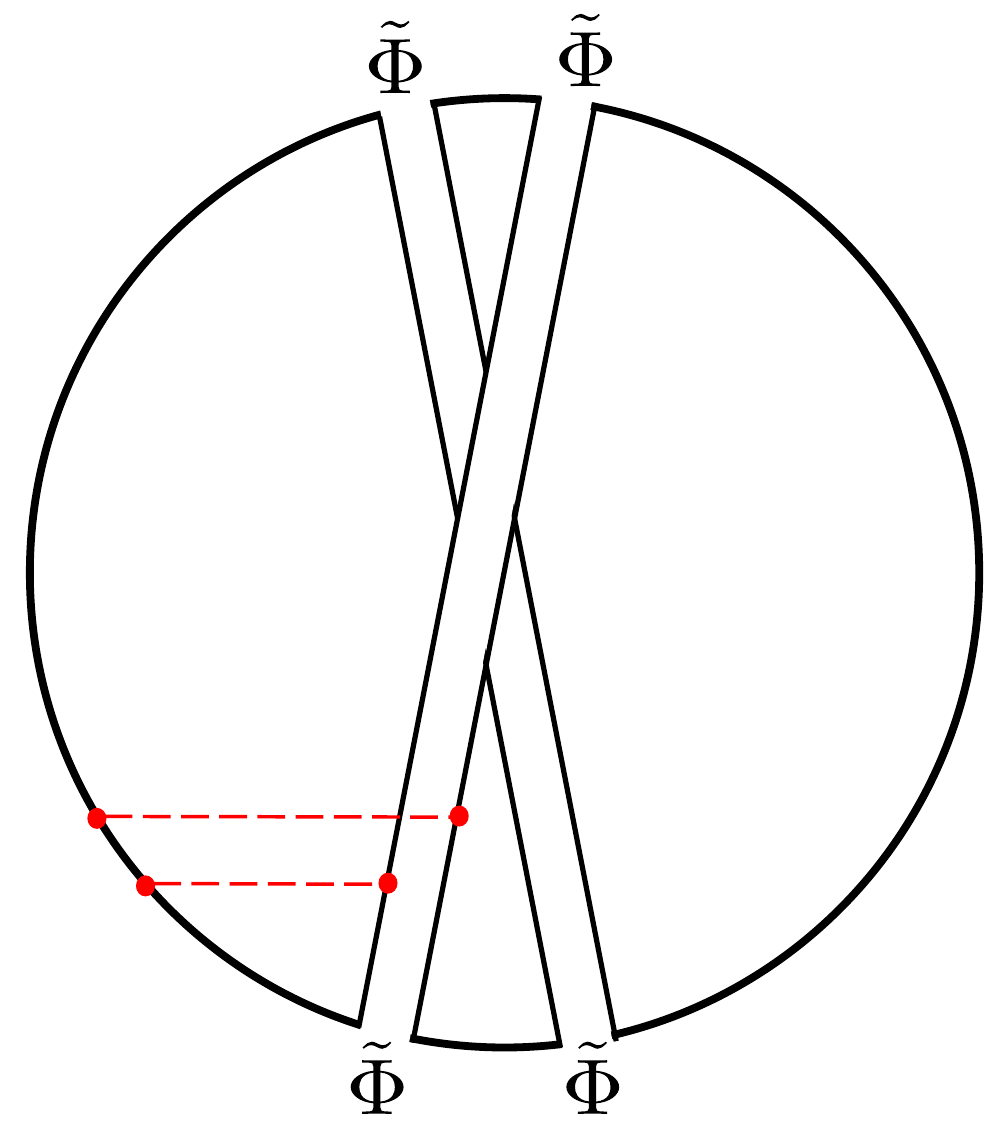}
\caption{The gluon exchange diagrams for the non-planar Wick contractions of the length-$2$ operators. For each propagator, one can attach a gluon from either side and the contributions cancel out with each other.}
\label{fig:figd5}
\end{figure}
\begin{figure}[h]
\centering
\includegraphics[clip,height=7cm]{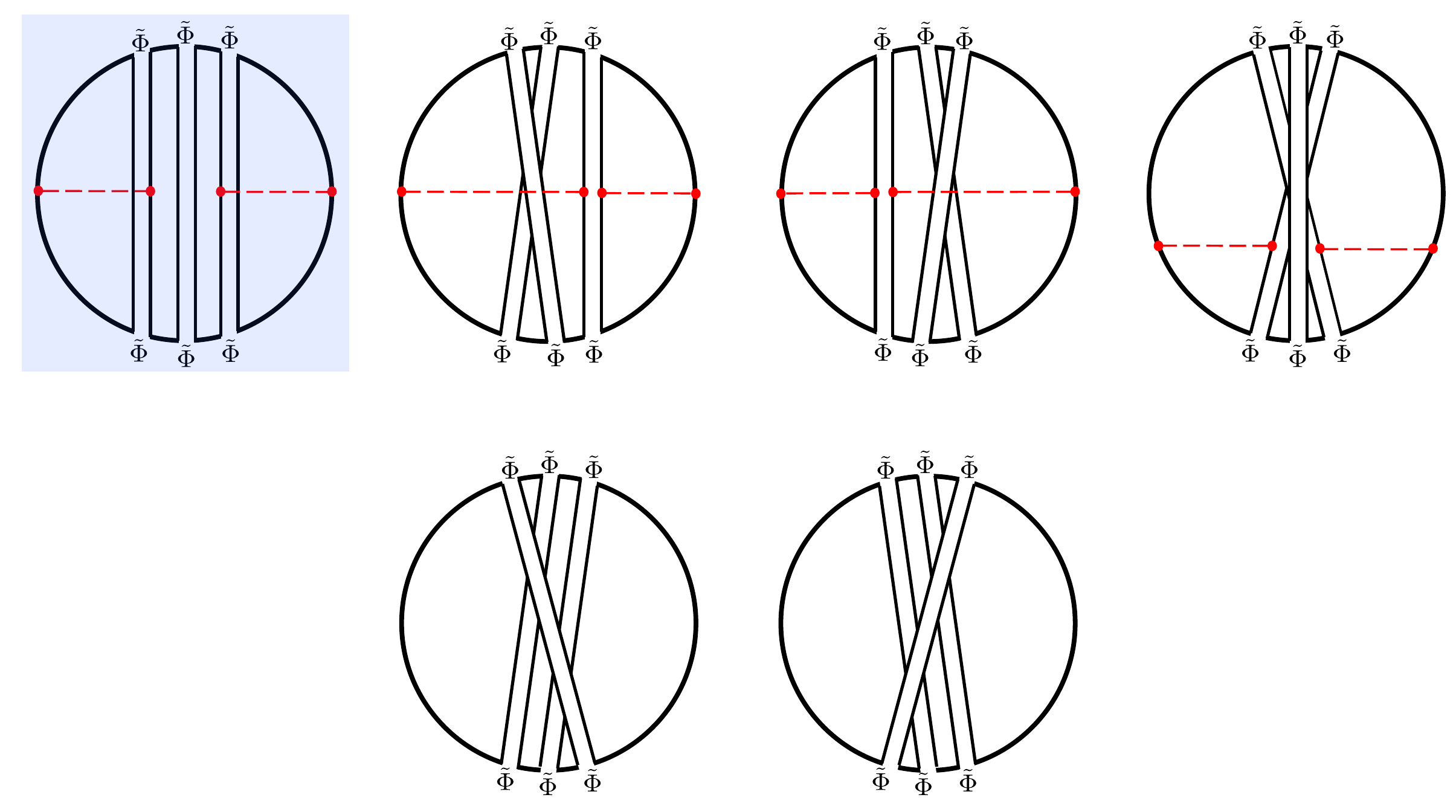}
\caption{Non-planar corrections to the two-point function of the length-$3$ operators. For each diagram, we showed the gluon exchange diagrams that are not cancelled. (We drew several gluon exchanges in one diagram but they should be understood as separate diagrams.) The first diagram comes from the planar Wick contraction and comes with a minus sign. The last two diagrams are X-shaped diagrams whose contributions add up to zero.}
\label{fig:figd6}
\end{figure}

For the length $2$ operators, there are two tree-level Wick contractions: The one is planar and the other is non-planar. For the planar Wick contraction, there are six different ways to attach a gluon propagator which makes the full diagram non-planar (see figure \ref{fig:figd4}). Taking into account the minus signs coming from the anti-symmetry of $f_{abc}$, we conclude that they sum up to $-({\tt planar})$. On the other hand, for the non-planar Wick contractions, one can attach a gluon propagator from either side of a scalar propagator and they completely cancel each other (see figure \ref{fig:figd5}). Therefore the final result is given by
\beq
({\tt nonplanar})|_{L=2}=-({\tt planar})=+\frac{2\pi^2g^2}{3}\comma
\eeq
which matches with \eqref{eq:nonplanar2ptlength2}.

Let us also analyze the length $3$ case.  At length $3$, there are one planar Wick contraction and five non-planar Wick contractions. For the planar Wick contraction, there are ten different ways to attach a gluon which makes the diagram non-planar. However, most of them cancel each other due to the minus signs coming from $f_{abc}$ leaving the two diagrams drawn in figure \ref{fig:figd6}, and we again obtain $-({\tt planar})$ in the end. Now, among the non-planar Wick contractions, two of them have the same topology (``X-shape'') as the non-planar contraction for the length $2$ operator and therefore their contributions add up to zero. For the remaining three diagrams, the cancellation is not complete and we are left with the gluon exchanges drawn in figure \ref{fig:figd6}. As a result, each of these three diagrams produces the same contribution as the planar diagram, namely $+({\tt planar})$. Summing up all, we finally obtain
\beq
({\tt nonplanar})|_{L=2}=(3-1)({\tt planar})=+2({\tt planar})=-\frac{4\pi^2g^2}{3}\comma
\eeq
which is again consistent with what we got in \eqref{eq:nonplanar2ptlength2}.
 
 By inspecting a few more cases, one can conclude that the planar Wick contraction always produces $-({\tt planar})$ at the non-planar level while the non-planar contraction gives $+({\tt planar})$ unless the diagram is of the ``X-shape'' (in which case the contribution is zero). This leads to the following formula for the non-planar correction at one loop:
 \beq
 \begin{aligned}
 ({\tt nonplanar})|_{L}=&\left[(\text{\# of non-planar contractions})\right.\\
 &\left.-(\text{\# of X-shaped non-planar contractions})-1\right]\times ({\tt planar})\period
 \end{aligned}
 \eeq 
As computed in section \ref{subsec:nonplanarweak}, the number of non-planar Wick contractions is $\left(\begin{array}{c}L+2\\4\end{array}\right)$. On the other hand, the number of X-shaped non-planar contractions is $L-1$ since such contractions are specified by a point inside the operator at which we break the operator into two and reconnect. Therefore, we obtain
 \beq
 ({\tt nonplanar})|_{L}=\left[\left(\begin{array}{c}L+2\\4\end{array}\right)-L\right]({\tt planar})=-\frac{2\pi^2g^2}{3}\left[\left(\begin{array}{c}L+2\\4\end{array}\right)-L\right]\period
 \eeq 
 This matches with the result we computed from localization \eqref{eq:generalplanarnonplanar}.
\newpage
\bibliographystyle{utphys}
\bibliography{bulkdefectref}
\end{document}